# *On theta function expressions of cyclic products of fermion correlation functions in genus two*

A.G. Tsuchiya

Abstract

In e-Print 2211.09069, significant progress was made in decomposing simple products of fermion correlation functions, and in summing over spin structures of superstring amplitudes in genus two under cyclic constraints.

In this manuscript we consider part of the same subject using a framework in which one of the branch points of the genus two curve is fixed at infinity. This framework is a direct generalization of the popular one in the case of genus one. We address some of the issues that remained unresolved in our previous paper e-Print 2209.14633.

We show that the spin structures of the simple products of fermion correlation functions with cyclic conditions depend only on the Pe-function values at the half-periods of the genus two surface, for any number of factors in the products. Similar to the genus one case, we can provide basis functions to decompose the product. Consequently, the trilinear relations found in e-Print 2211.09069 can be derived from the known set of differential equations of genus two Pe-functions by simply setting the variables equal to the half-periods of the non-singular and even spin structures, as is the case for genus one.

Based on these considerations, we present a procedure for expressing the results of decomposed formulae in terms of the unique genus two theta function. However, it should be noted that a general formula for the expression of the results in terms of the theta function for the product of an arbitrary number of the fermion correlation functions is not yet derivable, as outlined in the text. It is demonstrated that concrete expressions of both the spin structure dependent and independent parts can be derived and simplified for analysis using the logic of the derivations of the classical solutions to the Jacobi inversion problem and their modifications, which are given in this manuscript. The present study involves explicit theta function calculations of two, three, and four point functions, and discusses a realistic approach for calculating six point function.

## 1. Introduction

This manuscript focuses on the calculations of fermion field contributions to $N$-point scattering amplitudes in 2-loop (genus 2) in superstring theories. Throughout the paper, the superstring theory model is either Type I or Type II, and the external states are massless bosons. The RNS formalism is used for the calculations. Only the parity-even part is considered. The corresponding vertex operator is

$$V(\zeta,k) = ( \zeta \cdot \dot{X} - \zeta \cdot \psi \ k \cdot \psi ) \, e^{k \cdot X} \tag{1.1}$$

where $\zeta$ and $k$ are the polarization and momentum vectors, respectively, and $X$ and $\psi$ are the boson and fermion fields.

At the one-loop (genus one, torus) level, it was known that using the RNS formalism for amplitude calculations clarifies and simplifies some of its properties. For example, K. Amano investigated the divergence of the mass shift of massive states ( two-point massive amplitudes) in the Type II model, calculated using the RNS formalism in[1] in 1987, and discovered independently of [2]and [3] that the on-shell massless four-point amplitudes of the Type II closed superstring model generally diverge unless appropriate analytic continuations are performed with respect to external momentum, even when the integration area is restricted to the fundamental regions. He then provided a prescription to make them finite. His results are summarized in his paper [4].

Also, independently of [5,6], K. Amano and others also derived that, in closed strings, the right-movers of the boson fields and the left-movers have non-zero Wick contractions[1,7].

Consequently, there are various advantages to studying the properties of string amplitudes using the RNS formalism. Conversely, it has been suggested that a disadvantage of the RNS formalism is the need to perform a complicated sum over the spin structures for large values of $N$. This issue at the one loop level in the RNS formalism can be summarized as follows.

When calculating the trace of the mode expansions of the string fields in the vertex operators, the correlation functions on the torus naturally emerge. For example, for the fermion field, the following three types of function emerge:

$$S_\delta(\beta) = \frac{{\theta_1}^{(1)}(0)\theta_{\delta+1}(\beta)}{\theta_{\delta+1}(0)\theta_1(\beta)} \tag{1.2}$$

for $\delta = 1,2,3$ . The notations are summarized in Appendix A. The term “spin structure dependence” is used frequently below, but essentially refers to the $\delta$ dependence ( boundary condition dependence) of a function.

When calculating the amplitudes, the result of the Wick contractions as the contributions of the fermion fields appears as the sum of the following simple products of the correlation functions：

$(contractions\ of\ external\ momentum\ and\ polarization\ vectors) * \prod_{i=1}^{L} S_\delta(z_{i+1}-z_i)$

where $L$ is an integer taking the values $L = 2,3, \ldots . N$. In what follows, we will often denote this $L$ as $N$ for simplicity. In this expression, $z_i$ are the insertion points of the vertex operators.

The overall structure of the amplitude contributed by the fermion fields is obtained by multiplying the ghost and super-ghost contributions under the GSO projection to the above simple products, and taking the sum over all even spin structures. In all of spin sum processes, the contractions of the boson field part remain unchanged, and these boson field contributions are not discussed in this paper. The contributions from ghosts and super-ghosts are referred to in this paper as partition functions, which depend only on the moduli of the Riemann surface and not on the vertex insertion points. We need to perform a spin sum over each simple product of the fermion correlation functions.

In the case of free fermions, the original Fay's formula relates the determinant of the fermion correlation function to the bosonized form of the correlation functions. However, this and any other identities do not provide a transparent method of calculating the actual string amplitudes for large $N$ at the one-loop level, since each simple product of fermion correlation functions is multiplied by very complicated contractions of momentum and polarization vectors.

Initially, constructing a general method of performing the spin sum for all cases seemed very difficult. However, an important exception exists in the calculation of superstring scattering amplitudes, and the external massless boson amplitudes in the Type I and Type II we are discussing belong to this exception.

In genus one calculations of external massless boson amplitudes, a Wick contraction expressed as a simple product, $\prod_{i=1}^{N} S_\delta(z_{i+1}-z_i)$ , is a function of the differences of the vertex insertion points $z_{i+1}-z_i \ \ for\ \ i = 1,2, \ldots N$ . This gives zero contribution if the variables do not satisfy the following cyclic condition： $z_{N+1} = z_1$ . The same situation occurs in the calculation of the Kac-Moody algebra in one-loop in the heterotic string theory.

We define $\beta_1 \equiv z_1 - z_2, \quad \beta_2 \equiv z_3 - z_2. \quad \ldots., \ \beta_N \equiv z_N - z_1$ , and hence

$$\sum_{i=1}^{N} \beta_i = 0. \tag{1.3}$$

This observation is implicitly used in [7] and explicitly described in the first version of [8]. As noted in eq.(1.4) of [8], the following decomposition is possible for the simple product with a cyclic condition for any $N$, which was given in the Appendix of [7]:

$$\prod_{i=1}^{N} S_\delta(z_{i+1}-z_i) = \sum[elliptic\ functions\ of\ z_1,\dots,z_N,\ V] \times [polynomials\ of\ the\ branch\ points\ e_\delta\ ] \qquad (1.4)$$

In such cases, the spin structure sum can be performed using only the simple algebraic summation of branch points on the curve, after multiplying the partition functions on the amplitudes for both genus 1 and genus 2. We will see some examples of this later. This decomposition should be performed one step before taking the spin sum of the amplitudes.

Throughout this paper, we refer to the boundary condition-independent parts (or factors) "$elliptic\ functions\ of\ z_1,\dots,z_N,$" in eq.(1.4) as spin structure-independent parts (or factors), and the boundary condition-dependent parts "$polynomials\ of\ the\ branch\ points\ e_\delta$" as spin structure-dependent parts. We also refer to terms containing only spin structure-independent parts as spin structure independent terms.

Initially, in the case of genus one, the elliptic function parts in eq.(1.4) , which were obtained in the Appendix of [7] and represent the spin structure independent part for arbitrary $N$,were given as ratios of two determinants of elliptic functions of $z_1,\dots,z_N$. However, these results were inconvenient for practical use and had to be rewritten as polynomials of derivatives of theta functions. Moreover, for $N>7$, the pole structure considerations used in [7] to rewrite the spin structure-independent parts as polynomials of elliptic functions were incomplete.

For genus one, a useful and simple method of obtaining polynomials of elliptic functions of $z_1,\dots,z_N$ is already known, based on contributions from many people.

We use

$$\frac{\theta_1(\beta+\Omega_\delta)}{\theta_1(\Omega_\delta)} = \exp(-\pi i\, a_\delta\, \beta)\,\frac{\theta_{\delta+1}(\beta)}{\theta_{\delta+1}(0)} \qquad (1.5)$$

where the variables $\Omega_\delta$ are the half periods at the genus 1 defined as :

$$\Omega_1 = \frac{1}{2},\quad \Omega_2 = -\frac{1+\tau}{2},\quad \Omega_3 = \frac{\tau}{2} \quad . \qquad (1.6)$$

And $a_\delta$ are the following constants:

$$a_1 = 0,\quad a_2 = -1,\quad a_3 = 1 \quad . \qquad (1.7)$$

Then,

$$\frac{\theta_1^{(1)}(0)\theta_1(\beta+\Omega_\delta,\tau)}{\theta_1(\beta,\tau)\theta_1(\Omega_\delta,\tau)} = \exp(-\pi i\, a_\delta\, \beta)\,\frac{\theta_1{}^{(1)}(0)\theta_{\delta+1}(\beta)}{\theta_{\delta+1}(0)\theta_1(\beta)} \qquad (1.8)$$

By replacing $\Omega_\delta$ with an arbitrary parameter $\alpha \in C^1$ on the left-hand side of (1.8), we consider the product

$$\Pi_{i=1}^{N} F(\beta_i,\alpha,\tau) = \Pi_{i=1}^{N} \frac{\theta_1^{(1)}(0)\theta_1(\beta_i+\alpha,\tau)}{\theta_1(\beta_i,\tau)\theta_1(\alpha,\tau)} \quad , \tag{1.9}$$

Then we can say that

$$\Pi_{i=1}^{N} S_\delta(\beta_i) = \Pi_{i=1}^{N} F(\beta_i,\Omega_\delta,\tau) = \Pi_{i=1}^{N} \frac{\theta_1^{(1)}(0)\theta_1(\beta_i+\Omega_\delta,\tau)}{\theta_1(\beta_i,\tau)\theta_1(\Omega_\delta,\tau)} \tag{1.10}$$

because under the cyclic condition (1.3) the exponential factor in (1.8) becomes 1.
The product $\Pi_{i=1}^{N} F(\beta_i,\alpha,\tau)$ can be regarded as an elliptic function of the parameter of $\alpha$ because of the condition $\sum_{i=1}^{N}\beta_i = 0$. Each factor $\frac{\theta_1^{(1)}(0)\theta_1(\beta_i+\alpha,\tau)}{\theta_1(\beta_i,\tau)\theta_1(\alpha,\tau)}$ has the form of a Kronecker-Eisenstein series, which is investigated in [9] . Dolan-Goddard [10] found an accurate method of expressing the product $\Pi_{i=1}^{N} \frac{\theta_1^{(1)}(0)\theta_1(\beta_i+\alpha,\tau)}{\theta_1(\beta_i,\tau)\theta_1(\alpha,\tau)}$ in Laurant expansion form by correctly incorporating the effect of the factor $\theta_1(\alpha,\tau)$ in the denominator of the product. However, to efficiently calculate the spin sum in superstring amplitudes for arbitrary $N$ at genus 1, it is more convenient to slightly modify their method as follows.

Since setting the parameter $\alpha$ equal to the half-period $\Omega_\delta$ at the end of the decomposition calculations before summing over spin structures is a natural procedure, we expand the product $\Pi_{i=1}^{N} F(\beta_i,\alpha,\tau)$ in (1.9) by the derivatives of Weierstrass Pe function $P(\alpha)$. Using the set of total $N$ number of basis functions $1, P(\alpha), P^{(1)}(\alpha), P^{(2)}(\alpha), \dots P^{(N-2)}(\alpha)$ in the form of a finite sum, we can write the product as

$$\Pi_{i=1}^{N} F(\beta_i,\alpha,\tau) = \sum_{M=0}^{N} h_{N,M}(\beta_i) P^{(N-2-M)}(\alpha) \tag{1.11}$$

where $h_{N,M}(\beta_i)$ are the coefficients to be determined and defined in this equation. The $P^{(k)}(\alpha)$ are derivatives of the Pe function. In eq.(1.11), we define

$$P^{(-2)}(\alpha) \equiv 1,\ P^{(-1)}(\alpha) \equiv 0,\ h_{N,N-1} \equiv 0,\ P^{(0)}(\alpha) \equiv P(\alpha)\,.$$

In genus one, expanding the product $\Pi_{i=1}^{N} \frac{\theta_1^{(1)}(0)\theta_1(\beta_i+\alpha,\tau)}{\theta_1(x_i,\tau)\theta_1(\alpha,\tau)}$ by the Pe function and then setting $\alpha$ equal to the half periods is nearly unique procedure for transparent spin sum calculations of the amplitudes for large $N$ values. We will try to perform an analogous process for genus 2 in the next chapter. In genus 1, one branch point, $e_\delta$ , corresponds to one even half period $\Omega_\delta \in C^1$, $\delta = 1,2,3$.

In genus 2, the fermion correlation function is written as $S_\delta(z,w) = \frac{\theta[\delta](A(z-w))}{\theta[\delta](0)E(z,w)}$,

where $E(z,w) = \frac{\theta[\nu](A(z-w))}{h_\nu(z)h_\nu(w)}$ is the prime form,

$A(z-w) \equiv (\int_w^z \omega_i) \in C^2$ represents the Abel map, and $\omega_i$ are holomorphic 1-forms, $i = 1,2$.

The vertex insertion points of the external particles (massless bosons) are $z_1, z_2, \ldots\ldots z_N$ .

We consider the product $S_\delta(z_1,z_2)S_\delta(z_2,z_3)\ldots\ldots S_\delta(z_N,z_{N+1}) = \Pi_{i=1}^N \frac{\theta[\delta](A(z_i-z_{i+1}))}{\theta[\delta](0)E(z_i,z_{i+1})}$. The cyclic condition is expressed as the constraint $z_{N+1} = z_1$.

In genus 2, we will use the same variables $z_i$, $\alpha$, $\beta_i$, $\Omega_\delta$ in the genus 1 case for the same meaning, but the last three of these belong to $C^2$ .

The theta function $\theta[\delta](A(z_i - z_{i+1}))$ has non-singular and even characteristics $\delta$. Each of these corresponds to a choice of two branch points $e_i$ , $e_j$ from 5 branch points $e_1, e_2, \ldots e_5$. We define the following product with a $C^2$ parameter $\alpha$ :

$$\Pi_{i=1}^N F_g(u_i,\alpha) = \Pi_{i=1}^N \frac{\theta_R(\beta_i+\alpha)}{\theta_R(\alpha)E(z_i,z_{i+1})} \qquad .$$

and will try to expand it using genus 2 Pe functions. Here, $\theta_R$ is the genus 2 theta function whose characteristics are Riemann constants, the variables $\beta_i$ are defined as $A(z_i - z_{i+1})$, as an analogue of the vertex insertion point differences $z_i - z_{i+1}$ , as defined in the first subsection of section 2-2 , eq.(2.28) :

$$\beta_i = A(z_i - z_{i+1}) = \int_{(x_{i+1},y_{i+1})}^{(x_i,y_i)} \omega = \int_\infty^{(x_i,y_i)} \omega - \int_\infty^{(x_{i+1},y_{i+1})} \omega$$

$\theta_R$ is a natural generalization of the theta function $\theta_1$ in genus 1.

In genus 2, a natural generalization of the half-period corresponding to the non-singular and even spin structure can be given as

$\int_\infty^{(e_i,0)} \omega + \int_\infty^{(e_j,0)} \omega \in C^2$ as in eq.(2.27) , which will be denoted as $\Omega_\delta \in C^2$, for any pair of branch points $e_i$ and $e_j$ out of 5 branch points, giving a total 10 half periods.

A formula similar to eq.(1.8) can also be found in eq.(2.30) :

$$\frac{\theta[\delta](u)}{\theta[\delta](0)E(z,w)} = \exp\left(2\pi i\left(\sum_{k=1}^g a_{i_k}\right)\cdot u\right) \frac{\theta_R(u+\Omega_\delta)}{\theta_R(\Omega_\delta)E(z,w)} \qquad .$$

This was given in eq.(2.14) of [11] as Proposition 1 . In this paper, we will also refer to this equation as Proposition 1.

We would like to expand the product of the genus 2 theta function, the characteristics of which are the Riemann constants $\Pi_{i=1}^N \frac{\theta_R(\beta_i+\alpha)}{\theta_R(\alpha)E(z_i,z_{i+1})}$ by a proper set of basis functions. In this expansion, the half-periods $\Omega_\delta$ are replaced by a parameter

$\alpha \in C^2$.

In Chapter 2, we will see that the classical genus 2 Pe functions can be basis functions of expanding this simple product again, using the method of deriving addition formula for the ratios of sigma functions in the hyper-elliptic case in higher genus.[12,13]

As for the spin structure dependent parts, we can use the similar methods to those in genus one, where the spin sum is reduced to the algebra of branch points for higher values of $N$.

The main focus of this paper is is to clarify how the logic of deriving the modified version of Jacobi inversion theorem can be applied to determine the forms and properties of the spin structure independent parts of the decomposition formula for the simple products of fermion correlation functions in genus 2, in terms of theta functions. It is already known that ( see e.g. [14]), in the hyperelliptic higher genus case, the standard method of proving the Jacobi inversion theorem can be used to obtain the higher-genus generalized formula of $P(\Omega_\delta) = e_\delta$ in the genus 1 case. As is well known, $P(\Omega_\delta) = e_\delta$ can also be interpreted as a result of the inversion theorem in genus one. (see Chapter 5 for an example.) The theorem is applied to the spin structure dependent parts in genus 2 case and a similar analysis can be performed on these parts as in genus one. In this paper, we will prove that a modified (involuted) version of the inversion theorem logic, as described in Chapter 4-2, can be applied to theta function expressions in the decomposed formula for the spin structure-independent parts in genus 2. This result also demonstrates the close correspondence between our findings and the hyperelliptic representation of simple products with cyclic conditions discussed in [15].

Remarkably, for any genus, the decomposition such as (1.4) has been accomplished in [16] via a descent procedure which systematically separates the dependence of the cyclic product of an arbitrary number of fermion correlation functions on the points of the Riemann surface from the dependence on the spin structure.

In [16], a general method for determining the spin structure independent parts that correspond to the $V_{N-2K}$ functions (the coefficients of the expansion of the genus 1 expansion, as shown below ), which includes the full contributions of the moduli of the Riemann surface, has been established.

However, this paper does not establish a general method for determining the spin structure independent parts that correspond to the $V_{N-2K}$ functions in terms of theta functions for an arbitrary number of products of fermion correlation functions by adopting direct generalization of the case of genus one, for the genus 2. This is because the basis functions in the form of Pe-functions we prepare and the calculation method we use are for the periodic functions on the genus 2 Jacobian variety, rather than functions restricted to the genus 2 Riemann surface embedded in the Jacobian variety. In Chapter 4, we demonstrate a direct generalization of the genus one calculation method combined with an established result given in [15] that can be used to obtain spin structure independent parts in terms of theta functions for $N = 2,3,4$ . This method can be applied to calculate the $N = 6$ case as described there.

As for the genus 1 case, the result of the decomposition expressed by the unique odd theta function $\ln\theta_1$ under the cyclic condition is given by ( eq.(1.11), (1.12), (1.13) of [11] )

$$\prod_{i=1}^{N} S_\delta(\beta_i) = V_N - \sum_{K=2}^{\left[\frac{N}{2}\right]} G_{2K}\cdot V_{N-2K} + \sum_{K=1}^{\left[\frac{N}{2}\right]} \frac{1}{(2K-1)!} P^{(2K-2)}(\Omega_\delta) V_{N-2K} \tag{1.12}$$

where

$$\beta_i = z_i - z_{i+1},$$

$$V_p(\beta_1,\beta_2,\ldots\ldots,\beta_N) \equiv \frac{1}{p!}\frac{\partial^p}{\partial\alpha^p}\exp\Big(\sum_{i=1}^{N}\ln\theta_1(\beta_i+\alpha) - \sum_{i=1}^{N}\ln\theta_1(\beta_i) + N\sum_{k=1}^{\infty}\frac{\alpha^{2k}}{2k}G_{2k}(\tau)\Big)\Big|_{\alpha=0} \tag{1.13}$$

and $V_p \equiv V_p(\beta_1,\beta_2,\ldots\ldots,\beta_N)$

Here we used the notations

$$P^{(n)}(\alpha) \equiv \frac{\partial^n}{\partial\alpha^n}P(\alpha), \qquad P^{(0)}(\alpha) \equiv P(\alpha),\ V_0(\beta_1,\beta_2,\ldots\ldots,\beta_N) \equiv 1$$

and $G_{2k}(\tau)$ are the holomorphic Eisenstein series.

The terms $N\sum_{k=1}^{\infty}\frac{\alpha^{2k}}{2k}G_{2k}(\tau)$ in the right-hand side of eq.(1.13) originate from the factors $\prod\theta_1(\alpha,\tau)$ in the denominator of the product $\Pi_{i=1}^{N}\frac{\theta_1^{(1)}(0)\theta_1(\beta_i+\alpha,\tau)}{\theta_1(\beta_i,\tau)\theta_1(\alpha,\tau)}$ which Dolan-Goddard noticed to give a precise Laurant expansion of this product. In eq.(1.12), $V_{N-2K}$ is exactly the same as Dolan-Goddard's result for the coefficients $H_{N,N-2K}$ given in [10]. On each of $V_{N-2K}$ , a spin structure dependent factor $\frac{1}{(2K-1)!}P^{(2K-2)}(\Omega_\delta)$ is multiplied. Eq.(1.12) was proved in [17], but a much simpler and better proof is given in Chapter 1 of [11]. As described there, (1.12) can also be expressed as (1.14) below as follows :

The spin structure independent terms $\sum_{K=2}^{\left[\frac{N}{2}\right]} G_{2K}\cdot V_{N-2K}$ on the right-hand side of eq.(1.12) appear naturally in the proof of eq.(1.12).

When we write the right-hand side of (1.12) as

$$V_N + P(\Omega_\delta)\cdot V_{N-2} + \sum_{K=2}^{\left[\frac{N}{2}\right]} \frac{1}{(2K-1)!}\left[P^{(2K-2)}(\Omega_\delta) - (2K-1)!\,G_{2K}\right]\cdot V_{N-2K}$$

and see the formula

$$P^{(2n-2)}(\alpha) = \frac{(2n-1)!}{\alpha^{2n}} + (2n-1)!\,G_{2n}(\tau) + O(\alpha)\ldots\ldots,$$

we see that the terms $\sum_{K=2}^{\left[\frac{N}{2}\right]} G_{2K}\cdot V_{N-2K}$ and the "$\alpha$-independent terms" $(2n-1)!\,G_{2n}(\tau)$ in the expansions of $\sum_{K=1}^{\left[\frac{N}{2}\right]} \frac{1}{(2K-1)!} P^{(2K-2)}(\alpha) V_{N-2K}$ (before setting $\alpha$ equal to $\Omega_\delta$) all cancel each other out.

Then, under the cyclic condition, we can write the decomposition of the product $\Pi_{i=1}^{N} S_\delta(z_i - z_{i+1})$ in a quite simple form, as described in eqs.(2.52) and (2.53) of [11]:

$$\Pi_{i=1}^{N} S_\delta(z_i - z_{i+1}) = V_N - \sum_{K=1}^{\left[\frac{N}{2}\right]} \frac{1}{(2K-1)!} V_{N-2K}\cdot D^{2K}\ln\theta_1(\Omega_\delta) \qquad (1.14)$$

where

$$D^{2K}\ln\theta_1(\Omega_\delta) \equiv \frac{\partial^{2K}}{\partial\alpha^{2K}}\ln\theta_1(\alpha\ ,\tau)\Big|_{\alpha=\Omega_\delta} + (2K-1)!\,G_{2K} \qquad (1.15)$$

( $G_2$ is regularized as $G_2 = 2\eta_1$ )

$D^{2K}$ means that it cancels the $\alpha$-independent term from the expansion form of $\frac{\partial^{2K}}{\partial\alpha^{2K}}\ln\theta_1(\alpha\,,\tau)$ for $K > 0$ . For $K = 1$ , it cancels the " bad behavior constant" $G_2(= 2\eta_1)$ from the expansion form of $\frac{\partial^2}{\partial\alpha^2}\ln\theta_1(\alpha\,,\tau)$ , which produces $-P(\alpha)$ .

Interestingly, not only the spin structure dependent parts, but also all the spin structure independent parts $V_{N-2K}$ , can be written using $D^k$ only, as in eqs.(1.35), (1.41) in [11] :

$$V_p(\beta_1,\beta_2,\ldots\ldots,\beta_N) = \frac{1}{p!}\frac{\partial^p}{\partial\alpha^p}\exp\Big(\sum_{k=1}^{\infty}\frac{\alpha^k}{k!}\sum_{i=1}^{N} D^k\ \ln\theta_1(\beta_i)\Big)\Big|_{\alpha=0} \qquad (1.16)$$

where $G_k = 0$ if $k$ is an odd number in the case of spin structure independent part. The Dolan-Goddard factor $[\,\theta_1(\alpha,\tau)]^N$ in the denominator of the product $\Pi_{i=1}^{N}\frac{\theta_1^{(1)}(0)\theta_1(\beta_i+\alpha,\tau)}{\theta_1(\beta_i,\tau)\theta_1(\alpha,\tau)}$ produces the term $(k-1)!\,G_k$ inside the factor $D^k\ln\theta_1(\beta_i)$ in (1.16).

In the external massless boson amplitudes of Type I and II theories, the spin-structure independent terms will all be zero after the spin sum. However, in the heterotic string amplitudes, due to the forms of partition functions, the terms $\sum_{K=2}^{\left[\frac{N}{2}\right]} G_{2K} \cdot V_{N-2K}$ give non-zero contributions even to four-point massless boson amplitudes.

As previously mentioned, for spin sums where the periodic condition is satisfied, calculations are performed using the algebra of branch points in the part that depends on moduli only. Define

$$f(K) \equiv \frac{(e_2-e_3)e_1^K+(e_3-e_1)e_2^K+(e_1-e_2)e_3^K}{(e_1-e_2)(e_2-e_3)(e_3-e_1)} \tag{1.17}$$

where $K$ is a non-negative integer.

The trivial identity $f(0)=0$ plays an equivalent role to Jacobi's formula $(\theta_3(0))^4=(\theta_2(0))^4+(\theta_4(0))^4$. This identity as well as $f(1)=0$ provides a proof of the non-renormalization theorem which states that the zero, two, three point amplitudes of the external massless bosons vanish for Type I and II string theories. $f(2)=1$ is sufficient to sum over four-point Type I and II superstring amplitudes of massless external bosons. $f(3)=e_1+e_2+e_3=0$ leads to the conclusion that, for $N=4,5,6,7$ point amplitudes of the external massless bosons of Type I and II superstrings, the results of the branch point algebra in the spin sum calculations, that is $f(K)$, give only $C$ numbers while the spin structure independent parts $V_{N-2K}$ in (1.12) are left unchanged. Therefore, they do not introduce new holomorphic Eisenstein series into the integrand of the amplitudes. For $K \geq 4$, $f(K)$, which corresponds to the external massless bosons amplitudes for $N>7$, the results of the spin sum gives polynomials of $g_2$ and $g_3$, which can be converted to polynomials of holomorphic Eisenstein series.

We can derive a recurrence formula for $f(K)$ for any integer $K \geq 2$ as

$$f(K+1)=15G_4\, f(K-1)+35G_6\, f(K-2) \qquad \left(=\frac{g_2}{4}\, f(K-1)+\frac{g_3}{4}\, f(K-2)\right) \tag{1.18}$$

In Appendix A , we list some of the results of the spin structure sum for the external massless states of Type I and II superstring theories. Calculations involving rewriting $P^{(2K-2)}(\Omega_\delta)$ into polynomials of $e_\delta$ become cumbersome for large values of $K$ .

In genus 2, as one of the remarkable results of [15], trilinear relations are found that reduce the spin structure dependence of a cyclic product of fermion correlation functions to that of the four-point function. It is also noted that trilinear relations at genus 1 can be obtained by setting the variables of a fundamental differential equation of Pe function to half periods [18]. It is therefore expected that, in genus 2, trilinear relations in our notation can be derived by setting the variables $\alpha$ inside genus 2 Pe functions equal to half periods of even spin structures $\Omega_\delta$. Chapter 3 demonstrates that this is indeed the case in a natural way. The differential equations of the Pe functions on a hyperelliptic surface have been studied in detail and the necessary and sufficient number of equations required to derive the 10 trilinear relations are known.

In the next chapters, we will see some similarities between the genus 1 and the genus 2 cases, but there are also differences.

In genus one, as is well known, the theta constants and the coefficients of the curve can be expressed using the holomorphic Eisenstein series, which makes the argument simple and transparent. In genus 2, such expressions do not exist at least not yet. The poles and zeros in theta function expressions in genus 2 have to be discussed using the language of divisors, and a naïve generalization of the genus 1 analysis does not work. This is related to the fact that the infinite product representation of sigma function does not exist in genus 2.

In genus 1, the curve is defined as

$$y^2 = 4\prod_{k=1}^{3}(x-e_k) = R(x) = 4x^3+\mu_1 x^2+\mu_2 x^1+\mu_3 \tag{1.19}$$

and the following relations are well-known ：

$$\mu_1 = e_1+e_2+e_3 = 0$$

$$\mu_2 = 4(e_1e_2+e_2e_3+e_3e_1) = -\frac{3}{2}(\pi)^4[\,\theta_2^8(0)+\theta_3^8(0)+\theta_4^8(0)]$$

$$= -(2\pi)^4\Big(\frac{1}{12}+20\sum_{n=1}^{\infty}\frac{n^3q^n}{1-q^n}\Big) = -g_2 = -60G_4$$

$$\mu_3 = -4e_1e_2e_3 = -\frac{4}{27}(\pi)^6[\,\theta_4^4(0)-\theta_2^4(0)][\,2\theta_3^8(0)+\theta_2^4(0)\theta_4^4(0)]$$

$$= -(2\pi)^6\Big(\frac{1}{216}+\frac{7}{3}\sum_{n=1}^{\infty}\frac{n^5q^n}{1-q^n}\Big) = -g_3 = -140G_6$$

$$G_2 = 2\eta_1 = \varsigma(\omega_1) = (2\pi)^2\Big(\frac{1}{48}-\frac{1}{2}\sum_{n=1}^{\infty}\frac{nq^n}{1-q^n}\Big) = -\frac{1}{6}\frac{\theta_1^{(3)}(0)}{\theta_1^{(1)}(0)} \qquad q = \exp(2i\pi\tau)$$

In genus 2, the coefficients $\mu_i$ of the curve are represented by theta constants using Thomae formula. However, it is not possible to represent them using "generalized holomorphic Eisenstein series in genus 2 ". In this paper, we do not attempt to express $\mu_i$ in terms of theta constants; and the coefficients $\mu_i$ will be included in the calculation results.

## 2. Notations and method in genus 2

### 2-1 Notations

In the context of hyperelliptic curves, there exists a set of 2g+2 branch points, which are denoted $e_1, e_2, \ldots\ldots e_{2g+2}$. In genus one, there are four branch points and it is standard to fix the branch point $e_4$ at infinity. In this paper, the value of $e_{2g+2}$ is fixed at ∞. We sometimes use g as the genus of the Riemann surface, and in this paper in almost all cases $g = 2$.

The branch points $e_1, e_2, \ldots\ldots e_6$ are arranged from left to right, and there are cuts between $e_1$ and $e_2$; $e_3$ and $e_4$ ; $e_5$ and $e_6 = \infty$ . (Please see Figure 1 of [19].)

We again note that in genus 2 we will use the same variables $z_i,\ \alpha,\ \beta_i,\ \Omega_\delta$ in the genus 1 case for the same meaning, but $\alpha,\ \beta_i,\ \Omega_\delta$ belong to $C^2$ .

The Jacobi theta function in genus $g$ is defined in a standard notation:

$$\theta\begin{bmatrix} a \\ b \end{bmatrix}(u,\ \Omega) = \sum_{n \in Z^g} \quad \exp\{\, i\pi(n+a)^t T(n+a) + 2\pi i(n+a)(u+b)\} \qquad (2.1)$$

$$u \in C^g \qquad a, b \in \frac{1}{2} Z^g \qquad . \qquad (2.2)$$

The theta function is called odd or even depending on whether the inner product $4a \cdot b$ is even or odd. The matrix $T$ is defined in (2.9) below.

For genus 2, the curve is expressed in this paper as

$$y^2 = \prod_{k=1}^{5}(x - e_k) = R(x) = x^5 + \mu_1 x^4 + \mu_2 x^3 + \ldots + \mu_5 \qquad (2.3)$$

Please note that the coefficient of the term $x^5$ is 1, not 4.

Let $A_1, A_2$ and $B_1, B_2$ be a canonical homology basis. We choose canonical holomorphic differentials of the first kind $\omega_1, \omega_2$ and associated meromorphic differentials of the second kind $r_1, r_2$. . The periods are given as

$$M_{IJ} = \frac{1}{2} \oint_{A_I} \omega_J \qquad \widehat{M_{IJ}} = \frac{1}{2} \oint_{B_I} \omega_J \qquad (2.4)$$

$$\eta_{IJ} = -\frac{1}{2} \oint_{A_I} r_J \qquad \hat{\eta}_{IJ} = -\frac{1}{2} \oint_{B_I} r_J \qquad . \qquad (2.5)$$

In order to utilize some of the results summarized in [14], the same notations (2.4), (2.5) as those in [14] are adopted. Please note that, in both [14] and in this manuscript, the canonical holomorphic differentials of the first kind $\omega_1, \omega_2$ are not generally normalized. However, the variables in the theta functions are always multiplied by $(2M)^{-1}$ where $M$ is a matrix defined in (2.4). ( Please refer to e.g. eq.(II.21), (II.27), (II39) in [14]. )

In the following, for genus 2, the following notations are adopted for the holomorphic differentials of the first kind $\omega_i$ ( $i = 1{,}2$) and the differentials of the second kind $r_i$ ( $i = 1.2$) :

$$\omega_1 = \frac{dx}{2y} \quad , \quad \omega_2 = \frac{xdx}{2y} \tag{2.6}$$

$$r_1 = -(\mu_2 x \ + 2\,\mu_1 x^2 + 3x^3\,)\frac{dx}{2y} \ , \qquad r_2 = -x^2\frac{dx}{2y} \tag{2.7}$$

**(classification of spin structures)**

Consider the abelian image of each of the branch points as

$$\Omega_j \equiv (2M)^{-1}\int_\infty^{(e_j,0)} \omega_I \in \ C^g \ , \tag{2.8}$$

where the upper integration bound $(e_j, 0)$ refers to the set of $x$ and $y$ co-ordinate of the branch point and the lower limit $(\infty,\infty)$ is abbreviated as $\infty$ for ease of notation. This constitutes a fundamental instrument for the consideration of half periods in genus two Riemann surface. It can be demonstrated that each $\Omega_j$ is the $g$ component vector for all of holomorphic one-forms $\omega_I$, $I = 1{,}2{,}\ldots g$. The index $j$ is attached to each of the branch points $e_j$ , $j = 1{,}2{,}\ldots 2g+1$. For $j = 2g+2$ , the integral is zero. The results of the integrals can be expressed as follows :

$$\Omega_j \ = \ (2M)^{-1}\int_\infty^{(e_j,0)} \omega_I \ = \ a_j \ T \ + \ b_j \quad , \quad T \equiv M^{-1}\widehat{M} \tag{2.9}$$

Both of $a_j$ and $b_j$ for each $j$ are vectors of dimension $g$ with all elements either of 0 or $\frac{1}{2}$ . $\Omega_j$ are also dimension $g$ vectors whose components depends on holomorphic 1 forms $\omega_I$ for $I = 1{,}2{,}\ldots g$. Strictly speaking $\Omega$ should be written as $\Omega_j^I$ .However, we will sometimes abbreviate the index $I$ .

In genus 2, we will sometimes write $\Omega_j \ = (\,\Omega_j^1,\, \Omega_j^2) \ \in \ C^2$ ,

where $\Omega_j^1 = \ (2M)^{-1}\int_\infty^{(e_j,0)} \omega_1 \ \in \ C^1$ , $\quad \Omega_j^2 = \ (2M)^{-1}\int_\infty^{(e_j,0)} \omega_2 \ \in \ C^1$ . (2.10)

Please note also that the result of the integral is not written as $b_j \ T \ + \ a_j$ , but as $a_j \ T \ + \ b_j$ .

Since $a_j$ and $b_j$ are often used as characteristics of theta functions, we call the pair even or odd when the inner product $4a \cdot b$ is even or odd.

For example, in the case of $g = 2$, since there are $2g + 2 = 6$ branch points,

$$a_1 = \tfrac{1}{2}(0,\ 0) \ \ b_1 = \tfrac{1}{2}(1,\ 0)\ ; \quad a_2 = \tfrac{1}{2}(1,\ 0) \ b_2 = \tfrac{1}{2}(1,\ 0)\ ; \quad a_3 = \tfrac{1}{2}(1,\ 0) \ \ b_3 = \tfrac{1}{2}(0,\ 1)$$
$$a_4 = \tfrac{1}{2}(1,\ 1) \ \ b_4 = \tfrac{1}{2}(0,\ 1)\ ; \quad a_5 = \tfrac{1}{2}(1,\ 1) \ \ b_5 = \tfrac{1}{2}(0,\ 0)\ ; \quad a_6 = \tfrac{1}{2}(0,\ 0) \ \ b_6 = \tfrac{1}{2}(0,\ 0). \tag{2.11}$$

We are also interested in the following combinations of $\Omega_k$ :

Choose any of combination of 2 indices $i, j$ from the non-zero $2g + 1$ number of $\Omega_k$ in the case of genus $g$, and define

$$\Omega_{ij} = \Omega_i + \Omega_j = \ (a_i \ + a_j)\,T \ + \ (b_i \ + \ b_j\ ) \ . \tag{2.12}$$

Similarly, we define $\Omega_{ijk}$ , which is any three combinations of $\Omega_j$ ,
that is, $(a_i + a_j + a_k)T + (b_i + b_j + b_k)$ .
Also, we define $\Omega_{ijkl}$ , $\Omega_{ijklm}$ , …. .
We define the following sum, choosing only even numbers of $k$ in $\Omega_k$ and summing all of them in genus $g$ :

$$R \equiv \Omega_2 + \Omega_4 + \Omega_6 + \cdots \Omega_{2g} = \left(\sum_{i=1}^{g} a_{2i}\right)T + \left(\sum_{i=1}^{g} b_{2i}\right) . \tag{2.13}$$

$R$ is called the vector of Riemann constants with the base point $\infty$ .
We define $\Delta^a \equiv \sum_{i=1}^{g} a_{2i}$ , $\Delta^b \equiv \sum_{i=1}^{g} b_{2i}$ .
In genus 2, $R$ becomes

$$R = \Omega_2 + \Omega_4 = (a_2 + a_4)T + (b_2 + b_4) = \Delta^a T + \Delta^b \quad \text{and so} \quad \Delta^a = \left(0, \tfrac{1}{2}\right), \Delta^b = \left(\tfrac{1}{2}, \tfrac{1}{2}\right) . \tag{2.14}$$

In the following, the theta function in genus $g$ whose characteristics are Riemann constants will be denoted as $\theta_R(u)$. That is,

$$\theta_R(u) \equiv \theta\begin{bmatrix}\Delta^a \\ \Delta^b\end{bmatrix}(u) . \tag{2.15}$$

The function $\theta_R(u)$ is odd if $\frac{g(g+1)}{2}$ is odd, and even if $\frac{g(g+1)}{2}$ is even, for the genus $g$.

This function $\theta_R(u)$ appears in two different places in the below, one is in the definition of sigma-function and the other is in proposition 1 .
We denote that the characteristic of $\Omega_j$ , that is the set of vectors $a_j$ and $b_j$ , as $[\Omega_j]$. Also, the symbol $[\Omega_i]$, $[\Omega_{ij}]$, …. $[R]$ means their characteristics, a pair of vectors of dimension $g$ .

The characteristics $[\Omega_i]$, $[\Omega_{ij}]$, $[R]$, are not always odd or not always even, as can be checked directly from (2.11), (2.12), (2.13). However, if we do the following genus $g$ dependent procedures then we can say something more.
For example, in genus 2, if we add $R$ to $\Omega_i$ , $\Omega_{i,j}$ then the total of the characteristics will be as follows:
Characteristics $[\Omega_i + R]$, that is the pair of $a_i + \Delta^a$ *and* $b_i + \Delta^b$ : always odd for any value of $i$. (Total 6 cases, including $[R]$ itself which is $[\Omega_{2g+2} + R]$ )
Since $e_{2g+2} = e_6 = \infty$ , the value of any component in $[\Omega_6]$ is zero, so the characteristics are both zero vectors.
Characteristics $[\Omega_{ij} + R]$ : always even for all combinations of $i,j$ ( total $\binom{2g+1}{2} = 10$ cases.)

There are in total $2^{2g} = 16$ spin structures, $\frac{1}{2}(2^{2g} + 2^g) = 10$ are even and

$\frac{1}{2}(2^{2g}-2^{g})=6$ are odd.
In this sense, we say that $\Omega_i + R$ are non-singular and odd half-periods, and $\Omega_{ij} + R$ are non-singular even half periods at genus 2, although $[\Omega_i]$ is not always odd, and $[\Omega_{ij}]$ is not always even. For even characteristics $\delta$, the word non-singular means that the value of the theta function with zero characteristic at zero is not vanishing, $\theta_\delta(0) \neq 0$ .
The focus of this study is exclusively on non-singular even half-periods.
In this context, 10 $\Omega_{ij} + R$ represent the $g=2$ extension of half periods $\Omega_\delta \in C^1$, $\delta = 1,2,3$ in genus 1.

It is customary to denote any one choice of two branch points $i,j$ by $\delta$ .
In the following discussion, $\Omega_\delta$ or $\Omega_{ij}$ will be used to denote the non-singular and even half-periods of genus 2 surface. Whilst the alternative of referring to $\Omega_\delta + R$ as half-periods instead of $\Omega_\delta$ is arguably more accurate, later the latter definition will prove to be more convenient.

The above argument on the classifications of spin structures and theta functions can be extended to any genus g in the hyperelliptic case using the Riemann's theorem. For the details, please see e.g. [19] .

**(definitions of functions )**

The sigma-function $\sigma(u)$ is defined by

$$\sigma(u) = \mathrm{c}\cdot\exp\left(\frac{1}{2}\ {}^{t}u\eta M^{-1}u\right)\frac{\theta_R((2M\ \ )^{-1}u,\tau)}{\partial_{u_1}\theta_R(0,\tau)}\ \ ,\ \ \tau = M^{-1}\widehat{M} \tag{2.16}$$

where $u=(u_1,u_2)$ , an arbitrary $C^2$ variable. $\eta, M$ are the matrices defined in (2.4), (2.5).
The characteristics of the theta function in the definition of the $\sigma(u)$ are Riemann constants, which is defined in eq.(2.15). $\sigma(u)$ is normalized divided by $\partial_{u_1}\theta_R(0,\tau)$.
Please see page7 of [20].
The c is the over-all constant, which we will set equal to +1 here.
This is to make the signature of the calculation results shown in later the same as the results in [15].
The $\zeta$ functions are defined as

$$\zeta_I(u) = +\frac{\partial}{\partial u_I}\ln\sigma(u) \tag{2.17}$$

and the Pe functions are defined as

$$P_{JK}(u) = -\frac{\partial^2}{\partial u_J \partial u_K}\ln\sigma(u) = -\frac{\partial}{\partial u_J}\zeta_K(u),\ \ P_{I_1 I_2 \dots I_N}(u) = \frac{\partial^{N-2}}{\partial u_{I_3}\partial u_{I_4}\dots\partial u_{I_{N1}}}P_{I_1 I_2}(u) \tag{2.18}$$

Also, note "Corollary of the main formula" in [14], II.54

$$\frac{F(x_r,x)+2y_r y}{(x_r-x)^2} = \sum_{i=1}^{g}\sum_{j=1}^{g} P_{ij}\left(\int_{\infty}^{P}\omega - u\right)x_r^{i-1}x^{j-1} \tag{2.19}$$

where $u = \sum_{k=1}^{g}\int_{\infty}^{P_k}\omega$

This notation means that $u$ is a set of $g$ number of $C^1$ variables $u_1 = \sum_{k=1}^{g}\int_{\infty}^{P_k}\omega_1$ and $u_2 = \sum_{k=1}^{g}\int_{\infty}^{P_k}\omega_2$ , ... and write $u = (u_1, u_2, \dots u_g)$.

We used the coordinates $P = (x, y)$ , $P_k = (x_k, y_k)$ in eq.(2.19).

For example, in the genus two, Pe functions in the eq.(2.19) have the following two variables :

$$\int_{\infty}^{P}\omega_1 - \int_{\infty}^{(x_1,y_1)}\omega_1 - \int_{\infty}^{(x_2,y_2)}\omega_1 \quad \text{and} \quad \int_{\infty}^{P}\omega_2 - \int_{\infty}^{(x_1,y_1)}\omega_2 - \int_{\infty}^{(x_2,y_2)}\omega_2$$

In eq.(2.19), the following $F$ is called Kleinian 2-polar, and in the genus 2 it is defined as

$$F(x,z) = (x+z)x^2z^2 + 2\mu_1 x^2z^2 + \mu_2 (x+z)xz + 2\mu_3 xz + \mu_4 (x+z) + 2\mu_5 \tag{2.20}$$

where the coefficients $\mu_i$ ( $i = 1,2,..5$ ) are defined in (2.3) as

$$y^2 = \prod_{k=1}^{5}(x-e_k) = R(x) = x^5 + \mu_1 x^4 + \mu_2 x^3 + \dots + \mu_5 \quad .$$

The Taylor expansion of the sigma function at genus two is known [20] as follows:

$$\sigma(\alpha_1,\alpha_2) = \sum_{\substack{k,l\geq 0\\ k+l=odd}} A_{k,l}\,\alpha_1^k\alpha_2^l$$

$$= \left(+\alpha_1 - \frac{1}{3}\alpha_2^3\right) + \frac{\mu_3}{6}\alpha_1^3 + (A_{5,0}\alpha_1^5 + A_{4,1}\alpha_1^4\alpha_2 + A_{3,2}\alpha_1^3\alpha_2^2 + A_{2,3}\alpha_1^2\alpha_2^3$$

$$+A_{1,4}\alpha_1\alpha_2^4 + A_{0,5}\alpha_2^5) + A_{k+l=7}\ terms\ + \ \dots\dots \tag{2.21}$$

where

$$A_{5,0} = +\frac{(\mu_3^2+2\mu_2\mu_4)}{5!},\ A_{4,1} = -\frac{8\mu_5}{4!},\ A_{3,2} = -\frac{\mu_4}{6},\ A_{2,3} = -\frac{\mu_3}{6},\ A_{1,4} = -\frac{\mu_2}{12},\ A_{0,5} = 0 \tag{2.22}$$

It should be noted that in the expansion forms in (2.20), (2.21),the convention that $\mu_1 = 0$ is adopted, as outlined in [20]. In the following, we sometimes employ general results in which $\mu_1$ is explicitly present. In such cases, it is sufficient to set $\mu_1 = 0$ in the results.

The sum of the first two terms in (2.21) $+\alpha_1 - \frac{1}{3}\alpha_2^3$ which are independent of any $\mu_i$ , is referred to as the Schur – Weierstrass polynomial at genus two.

The Jacobian ( Jacobian variety ) of genus 2 , $J$ , is defined by the following equation :

$$J \equiv C^2/\Lambda \quad \text{where} \quad \Lambda = M Z^2 + \widehat{M}\ Z^2 \ ; \tag{2.23}$$

where $M$ and $\widehat{M}$ are the matrices defined in (2.4).

## 2-2 Method

**( type of variables)**

Although any variables $u \in C^2$ can be used in the theta function (2.1), we define two specific types of variables here. This classification will prove useful later on, especially in the genus 2 case.

We will call a variable $\alpha = (\alpha_1, \alpha_2) \in C^2$ an $\alpha$-type variable in genus 2 if it has the following form :

$$\alpha_1 = \int_\infty^{(X_2,Y_2)} \omega_1 + \int_\infty^{(X_1,Y_1)} \omega_1 , \quad \alpha_2 = \int_\infty^{(X_2,Y_2)} \omega_2 + \int_\infty^{(X_1,Y_1)} \omega_2 \tag{2.24}$$

where $X_i$, $Y_i$ are $x$, $y$ coordinates of any pair of two points on the Riemann surface. Note $X_i \in C^1$, $Y_i \in C^1$.

We will sometimes write

$$\alpha = \int_\infty^{(X_2,Y_2)} \omega + \int_\infty^{(X_1,Y_1)} \omega \tag{2.25}$$

and when a function $f$ depends on $\alpha_1, \alpha_2$ we will write

$$f(\alpha_1, \alpha_2) \text{ or } f(\alpha) \text{ or simply } f\left( \int_\infty^{(X_2,Y_2)} \omega + \int_\infty^{(X_1,Y_1)} \omega \right) .$$

Later we will select two branch points $e_i$, $e_j$ ( each of these selections is denoted as $\delta$ ) out of 5 branch points, and define

$$\alpha_{\delta 1} \equiv \int_\infty^{(e_i,0)} \omega_1 + \int_\infty^{(e_j,0)} \omega_1, \quad \alpha_{\delta 2} \equiv \int_\infty^{(e_i,0)} \omega_2 + \int_\infty^{(e_j,0)} \omega_2 , \tag{2.26}$$

We will always set $\alpha$ equal to $\Omega_\delta$ at the end of the calculations where

$$\Omega_\delta \equiv (\alpha_{\delta 1}, \alpha_{\delta 2}) \tag{2.27}$$

Differentiation of theta function is always with respect to $\alpha_1, \alpha_2$ , which will be denoted as $\partial_{\alpha_1}\theta(\alpha)$ , $\partial_{\alpha_2}\theta(\alpha)$ or simply $\partial_1\theta(\alpha)$ , $\partial_2\theta(\alpha)$ .

It is known that $\alpha = (\alpha_1, \alpha_2)$ is a point on the Jacobian variety.

We will call a variable $\beta = (\beta_1, \beta_2) \in C^2$ an $\beta$-type variable in genus 2 if it has the following form. Since there may be multiple $\beta$-type variables at the same time, we will write them using another index $i$ ,as

$$\beta_{i,1} = \int_\infty^{(x_i,y_i)} \omega_1 - \int_\infty^{(x_{i+1},y_{i+1})} \omega_1 = \int_{(x_{i+1},y_{i+1})}^{(x_i,y_i)} \omega_1 \tag{2.28}$$

$$\beta_{i,2} = \int_\infty^{(x_i,y_i)} \omega_2 - \int_\infty^{(x_{i+1},y_{i+1})} \omega_2 = \int_{(x_{i+1},y_{i+1})}^{(x_i,y_i)} \omega_2 \tag{2.29}$$

where $x_i$, $y_i$ are $x$, $y$ coordinates of a variable $z_i = (x_i, y_i)$ on the Riemann surface. We may sometimes abbreviate the index 1, 2 and write $\beta_i = (\beta_{i,1}, \beta_{i,2})$ :

$$\beta_i = A(z_i - z_{i+1}) = \int_{(x_{i+1},y_{i+1})}^{(x_i,y_i)} \omega = \int_\infty^{(x_i,y_i)} \omega - \int_\infty^{(x_{i+1},y_{i+1})} \omega \tag{2.30}$$

In the $\beta$-type variable, the $z_i$, $z_{i+1}$ are often the two insertion points of the vertex operators.

As in the case of the $\alpha$-type, differentiations on theta functions are always with respect to $\beta_{i,1}, \beta_{i,2}$ , which will be denoted as $\partial_{\beta_{i,1}}\theta(\beta_i)$ , $\partial_{\beta_{i,2}}\theta(\beta_i)$ or simply as $\partial_1\theta(\beta_i)$ , $\partial_2\theta(\beta_i)$ .

$\alpha$-type variables are essentially used in the spin structure-dependent parts, and $\beta$-type variables are used in the spin structure-independent parts of the calculation steps.

In the calculations, the mixture of these two types of variables and any other type will also appear.

**(logic)**

Suppose we choose any $g$ number of branch points $e_{i_1}, e_{i_2}, \dots e_{i_g}$ out of $2g+1$ branch points of arbitrary genus $g$ , with $e_{2g+2}$ fixed at $\infty$, and compute the integral (2.9). This determines one non-singular and even spin structure, which we denote $\delta$ :

$$\Omega_\delta = \Omega_{i_1 i_2 \dots i_g} = \left(\sum_{k=1}^{g} a_{i_k}\right)T + \left(\sum_{k=1}^{g} b_{i_k}\right) . \qquad (2.31)$$

Here $a_{i_k}, b_{i_k}$ are the numbers calculated by (2.6) for the selected branch points $e_{i_1}, e_{i_2}, \dots e_{i_g}$ .

The characteristic $[\, R + \Omega_\delta \,]$ will always be one of the characteristics of non-singular even spin structures.

We showed the following formula holds in eq.(2.14) of .[11] for an arbitrary genus $g$ in hyper-elliptic case and arbitrary variable $\mathrm{u} \in C^g$ , as Proposition 1 :

$$\frac{\theta[\delta](u)}{\theta[\delta](0)E(z,w)} = \exp\left(2\pi i\left(\sum_{k=1}^{g} a_{i_k}\right)\cdot \mathrm{u}\right) \frac{\theta_R(u+\Omega_\delta)}{\theta_R(\Omega_\delta)E(z,w)} . \qquad (2.32)$$

where $E(z,w)$ is the prime form, and $a_i$ are defined in eq.(2.11).

We consider the case that $u$ is of $\beta$ type Abel map. Under the condition $z_{N+1} = z_1$, exp part becomes 1, we have

$$\Pi_{i=1}^{N} S_\delta(z_i, z_{i+1}) = \Pi_{i=1}^{N} \frac{\theta_R(\beta_i + \Omega_\delta)}{\theta_R(\Omega_\delta)E(z_i, z_{i+1})} . \qquad (2.33)$$

where $\beta_i$ are defined in (2.28).

The $x$ and $y$ coordinates of $z_i$ are defined as $x_i, y_i$. Each of $x_i, y_i$ is a c-number.

Setting $g=2$ and replacing $\Omega_\delta$ with the parameter $\alpha$, $\alpha = (\alpha_1, \alpha_2) \in C^2$ , in the similar manner to the genus one case, we have a product $\Pi_{i=1}^{N} \frac{\theta_R(\beta_i+\alpha)}{\theta_R(\alpha)E(z_i, z_{i+1})}$ . We then try to decompose this product by the following logic.

Under the cyclic condition, the product of the prime forms becomes holomorphic one- form. In genus 2, we also set the odd theta function in each of the prime form equal to $\theta_R(\beta_i)$.

In order to make use of some properties of the genus 2 sigma function, we first try to rewrite the product in terms of the sigma function defined in eq.(2.16).

By the explicit form of the Taylor expansion of the sigma function eq.(2.21),

$$\partial_2\,\sigma(0) \equiv \partial_{\alpha_2}\,\sigma(0) = 0 \ . \tag{2.34}$$

$\alpha = 0$ means $\alpha_1 = \alpha_2 = 0$. Since the inside of the exponential function in front of the definition of the sigma function is quadratic, it follows that

$$\partial_2\theta_R(0) = 0 \tag{2.35}$$

Therefore, the product of the prime form $E(z_i, z_{i+1})$ is equal to

$$\prod_{i=1}^{N} E(z_i, z_{i+1}) = \frac{\prod_{i=1}^{N}\theta_R(A(z_i - z_{i+1}))}{(\partial_1\theta_R(0))^N \prod_{i=1}^{N}\omega_1(z_i)} \ . \tag{2.36}$$

Considering the normalization of the sigma function, and using the cyclic condition again, we have

$$\Pi_{i=1}^{N}\frac{\theta_R(A(z_i - z_{i+1}) + \alpha)}{\theta_R(\alpha)E(z_i, z_{i+1})} = \prod_{i=1}^{N}\frac{\sigma(\beta_i + \alpha)}{\sigma(\beta_i)\sigma(\alpha)} \cdot \prod_{i=1}^{N}\omega_1(z_i) \tag{2.37}$$

On the right-hand side of eqs.(2.36) or (2.37), the product of the prime form becomes the product of the holomorphic one-form $\prod_{i=1}^{N}\omega_1(z_i)$ which does not include $\omega_2(z_i)$ due to eq.(2.35).

We will see later how $\omega_2(z_i)$ can be restored by utilizing a modified version of the Jacobi inversion formulae in Chapter 4, and how some of the known results can be reproduced. Even among hyperelliptic curves, only the genus 2 case can have a form as simple as eq.(2.37), in which only one type of holomorphic one-form $\omega_1(z_i)$, is used in the prime form. This can be seen by considering the forms of Schur–Weierstrass polynomials in higher genus.

For a while we will focus on the product of the sigma functions, $\prod_{i=1}^{N}\frac{\sigma(\beta_i + \alpha)}{\sigma(\beta_i)\sigma(\alpha)}$ in (2.37).

We represent the product $\Pi_{i=1}^{N}\frac{\theta_R(\beta_i + \alpha)}{\theta_R(\alpha)E(z_i, z_{i+1})}$ by the product of sigma functions because the sigma function in genus 2 has the following properties for any variable $u \in C^2$ [12] [13] :

1) $$\sigma(u + l) = \chi(l)\sigma(u)\exp L\left(u + \tfrac{1}{2}l, l\right), \tag{2.38}$$

where $l \in C^2$ is a point on Jacobian variety $J = C^2/\Lambda$, which is decomposed as

$$l = Ml_1 + \widehat{M}\ l_2 \in \Lambda \qquad l_1,\ l_2 \in Z^2 \tag{2.39}$$

( $M, \widehat{M}$ are the matrixes defined in eq.(2.4) )

Also, $\chi(l)$ and $L\left(u+\frac{1}{2}l\,,l\right)$ are defined by

$$L\left(u+\frac{1}{2}l,l\right) = {}^{t}(u+\frac{1}{2}l)\cdot(\hat{\eta}l_1+\hat{\eta}\,l_2) \tag{2.40}$$

where $\hat{\eta}$ is defined in eq. (2.5) .

$$\chi(l) = \exp\left(2\pi i\left({}^{t}l_1\cdot\Delta_b - {}^{t}l_2\cdot\Delta_a+\frac{1}{2}\,{}^{t}l_1\quad l_2\right)\right) \in \{-1,1\} \tag{2.41}$$

2) $\sigma(u)$ in genus 2 is an odd function with a zero of order one along the pullback of the theta divisor $\Theta$ to $C^2$.

Then, the product $\prod_{i=1}^{N}\frac{\sigma(\beta_i+\alpha)}{\sigma(\beta_i)\sigma(\alpha)}$ becomes

$$\prod_{i=1}^{N}\frac{\sigma(\beta_i+\alpha+l)}{\sigma(\beta_i)\sigma(\alpha+l)} = \prod_{i=1}^{N}\frac{\sigma(\beta_i+\alpha)}{\sigma(\beta_i)\sigma(\alpha)}\ \prod_{i=1}^{N}\frac{\exp L(\beta_i+\alpha+\frac{1}{2}l,l)}{\exp L(\alpha+\frac{1}{2}l,l)} \tag{2.42}$$

Under the cyclic condition $\sum_{i=1}^{N}\beta_i=0$, the last factor of the right-hand side of (2.42) is unity, and then

$$\prod_{i=1}^{N}\frac{\sigma(\beta_i+\alpha+l)}{\sigma(\beta_i)\sigma(\alpha+l)} = \prod_{i=1}^{N}\frac{\sigma(\beta_i+\alpha)}{\sigma(\beta_i)\sigma(\alpha)} \tag{2.43}$$

That is, $\prod_{i=1}^{N}\frac{\sigma(\beta_i+\alpha)}{\sigma(\beta_i)\sigma(\alpha)}$ can be regarded as a periodic function on Jacobian variety $J$ .

Next, define $f$ as $f = \prod_{i=1}^{N}\frac{\sigma(\beta_i+\alpha)}{\sigma(\beta_i)\sigma(\alpha)}$ as a function of $\alpha$, and regard the $\beta_i$ as points in $C^2$. Let $D$ be the divisor of $f$. The contribution of the denominator to the divisor is $-N\Theta$ , where $\Theta$ is the theta divisor. The Riemann theorem on the Riemann constant states the following:

Denote the divisor of $\theta_R(z-c)$ as $d$. The image of the Abel map of $d$ is equal to $c$ , that is, $A(d)=c$ .

Then, under the cyclic condition, the image of the Abel map of $D$, the divisor of $f$, is equal to $A(D)=\sum\beta_i \ =0$ .

Also, since each factor of $\sigma(\beta_i+\alpha)$ , $\sigma(\alpha)$ has order one zero point, the degree of $D$ is zero. Therefore, by Abel's theorem, $D$ can be a divisor of a meromorphic function.

The logic above is known to be used to derive the "Addition formula" ,which states that, for genus $g$ hyper-elliptic curves, the product $\frac{\sigma(\alpha+\beta)\,\sigma(\alpha-\beta)}{\sigma(\alpha)^2\,\sigma(\beta)^2}$ can be expanded using the Pe functions. [12][13][21] . It states that the product with periodicity on the Jacobian variety and theta divisor in its denominator can be expanded using Pe-functions, as in the way in genus one case.

The objective pursued in genus one and two was as follows :

In genus one, by seeing the fact that $\Pi_{i=1}^{N} S_\delta(\beta_i) = \Pi_{i=1}^{N} \frac{\theta_1^{(1)}(0)\theta_1(\beta_i+\Omega_\delta,\tau)}{\theta_1(\beta_i,\tau)\theta_1(\Omega_\delta,\tau)}$ under the cyclic condition, we replaced the half-periods $\Omega_\delta$ by a C-number parameter $\alpha$ ,

and we regard $\Pi_{i=1}^{N} \frac{\theta_1^{(1)}(0)\theta_1(\beta_i+\alpha,\tau)}{\theta_1(\beta_i,\tau)\theta_1(\alpha,\tau)}$ as an elliptic function of $\alpha \in C^1$ under the cyclic condition. Elliptic functions are defined as meromorphic functions of a single complex variable $\alpha$, which are doubly periodic. In accordance with the basic theory of elliptic functions, the product was expanded using the derivatives of the Pe-functions. Following the acquisition of all coefficients of the expansion, the parameter $\alpha$ is set equal to the half period $\Omega_\delta$ . This process enables the subsequent derivation of the decomposition formula.

In genus two, for even spin structures, there exist $\Omega_\delta \in C^2$ which are the natural generalization of the half-periods of the case of genus one, $\Omega_\delta \in C^1$ . Under the cyclic condition, since it can be posited that Proposition 1 provides

$$\Pi_{i=1}^{N} S_\delta(z_i,\ z_{i+1}) = \Pi_{i=1}^{N} \frac{\theta_R(\beta_i+\Omega_\delta)}{\theta_R(\Omega_\delta)E(z_i,z_{i+1})} = \prod_{i=1}^{N} \frac{\sigma(\beta_i+\Omega_\delta)}{\sigma(\beta_i)\sigma(\Omega_\delta)} \cdot \prod_{i=1}^{N} \omega_1(z_i) ,$$

we replace $\Omega_\delta \in C^2$ by a parameter $\alpha$ as $\prod_{i=1}^{N} \frac{\sigma(\beta_i+\alpha)}{\sigma(\beta_i)\sigma(\alpha)} \cdot \prod_{i=1}^{N} \omega_1(z_i)$ and regard $\prod_{i=1}^{N} \frac{\sigma(\beta_i+\alpha)}{\sigma(\beta_i)\sigma(\alpha)}$ as an Abelian function of $\alpha \in C^2$ . In this context, Abelian functions at genus 2 , as defined within classical terminology, refer to meromorphic functions of double complex variables $\alpha \in C^2$ ,which possess quadruple periodicity in total. It was demonstrated that, by employing the properties of sigma functions, the product $\prod_{i=1}^{N} \frac{\sigma(\beta_i+\alpha)}{\sigma(\beta_i)\sigma(\alpha)}$ is a periodic function on the Jacobian variety. This finding indicates that the product is Abelian. Provided that the cyclic condition is satisfied, it can be deduced that the $\beta_i$ 's are not subject to any contributions to the divisor in total. It was established that the divisor in question complies with the Abel theorem and can be considered a divisor of a meromorphic function.

For a divisor $D$ on $J$ , let $l(D)$ be the dimension of $\mathcal{L}(D)$ which is the space of functions $F$ on $J$ such that $(F) + D$ is an effective divisor .

As asserted by Grant[13], it has been established that the dimension is $l(N\Theta) = N^2$ in genus 2, where $\Theta$ is the theta divisor. In genus $g$, in the hyper-elliptic cases, it has been established that the dimension of the function space $l(N\Theta)$ is equal to $N^g$, as articulated in Mumford's work[22].

In genus 2 , the concrete basis functions for small $N$ are given in [13] as

$$\mathcal{L}(\Theta) = \mathbb{C}\,1$$

$$\mathcal{L}(2\Theta) = \mathcal{L}(\Theta) \oplus \mathbb{C}P_{11} \oplus \mathbb{C}P_{12} \oplus \mathbb{C}P_{22}$$

$$\mathcal{L}(3\Theta) = \mathcal{L}(2\Theta) \oplus \mathbb{C}\,P_{111} \oplus \mathbb{C}P_{112} \oplus \mathbb{C}P_{122} \oplus \mathbb{C}P_{222} \oplus \mathbb{C}\,P$$

where $P = P_{11}P_{22} - P_{12}^2$

$$\mathcal{L}(4\Theta) = \mathcal{L}(3\Theta) \oplus \mathbb{C}\,P_{11}{}^2 \oplus \mathbb{C}P_{11}P_{12} \oplus \mathbb{C}P_{11}P_{22} \oplus \mathbb{C}\,P_{12}P_{22} \oplus \mathbb{C}P_{22}^2 \oplus \mathbb{C}\frac{\partial P}{\partial \alpha_1} \oplus \mathbb{C}\frac{\partial P}{\partial \alpha_2}$$

$$\mathcal{L}(5\Theta) = \mathcal{L}(4\Theta) \oplus \mathbb{C}P_{22}\,\mathcal{L}(3\Theta)/\mathcal{L}(2\Theta) \oplus \mathbb{C}PP_{11} \oplus \mathbb{C}PP_{12} \oplus \mathbb{C}\,P_{111}P_{11} \oplus \mathbb{C}P_{111}P_{12}$$

$$\frac{\partial P}{\partial \alpha_1} = P_{11}P_{122} + P_{111}P_{22} - 2P_{12}P_{112}\ , \quad \frac{\partial P}{\partial \alpha_2} = P_{11}P_{222} + P_{112}P_{22} - 2P_{12}P_{122} \tag{2.44}$$

$$P_{IJ}(\alpha) = \frac{\sigma_I(\alpha)\sigma_J(\alpha) - \sigma_{IJ}(\alpha)\sigma(\alpha)}{\sigma(\alpha)^2} \tag{2.45}$$

$$P_{IJK}(\alpha) = \frac{\{\sigma_{IJ}(\alpha)\sigma_K(\alpha) + \sigma_{JK}(\alpha)\sigma_I(\alpha) + \sigma_{KI}(\alpha)\sigma_J(\alpha)\}\sigma(\alpha) - \sigma_{IJK}(\alpha)\sigma^2(\alpha) - 2\sigma_I(\alpha)\sigma_J(\alpha)\sigma_K(\alpha)}{\sigma(\alpha)^3} \tag{2.46}$$

$\mathcal{L}(2\Theta)$, for instance, upon examination of (2.44), it becomes evident that the denominator of $P_{IJ}(\alpha)$ possesses a divisor $2\Theta$ . It has been established that neither $P_{11}(\alpha)$, $P_{12}(\alpha)$, nor $P_{22}(\alpha)$ can be expressed as a linear combination of the remaining two.

$\mathcal{L}(3\Theta)$ is regarded in a similar manner. It is evident that both $P_{12}^2$ and $P_{11}P_{22}$ are products with $\sigma(\alpha)^4$ in the denominator. However, when the combination $P_{11}P_{22} - P_{12}^2$ is calculated by (2.45) (2.46), the numerator has an over-all factor of $\sigma(\alpha)$ , and the degree of the denominator is reduced by one to $\sigma(\alpha)^3$ . Therefore, it can be deduced that $P$ belongs to $\mathcal{L}(3\Theta)$ , rather than $\mathcal{L}(4\Theta)$. In genus 2 and higher, it is necessary to utilize divisor language, which necessitates the management of such matters.

(In [13] , the first line of page 100 , it is described that $P = P_{11}P_{22} - P_{11}^2$ but this is typo and it should be $P = P_{11}P_{22} - P_{12}^2$ .)

It is demonstrated in Chapter3, the differential equations eqs.(3.15) - (3.19), that any derivatives terms of the type $P_{ABCD}$ , and also any higher derivatives, can be expressed as at most degree 2 polynomials of $P_{EF}$ .

It can be demonstrated that all terms of the form $P_{ABC}P_{DEF}$ can be expressed in terms of at most degree 3 polynomials of $P_{GH}$ by means of the differential equations (3.5) - (3.14)

It is also evident that the odd basis functions, such as $P_{ABC}$ , will all be zero after setting the parameter $\alpha$ equal to the half period, $P_{ABC}(\Omega_\delta) = 0$ . This assertion is applicable to any of the type $P_{I_1 I_2 \dots I_k}$ , with odd $k$, on account of the differential equations (3.15) - (3.19)

Furthermore, as will be demonstrated subsequently, in the present context, calculations employing odd basis functions do not exert an influence on the results of calculations employing even basis functions. ( See Chapter 4. )

For example, as for $\mathcal{L}(3\Theta)$, in our calculations, it will turn out that $\mathcal{L}(2\Theta) \oplus \mathbb{C}\, P$ are enough to have correct results.

With regard to $\mathcal{L}(4\Theta)$ , the following will suffice for the calculations : $\mathcal{L}(3\Theta) \oplus \mathbb{C}\, P_{11}^{\,2} \oplus \mathbb{C} P_{11} P_{12} \oplus \mathbb{C} P_{11} P_{22} \oplus \mathbb{C}\, P_{12} P_{22} \oplus \mathbb{C} P_{22}^2$

In view of the aforementioned considerations, it can be posited that, the product $\Pi_{i=1}^{N} \frac{\theta_R(A(z_i - z_{i+1}) + \Omega_\delta)}{\theta_R(\Omega_\delta) E(z_i, z_{i+1})}$ , in which the parameter $\alpha$ is set equal to $\Omega_\delta$ , can be expressed as a linear combination of the polynomials of the products of the type

$$P_{AB}(\Omega_\delta),\ \ P_{AB}(\Omega_\delta)\, P_{CD}(\Omega_\delta),\ \ P_{AB}(\Omega_\delta)\, P_{CD}(\Omega_\delta)\, P_{EF}(\Omega_\delta), \dots \quad .$$

with appropriate coefficients which are functions of the vertex insertion points under the cyclic condition.

## 3. Trilinear relations from the differential equations of Pe functions

The trilinear relations found in [15] are essential for simplifying the cyclic product of any number of Szegö kernels and their spin sums in genus 2. This discovery, which is based on the consideration of the structure of the ring of the coefficients of the curve, has made the higher-point amplitudes calculations in string theory in genus 2 very transparent.

Having seen that the product (2.37) can be expanded using Pe-functions, we expect the trilinear relations to be obtained by setting the variables in the corresponding Pe function differential equations equal to half the periods in genus 2. This is noted for the genus one case in [18]. In this chapter, we will demonstrate that this is also the case for the genus two in a natural way.

In genus 1, there is only one fundamental differentiation equation for Pe, which takes the form

$$\{P^{(1)}(z)\}^2 = 4\{P(z)\}^3 - g_2 P(z) - g_3$$

Or this can be written as

$$P_{111}(z)\, P_{111}(z) \quad = \quad 4[P_{11}(z)]^3 - \ g_2 P_{11}(z) - g_3\, . \qquad (3\text{-}1)$$

Differentiating one more time, we have

$$P^{(2)}(z) = 6\{P(z)\}^2 \ - g_2/2$$

Or equivalently

$$P_{1111}(z) = \ 6[P_{11}(z)]^2 - \ g_2/2 \ . \tag{3-2}$$

In genus 2 , there are 15 fundamental Pe function differential equations in total, as shown below: eqs. (3.5)-(3-19) The variables in the all of equations (3.5)-(3-19) below are arbitrary.

The first 10 differential equations are of the type

$$P_{ABC}P_{DEF} = \ P_{GH}P_{IJ}P_{KL} + lower\ degree\ terms\ of\ P \tag{3-3}$$

where the left-hand side is a degree two monomial of the derivatives of Pe functions with three indices, and the right-hand side is a sum of the products of degree three or lower monomials of the derivatives of the Pe functions with two indices. These ten equations correspond to eq. (3-1) of the genus one case.

The rest of 5 differential equations are of the type

$$P_{ABCD} = \ P_{EF}\,P_{GH} + lower\ degree\ terms\ of\ P \tag{3-4}$$

which correspond to the eq.(3-2) of the genus one case.

Each of the indices of $A, B, C, \dots\dots$ takes the values of 1 and 2. There are 10 different combinations of the type $P_{ABC}P_{DEF}$ , and 10 different combinations of the type $P_{GH}P_{IJ}P_{KL}$.

The list of the differential equations is as follows: it is described in [21], Appendix I, or in [12]. All equations should be written as $P_{222}(u)^2 = 4P_{11}(u) + 4\,\mu_3 + 4\,P_{12}(u)P_{22}(u) + \cdots$ as functions of an arbitrary $u \in C^2$ number, but we abbreviate the variable $u$.

Note that in our notations, $\mu_1 = 0$. In the following, we list up differential equations in general notation. As described below eq.(2.22), $\mu_1$ can be set to 0 in all of the equations in all calculations in later chapters.

$$P_{222}{}^2 = 4P_{11} + 4\,\mu_3 + 4\,P_{12}P_{22} + 4\mu_2 P_{22} \ + 4P_{22}^3 + 4\mu_1 P_{22}^2 \tag{3-5}$$

$$P_{122}{}^2 = 4\mu_5 \ - 4\,P_{11}P_{12} + 4\,P_{12}^{\ 2}P_{22} \ + 4\,\mu_1 P_{12}^2 \tag{3-6}$$

$$P_{122}P_{222} = 2\,\mu_4 + 2P_{12}^2 + 2\mu_2 P_{12} \ - 2\,P_{22}P_{11} + 4P_{12}P_{22}^2 + 4\mu_1 P_{22}P_{12} \tag{3-7}$$

$$P_{112}P_{222} = 2\,\mu_2 P_{22}P_{12} + 2P_{12}^{\ 2}\,P_{22} + 2P_{22}^{\ 2}P_{11} - 2\,\mu_4 P_{22} + 4\mu_3 P_{12} + 4P_{11}P_{12} \tag{3-8}$$

$$P_{112}{}^2 = -4\mu_4\,P_{12}P_{22} + 4\mu_3\,P_{12}^{\ 2} + 4\,P_{12}^{\ 2}P_{11} + \ 4\mu_5\,P_{22}^2 \tag{3-9}$$

$$P_{112}P_{122} = 2\,P_{22}P_{11}P_{12} + 2P_{12}^3 \ + 2\mu_2 P_{12}^2 \ + 2\,\mu_4 P_{12} - 4\mu_5 P_{22} \tag{3-10}$$

$$\begin{aligned} P_{222}P_{111} = &-4\mu_4\,\mu_1 P_{22} + 8\mu_3\,\mu_1 P_{12} + 8\mu_1 P_{11}P_{12} - 4\mu_4\,P_{22}^2 + 8\mu_3 P_{12}P_{22} \\ &+6P_{22}P_{11}P_{12} - 2\,\mu_4 P_{12} - 2\mu_2^2 P_{12} - 2P_{12}^3 - 4\mu_2 P_{12}^2 \ - 2\mu_2\,P_{22}P_{11} \\ &-2\mu_4\,\mu_2 - 4P_{11}^{\ 2} - 4\mu_3 P_{11} \end{aligned} \tag{3-11}$$

$$\begin{aligned} P_{122}P_{111} = &\ 2P_{12}^{\ 2}\,P_{11} + 4\mu_4 P_{12}P_{22} - 8\mu_5\,P_{22}^2 \ + 4\mu_1\,\mu_4 P_{12} - 8\mu_1\,\mu_5 P_{22} \\ &-4\,\mu_5 P_{12} + 2\mu_2\,P_{11}P_{12} + 2P_{11}^{\ 2}P_{22} - 2\mu_4\,P_{11} - 4\mu_2\,\mu_5 \end{aligned} \tag{3-12}$$

$$P_{112}P_{111} = 4{P_{11}}^2 P_{12} - 8\mu_5 P_{11} - 4\mu_2 \mu_5 P_{22} + 2\mu_2 \mu_4 P_{12} - 2\mu_4 P_{22}P_{11}$$
$$+4\mu_3 P_{11}P_{12} - 4\mu_5 P_{22}P_{12} + 2\mu_4 {P_{12}}^2 - 8\mu_3\mu_5 + 2{\mu_4}^2 \qquad (3\text{-}13)$$
$${P_{111}}^2 = 4{P_{11}}^3 + 4\mu_4 P_{12}P_{11} + 4\mu_5 {P_{12}}^2 + 4\mu_3 {P_{11}}^2 - 16\mu_5 P_{22}P_{11} + 8\mu_5 \mu_2 P_{12}$$
$$+4(\mu_4\mu_2 - 4\mu_1\mu_5)P_{11} + 4(\mu_4^2 - 4\mu_5 \mu_3)P_{22}$$
$$+4({\mu_4}^2\mu_1 + \mu_2^2\mu_5 - 4\mu_1\mu_3 \mu_5) \qquad (3\text{-}14)$$

and

$$P_{1111} = 6P_{11}^2 - 8\mu_5\mu_1 + 2\mu_4\mu_2 - 12\mu_5 P_{22} + 4\mu_4 P_{21} + 4\mu_3 P_{11} \qquad (3\text{-}15)$$
$$P_{1112} = 6P_{11}P_{12} - 4\mu_5 - 2\mu_4 P_{22} + 4\mu_3 P_{21} \qquad (3\text{-}16)$$
$$P_{1122} = 4P_{21}^2 + 2P_{22}P_{11} + 2\mu_2 P_{21} \qquad (3\text{-}17)$$
$$P_{1222} = 6P_{22}P_{21} + 4\mu_1 P_{21} - 2P_{11} \qquad (3\text{-}18)$$
$$P_{2222} = 6P_{22}^2 + 2\mu_2 + 4\mu_1 P_{22} + 4P_{21} \qquad (3\text{-}19)$$

We set all the variables in eq.(3-5)-(3-14) equal to any of half periods of non-singular even spin structure $\Omega_\delta$ . Since $P_{ABC}(\Omega_\delta) = 0$ for all values of $A, B, C = 1,2$ , the left-hand sides of eq.(3-5) - (3-14) all become zero.

Therefore, all combinations of the degree 3 polynomial type $P_{AB}(\Omega_\delta)P_{CD}(\Omega_\delta)P_{EF}(\Omega_\delta)$ in the right-hand sides can be solved as an at most degree 2 polynomial of the type $P_{GH}(\Omega_\delta)\, P_{IJ}\,(\Omega_\delta) + \mathit{lower\ degree\ polynomialos\ of\ P}$ as follows

$$P_{22}(\Omega_\delta)^3 = -P_{11}(\Omega_\delta) - \mu_3 - P_{12}(\Omega_\delta)P_{22}(\Omega_\delta) - \mu_2 P_{22}(\Omega_\delta) - \mu_1 P_{22}(\Omega_\delta)^2 \qquad (3\text{-}20)$$
$$P_{12}(\Omega_\delta)^2 P_{22}(\Omega_\delta) = -\mu_5 + P_{11}(\Omega_\delta)P_{12}(\Omega_\delta) - \mu_1 P_{12}(\Omega_\delta)^2 \qquad (3\text{-}21)$$
$$P_{12}(\Omega_\delta)P_{22}(\Omega_\delta)^2 = -\frac{1}{2}\mu_4 - \frac{1}{2}P_{12}(\Omega_\delta)^2 - \frac{1}{2}\mu_2 P_{12}(\Omega_\delta)$$
$$+\frac{1}{2}P_{22}(\Omega_\delta)P_{11}(\Omega_\delta) - \mu_1 P_{22}(\Omega_\delta)P_{12}(\Omega_\delta) \qquad (3\text{-}22)$$
$$P_{22}(\Omega_\delta)^2 P_{11}(\Omega_\delta) = -\mu_2 P_{22}(\Omega_\delta)P_{12}(\Omega_\delta) + \mu_4 P_{22}(\Omega_\delta) - 2\mu_3 P_{12}(\Omega_\delta)$$
$$-3P_{11}(\Omega_\delta)P_{12}(\Omega_\delta) + \mu_5 + \mu_1 P_{12}(\Omega_\delta)^2 \qquad (3\text{-}23)$$
$$P_{12}(\Omega_\delta)^2 P_{11}(\Omega_\delta) = \mu_4 P_{12}(\Omega_\delta)P_{22}(\Omega_\delta) - \mu_3 P_{12}(\Omega_\delta)^2 - \mu_5 P_{22}(\Omega_\delta)^2 \qquad (3\text{-}24)$$
$$P_{11}(\Omega_\delta)^2 P_{22}(\Omega_\delta) = \mu_3 P_{12}(\Omega_\delta)^2 - 3\mu_4 P_{12}(\Omega_\delta)P_{22}(\Omega_\delta) + 5\mu_5 P_{22}(\Omega_\delta)^2$$
$$-2\mu_1 \mu_4 P_{12}(\Omega_\delta) + 4\mu_1 \mu_5 P_{22}(\Omega_\delta)$$
$$+2\mu_5 P_{12}(\Omega_\delta) - \mu_2 P_{11}(\Omega_\delta)P_{12}(\Omega_\delta) + \mu_4 P_{11}(\Omega_\delta) + 2\mu_2\mu_5 \qquad (3\text{-}25)$$
$$P_{11}(\Omega_\delta)^2 P_{12}(\Omega_\delta) = 2\mu_5 P_{11}(\Omega_\delta) + \mu_2 \mu_5 P_{22}(\Omega_\delta) - \frac{1}{2}\mu_2 \mu_4 P_{12}(\Omega_\delta) + 2\mu_3\mu_5 - \frac{1}{2}{\mu_4}^2$$
$$+\frac{1}{2}\mu_4 P_{22}(\Omega_\delta)P_{11}(\Omega_\delta) - \mu_3 P_{11}(\Omega_\delta)P_{12}(\Omega_\delta)$$
$$+\mu_5 P_{22}(\Omega_\delta)P_{12}(\Omega_\delta) - \frac{1}{2}\mu_4 {P_{12}}^2(\Omega_\delta) \qquad (3.26)$$
$$P_{11}(\Omega_\delta)^3 = -\mu_4 P_{12}(\Omega_\delta)P_{11}(\Omega_\delta) - \mu_5 P_{12}(\Omega_\delta)^2 - \mu_3 P_{11}(\Omega_\delta)^2 + 4\mu_5 P_{22}(\Omega_\delta)P_{11}(\Omega_\delta)$$
$$-2\mu_5 \mu_2 P_{12}(\Omega_\delta) - (\mu_4\mu_2 - 4\mu_1\mu_5)P_{11}(\Omega_\delta) - (\mu_4^2 - 4\mu_5 \mu_3)P_{22}(\Omega_\delta)$$
$$-({\mu_4}^2\mu_1 + \mu_2^2\mu_5 - 4\mu_1\mu_3 \mu_5) \qquad (3\text{-}27)$$

From (3-10) and (3-11), we have the following two more solutions :

$$P_{22}(\Omega_\delta)P_{11}(\Omega_\delta)P_{12}(\Omega_\delta) = \frac{1}{2}\mu_5 P_{22}(\Omega_\delta) + \frac{1}{2}\mu_4\ \mu_1 P_{22}(\Omega_\delta) - \mu_3\ \mu_1 P_{12}(\Omega_\delta) -\mu_1 P_{11}(\Omega_\delta)P_{12}(\Omega_\delta) + \frac{1}{2}\mu_4\ P_{22}(\Omega_\delta)^2 - \mu_3 P_{12}(\Omega_\delta)P_{22}(\Omega_\delta) + \frac{1}{4}\mu_2^2 P_{12}(\Omega_\delta) + \frac{1}{4}\mu_2 P_{12}(\Omega_\delta)^2 + \frac{1}{4}\mu_2 P_{22}(\Omega_\delta)P_{11}(\Omega_\delta) + \frac{1}{4}\mu_4\mu_2 + \frac{1}{2}P_{11}(\Omega_\delta)^2 + \frac{1}{2}\mu_3 P_{11}(\Omega_\delta) \quad (3\text{-}28)$$

$$P_{12}(\Omega_\delta)^3 = \frac{3}{2}\mu_5 P_{22}(\Omega_\delta) - \frac{1}{2}\mu_4\ \mu_1 P_{22}(\Omega_\delta) + \mu_3\ \mu_1 P_{12}(\Omega_\delta) + \mu_1 P_{11}(\Omega_\delta)P_{12}(\Omega_\delta) - \frac{1}{2}\mu_4\ P_{22}(\Omega_\delta)^2 + \mu_3 P_{12}(\Omega_\delta)P_{22}(\Omega_\delta) - \mu_4 P_{12}(\Omega_\delta) - \frac{1}{4}\mu_2^2 P_{12}(\Omega_\delta) - \frac{5}{4}\mu_2 P_{12}(\Omega_\delta)^2 - \frac{1}{4}\mu_2 P_{22}(\Omega_\delta)P_{11}(\Omega_\delta) - \frac{1}{4}\mu_4\mu_2 - \frac{1}{2}P_{11}(\Omega_\delta)^2 - \frac{1}{2}\mu_3 P_{11}(\Omega_\delta) \quad (3\text{-}29)$$

These can be considered trilinear relations found in [15] within our framework. An example of their usage will be given in Chapter 4-2. The 5 equations of type (3.4) have many other applications in Chapter 4.

The following facts are known, although we do not use these results in this paper： In [12], Chapter 9, describes how, in genus 2, the 3 equations (3-5), (3-6), (3-7) can be derived as the definition equations of Jacobian (Jacobian variety), from the corollary of the fundamental relation (2.19). All other equations (3-8) - (3-19), including all of (3-15) -(3-19) , are derived only from these 3 equations (3-5), (3-6), (3-7). In hyperelliptic cases, a generalized theory exists to obtain the corresponding definition equations of the Jacobian from (2.19), for the product of type $P_{ggi}P_{ggj}$ ( $g$ is the number of genus, and $i, j = 1, 2, .. g$. Total $\frac{g(g+1)}{2}$ number of equations ) and a method exists in each genus to obtain the other derived differential equations [12].

Define $N33(g)$ as the number of independent combinations of the symmetric tensors $P_{ABC}P_{DEF}$ where each of the indices $A, B, C, \ldots..$ takes the values $1,2, \ldots g$ . $g$ is equal to the genus.

Also define $N222(g)$ as the number of independent combinations of the symmetric tensors $P_{GH}P_{IJ}P_{KL}$ where each index takes the values $1,2, \ldots g$ .
For example $N33(1) = N222(1) = 1$ , and $N33(2) = N222(2) = 10$

In general, the formula for the number of combinations permitting duplicates immediately yields the following results：

$$N33(g) = \frac{1}{72}[g(g+1)(g+2)+6]g(g+1)(g+2) \quad (3\text{-}30)$$

$$N222(g) = \frac{1}{48}[g(g+1)+4][g+2]g(g+1) \quad (3\text{-}31)$$

Then

$$N222(g) - N33(g) = \frac{1}{144}(g-2)(g-1)^2 g^2 (g+1) \quad (3\text{-}32)$$

Therefore, except in the cases $g = 1$ or $g = 2$, $N222(g)$ is always greater than $N33(g)$. Genus 1 and 2 are the exceptional cases in which $N33(g) = N222(g)$. Therefore, even in the hyperelliptic case, deriving trilinear relations in the genus higher than 2 will not be so straightforward. However, this does not mean that trilinear relations or any other similar identities are impossible in higher genus.

# 4. Calculations

## 4-1 calculation methods

**( $\partial$ calculations, a direct generalizations of the method used in genus 1 )**

For genus 1, the decomposition formula calculations given in Chapter 1 of [11] were as follows.

We write the product $\Pi_{i=1}^{N} \frac{\theta_1^{(1)}(0)\theta_1(\beta_i+\alpha,\tau)}{\theta_1(\beta_i,\tau)\theta_1(\alpha,\tau)}$ as

$$\exp(\sum_{i=1}^{N} \ln \theta_1 (\beta_i + \alpha) - \sum_{i=1}^{N} \ln \theta_1 (\beta_i) - N \ln \theta_1 (\alpha) + N \ln \theta_1^{(1)}(0) ). \qquad (4.1.1)$$

Then, using

$$\ln \theta_1(\alpha) = \ln \sigma(\alpha) - \frac{G_2}{2}\alpha^2 + \ln\theta_1^{(1)}(0) = \ln \alpha - \sum_{k=1}^{\infty} \frac{G_{2k}}{2k} \alpha^{2k} + \ln\theta_1^{(1)}(0) ,$$

which is derived from the infinite product representation of the $\sigma$ function, (4.1.1) can be written as $\frac{1}{\alpha^N}\exp(W)$ , where

$$W = \sum_{i=1}^{N} \ln \theta_1 (\beta_i + \alpha) - \sum_{i=1}^{N} \ln \theta_1 (\beta_i) + N \sum_{k=1}^{\infty} \frac{\alpha^{2k}}{2k} G_{2k}(\tau) .$$

Alternatively, $W = \sum_{k=1}^{\infty} \frac{\alpha^k}{k!} ( \sum_{i=1}^{N} D^k \ln \theta_1 (\beta_i) )$ .

The derivatives of the logarithm of the infinite product formula of the $\sigma$ function naturally introduce holomorphic Eisenstein series into the results of the decomposition formula.

Equate $\frac{1}{\alpha^N}\exp(W) = \sum_{M=0}^{N} h_{N,M}(\beta_i)P^{(N-2-M)}(\alpha)$ (4.1.2)

as the definition of the expansion coefficients $h_{N,M}(\beta_i)$ ,

( as described below (1.11), $P^{(-2)}(\alpha) \equiv 1$, $P^{(-1)}(\alpha) \equiv 0$, $h_{N,N-1} \equiv 0$, $P^{(0)}(\alpha) \equiv P(\alpha)$ ).

Multiplying $\alpha^N$ on the both sides of (4.1.2), we have

$$\exp(W) = \alpha^N \sum_{M=0}^{N} h_{N,M}(\beta_i)P^{(N-2-M)}(\alpha) \qquad (4.1.3)$$

To calculate all the $h_{N,M}(\beta_i)$ coefficients, we only need to differentiate the both sides of (4.1.3) and set $\alpha$ equal to zero.

We will attempt to calculate the genus 2 case using the same logic. However, several problems arise. First, we set the expansion form as

$$\prod_{i=1}^{N} \frac{\sigma(\alpha+\beta_i)}{\sigma(\beta_i)\sigma(\alpha)} = H_0 + \sum_{I,J=1,2} H_{IJ}P_{IJ}(\alpha) + \sum_{I,J,K=1,2} H_{IJK}P_{IJK}(\alpha) + \ldots \qquad (4.1.4)$$

The definitions of the coefficients $H_{IJ}$, $H_{IJK}$, .... are to be determined. The $N^2$ number of expansion basis functions, $P_{IJ}(\alpha), P_{IJK}(\alpha)$, .... are prepared as (2.44) for each $N$ .

The right-hand side includes all basis functions with parameter is $\alpha$ for a fixed $N$ value, and $H_{IJK....}$ are the expansion coefficients which will be functions of $\beta_i$ .

In genus 2, the infinite product expansion formula of $\sigma$ function is not known. Instead we proceed in a rather naïve way as follows .

To calculate the coefficients $H_{IJK....}$ , multiply $[\sigma(\alpha)]^N$ on both sides of eq.(4.1-4) and write it as

$$\exp(\ \sum_{i=1}^{N}[\ln\sigma(\alpha+\beta_i)-\ln\sigma(\beta_i)])=\ H_0\ [\sigma(\alpha)]^N\ +H_{11}P_{11}(\alpha)[\sigma(\alpha)]^N+\cdots \quad (4.1.5)$$

Differentiating both sides with respect to $\alpha_1,\alpha_2$ total $k$ times $(k=0,1,\ldots.N)$ and set $\alpha_1=\alpha_2=0$ , we can calculate the coefficients $H_0$ ,$H_{IJ}$, $H_{IJK}$ ,.. . Note that, in calculating the coefficients, we don't set $\alpha=\Omega_\delta$, but set $\alpha=0$, as was the case for genus 1. From the weight considerations, $k$ should not exceed $N$ in order to obtain meaningful results.

The following basic explicit formulas are useful, using the notations

$$\partial_I=\partial_{\alpha_I}\ (I=1,2),\quad \sigma_{IJ}(\alpha)=\ \frac{\partial^2}{\partial_I\partial_J}\sigma(\alpha)\ ,$$

$$P_{IJ}(\alpha)=-\ \frac{\partial^2}{\partial_I\partial_J}\ln\sigma(\alpha)=\ \frac{\sigma_I(\alpha)\sigma_J(\alpha)-\ \sigma_{IJ}(\alpha)\sigma(\alpha)}{\sigma(\alpha)^2} \quad (4.1.6)$$

$$P_{IJK}(\alpha)=\frac{\partial}{\partial_K}P_{IJ}(\alpha)=\frac{\{\sigma_{IJ}(\alpha)\sigma_K(\alpha)+\sigma_{JK}(\alpha)\sigma_I(\alpha)+\sigma_{KI}(\alpha)\sigma_J(\alpha)\}\sigma(\alpha)-\ \sigma_{IJK}(\alpha)\sigma^2(\alpha)-2\sigma_I(\alpha)\sigma_J(\alpha)\sigma_K(\alpha)}{\sigma(\alpha)^3} \quad (4.1.7)$$

(and similar expressions for $P_{IJKL}(\alpha)$, and higher derivatives. )

We will expand the $P_{IJ}(\alpha)[\sigma(\alpha)]^N$, $P_{IJK}(\alpha)[\sigma(\alpha)]^N$, $P_{IJKL}(\alpha)[\sigma(\alpha)]^N,\ldots.$ using the Taylor expansion form of the sigma function, eq.(2.21).

Here we define $\partial_I\ \partial_J\ \partial_K \ldots$ operation as

$$\partial_I\ \partial_J\ \partial_K \ldots e_0^{L_N}\ \equiv\ \left[\ \partial_I\ \partial_J\ \partial_K \ldots \exp(\ \sum_{i=1}^{N}[\ln\sigma(\alpha+\beta_i)-\ln\sigma(\beta_i)])\right]_{\alpha=0} \quad (4.1.8)$$

where $\alpha=0$ means that we set $\alpha_1=\alpha_2=0$ after the differentiations, and

$$L_N=\sum_{i=1}^{N}[\ln\sigma(\alpha+\beta_i)-\ln\sigma(\beta_i)]. \quad (4.1.9)$$

We temporarily call this method a direct generalization of the genus 1 method, or $\partial$-calculation, because it is used repeatedly.

This calculation method is effective in genus one for arbitrary $N$, but not for genus two and higher values of $N$ .

In fact, the number of equations that can be obtained to determine the $H$ coefficients by differentiation, as in (4.1-8), for a fixed value of $N$ is $\frac{(N+1)(N+2)}{2}$, whereas the number of basis functions is $N^2$. For $N>3$, we cannot offer enough number of equations by only using this naïve extension method of genus one case.

The fundamental reasons for this situation are as follows:
For genus 2, the $N^2$ number of basis functions discussed in Chapter 2 are for the periodic functions on the Jacobian variety, which is a complex 2-dimensional torus. The $\sigma(\alpha)$ function can be defined on the entire Jacobian variety, and the equations obtained by the direct generalization of the genus 1 method, $\partial$- calculation, are correct.
On the other hand, when we proved in Chapter 2 that the product $\prod_{i=1}^{N}\frac{\sigma(\alpha+\beta_i)}{\sigma(\beta_i)\sigma(\alpha)}$ on the left-hand side of (4.1.4) is a periodic function of a Jacobian variety, we made use of the fact that the $\beta$-type variables with the cyclic condition do not affect the divisors in total. We proved its periodicity using the fact that the variables can only move over the Riemann surface which is a one-dimensional complex manifold embedded in the Jacobian variety, rather than over the entire Jacobian variety. In order to fully determine the expansion coefficients $H_{IJ}$, $H_{IJK}$, .... on the right-hand side of equation (4.1.4) for larger values of $N$ in genus 2, we may need to seek out additional restrictions (new equations) which will reflect the fact that the variables in the functions can only vary over the Riemann surface. This is a situation that differs significantly from the case where the genus is 1.

As examples of such additional equations, we will use established results given in [15] in addition to $\partial$- calculations in the section 4-4 for the $N = 4$ case to reinforce the situation. Combined these, we will calculate the case of $N = 2,3,4$ in the sections 4-2, 4-3, 4-4 and discuss the possibility of calculating $N = 6$ case in the section 4-5.

## 4-2 N=2

For the case $N = 2$ , the following classical formula has been known :

$$\frac{\sigma(\alpha+\beta)\,\sigma(\alpha-\beta)}{\sigma(\alpha)^2\,\sigma(\beta)^2} = P_{11}(\alpha) - P_{11}(\beta) + P_{12}(\alpha)P_{22}(\beta) - P_{12}(\beta)P_{22}(\alpha) \qquad (4.2.1)$$

where there are 4 $(= N^g)$ basis functions $1, P_{11}(\alpha), P_{12}(\alpha), P_{22}(\alpha)$.

Eq.(4.2.1) is a generalization of the well-known genus one result

$$\frac{\sigma(\alpha+\beta)\,\sigma(\alpha-\beta)}{\sigma(\alpha)^2\,\sigma(\beta)^2} = P_{11}(\alpha) - P_{11}(\beta) .$$

where there are 2 basis functions, 1 and $P_{11}(\alpha)$ $(= P(\alpha))$.
Usually, in the proof of (4.2.1), both of variables $\alpha$ and $\beta$ are of $\alpha$ type.
This identity holds even if the variable $\beta$ is of a different type to the $\alpha$ type. This can be seen by performing $\partial$- calculation in its simplest case, as follows.

$$\exp\left(\sum_{i=1}^{2}[\ln\sigma(\alpha+\beta_i) - \ln\sigma(\beta_i)]\right) = H_0\,[\sigma(\alpha)]^2 + H_{11}P_{11}(\alpha)[\sigma(\alpha)]^2$$
$$+ H_{12}P_{12}(\alpha)[\sigma(\alpha)]^2 + H_{22}P_{22}(\alpha)[\sigma(\alpha)]^2 \qquad (4.2.2)$$

$(\beta_1 = -\beta_2 = \beta)$

The following expansion, obtained from the Taylor expansion of the sigma function and definition formulas of Pe function, is convenient to use：

$$\sigma(\alpha)^2 = \alpha_1^2 + O(3)$$

$$\sigma(\alpha)^2 P_{11}(\alpha) = (1 + \tfrac{1}{2}\mu_3\alpha_1^2 + \cdots)^2 - (\mu_3\alpha_1 + \cdots)(\mu_3\alpha_1 + \cdots) = 1 + O(3)$$

$$\sigma(\alpha)^2 P_{12}(\alpha) = -\alpha_2^2 + O(3)$$

$$\sigma(\alpha)^2 P_{22}(\alpha) = 2\alpha_1\alpha_2 + O(3) \tag{4.2.3}$$

Here $O(3)$ are the polynomial terms of $\alpha_1, \alpha_2$ in which the sum of the degrees of $\alpha_1$ and $\alpha_2$ is 3 or greater. In eq.(4.2.2), simply set $\alpha_1 \to 0$ , $\alpha_2 \to 0$ and leave everything else as it is to obtain $H_{11} = 1$

We write the result of differentiating the left-hand side twice with respect to $\alpha_1$ as

$$\partial_1\partial_1\, e_0^{L_2} \equiv \frac{\partial^2}{\partial\alpha_1^2} \exp\left( \textstyle\sum_{i=1}^{2}[\ln\sigma(\alpha + \beta_i) - \ln\sigma(\beta_i)]\right) \Big|_{\alpha_1=\alpha_2=0} \tag{4.2.4}$$

Noting that $\zeta_1(= \partial_1 \ln\sigma)$ is odd, and $P_{11}(= -\partial_1\zeta_1)$ is even, we have

$$H_0 = \frac{1}{2}\, \partial_1\partial_1\, e_0^{L_2} = -\frac{1}{2}\big(P_{11}(\beta_1) + P_{11}(\beta_2)\big) = -P_{11}(\beta) \tag{4.2.5}$$

Similarly,

$$H_{12} = -\frac{1}{2}\, \partial_2\partial_2\, e_0^{L_2} = P_{22}(\beta) \tag{4.2.6}$$

$$H_{22} = \frac{1}{2}\, \partial_1\partial_2\, e_0^{L_2} = -P_{12}(\beta) \tag{4.2.7}$$

Therefore, eq.(4.2.1) is reproduced for any variable $\beta$ in the Pe functions.

We recall here classical formulae known as the solution to Jacobi's inversion problem at genus two, when the variable inside the Pe function is of the $\alpha$- type. The Pe functions, as defined in eq.(2.1), can also be expressed as follows：

$$P_{12}\left( \int_\infty^{(x_1,y_1)} \omega + \int_\infty^{(x_2,y_2)} \omega \right) = -x_1x_2 \tag{4.2.8}$$

$$P_{22}\left( \int_\infty^{(x_1,y_1)} \omega + \int_\infty^{(x_2,y_2)} \omega \right) = x_1 + x_2 \tag{4.2.9}$$

$$P_{11}\left( \int_\infty^{(x_1,y_1)} \omega + \int_\infty^{(x_2,y_2)} \omega \right) = \frac{F(x_1, x_2) - 2y_1y_2}{(x_1 - x_2)^2} \tag{4.2.10}$$

where $F(x_1, x_2)$ is the Kleinian 2-polar defined in eq.(2.20):

$$F(x,z) = (x+z)x^2z^2 + 2\mu_1\, x^2z^2 + \mu_2\,(x+z)xz + 2\mu_3\, xz + \mu_4\,(x+z) + 2\mu_5 \tag{4.2.11}$$

For convenience, we will use the notation $x_1$, $y_1$ for the x and y coordinates of the points in the $\alpha$-type variables rather than $X_1$, $Y_1$ . We do not anticipate any confusion regarding the x and y coordinates in the $\beta$-type variables.

How should these results be changed if the variables are of $\beta$-type, i.e. if we calculate string amplitudes where the variables in the spin structure-independent parts are the differences of the vertex insertion points ?

The answer will be as follows: If the result of the standard Jacobi inversion ( the right-hand sides of (4.2.8), (4.2.9) and (4.2.10) ) contains $y_2$ , then it should be replaced with $-y_2$ . If the results contain only $x_1, x_2$ , the result will not change.
That is

$$P_{12}\left( \int_{\infty}^{(x_1,y_1)} \omega - \int_{\infty}^{(x_2,y_2)} \omega \right) = -x_1 x_2 \tag{4.2.12}$$

$$P_{22}\left( \int_{\infty}^{(x_1,y_1)} \omega - \int_{\infty}^{(x_2,y_2)} \omega \right) = x_1 + x_2 \tag{4.2.13}$$

$$P_{11}\left( \int_{\infty}^{(x_1,y_1)} \omega - \int_{\infty}^{(x_2,y_2)} \omega \right) = \frac{F(x_1,x_2)+2y_1y_2}{(x_1-x_2)^2} \tag{4.2.14}$$

The signature in front of $2y_1y_2$ in the numerator of the right-hand side of (4.2.14) is a plus when the variable is of $\beta$-type, while it is a minus in the $\alpha$-type as in (4.2.10). These modified formulae can be validated only in genus 2. (See Chapter 5).
The proof will be given in Proposition 2 below.
These can be formally interpreted as follows. Using the involution point $\bar{P}(x_2, -y_2)$ we write the Pe function variable as

$$\int_{\infty}^{(x_1,y_1)} \omega_i - \int_{\infty}^{(x_2,y_2)} \omega_i = \int_{\infty}^{(x_1,y_1)} \frac{x^{i-1}}{2y} dx + \int_{\infty}^{(x_2,-y_2)} \frac{x^{i-1}}{2(-y)} dx \tag{4.2.15}$$

We treat $\bar{P}$ as $P$, and apply the standard formula for solving the inversion theorem formally.

This will be validated rigorously later, and then the eq. (4.2.1) can be modified as follows :

$$\frac{\sigma(\alpha+\beta)\,\sigma(\alpha-\beta)}{\sigma(\alpha)^2\,\sigma(\beta)^2} = P_{11}(\alpha) - P_{11}(\beta) + (x_1+x_2)P_{12}(\alpha) + x_1x_2P_{22}(\alpha) \tag{4.2.16}$$

where $P_{11}(\beta)$ is given by eq.(4.2.14), and represents the spin structure independent term of the $N=2$ case. Similar results are obtained in $N=3$, which are consistent with those in [15]. In the above, $x_1, x_2$ are the $x-$coordinates of the vertex insertion points, as $z_1 = (x_1, y_1)$, $z_2 = (x_2, y_2)$.
Further, since

$$0 = x_1^2x_2 + x_2^2x_1 - (x_1+x_2)x_1x_2 = -P_{12}(\beta)x_1 - P_{21}(\beta)x_2 - P_{22}(\beta)x_1x_2\,, \tag{4.2.17}$$

by adding this zero to the both side of eq.(4.2.16) we have

$$\frac{\sigma(\alpha+\beta)\,\sigma(\alpha-\beta)}{\sigma(\alpha)^2\,\sigma(\beta)^2} = P_{11}(\alpha) - P_{11}(\beta) + [P_{12}(\alpha) - P_{12}(\beta)](x_1 + x_2) + [P_{22}(\alpha) - P_{22}(\beta)]x_1x_2 \tag{4.2.18}$$

After multiplying $\prod_{i=1}^{2} \omega_1(z_i)$ on both sides of equation (4.2.17) to obtain $S_\delta(z_1, z_2)^2$ , the holomorphic one-form $\omega_1$ can be rewritten as $\omega_2$ when $x_1, x_2$ are multiplied :

$$x_1\omega_1(z_1) = \omega_2(z_1), \quad x_2\omega_1(z_2) = \omega_2(z_2) \tag{4.2.19}$$

by the definitions of $\omega_1$ and $\omega_2$ .

In general, when there are many variables $z_1$ $z_2, \ldots$ , we can replace

$$x_i\omega_1(z_i) = \omega_2(z_i), \quad \text{where} \quad z_i = (x_i,\ y_i)\ . \tag{4.2.20}$$

Then we have

$$\frac{\sigma(\alpha+\beta)\,\sigma(\alpha-\beta)}{\sigma(\alpha)^2\,\sigma(\beta)^2}\,\omega_1(z_1)\omega_1(z_2) = \sum_{I,J=1}^{2}[P_{IJ}(\alpha) - P_{IJ}(\beta)]\omega_I(z_1)\omega_J(z_2) \tag{4.2.21}$$

By setting $\alpha = \Omega_\delta$ and noting that $S_\delta(z_1, z_2)S_\delta(z_2, z_1) = -S_\delta(z_1, z_2)^2$,

$$-S_\delta(z_1, z_2)^2 = \sum_{I,J=1}^{2}[\,P_{IJ}(\Omega_\delta) - P_{IJ}(\beta)\,]\,\omega_I(z_1)\omega_J(z_2)\ . \tag{4.2.22}$$

Therefore, when the variables are expressed by (4.2.19), the formula resembles the one for genus one,

$$S_\delta(\beta_1)S_\delta(\beta_2) = -\,(S_\delta(\beta))^2 = e_\delta - P(\beta) = P(\Omega_\delta) - P(\beta) \tag{4.2.23}$$

where $\beta = \beta_1 = -\beta_2$ .

**Proposition 2 ( $\boldsymbol{\beta}$-type (modified) version of the inversion formula) :**

It is known that, when the variable in the theta function is of the form $\alpha = \int_\infty^{(x_1,y_1)} \omega + \int_\infty^{(x_2,y_2)} \omega$ , the following standard formulae for the solutions to Jacobi's inversion problem in genus 2 hold :

$$P_{11}(\alpha) = \frac{F(x_1,x_2) - 2y_1y_2}{(x_1 - x_2)^2}, \quad P_{12}(\alpha) = -x_1x_2\ , \quad P_{22}(\alpha) = x_1 + x_2$$

$$P_{111}(\alpha) = 2\,\frac{y_2\psi(x_1,\ x_2) - y_1\psi(x_2,\ x_1)}{(x_1 - x_2)^3}$$

where $\psi(x,\ z)$ is a polynomial of $x,\ z$ with coefficients of $\mu_i$ $(i = 1,2,..5)$ which will be defined later in the proof of the proposition in (4.2.33).

Also, $P_{112}(\alpha) = 2\,\frac{y_1x_2^2 - y_2x_1^2}{x_1 - x_2}$ , $P_{122}(\alpha) = 2\,\frac{y_2x_1 - y_1x_2}{x_1 - x_2}$ , $P_{222}(\alpha) = 2\,\frac{y_1 - y_2}{x_1 - x_2}$ ,

These will be denoted as Eqs. (4.2.29) – (4.2.36) below.

When the variable is in the form $\beta = \int_\infty^{(x_1,y_1)} \omega - \int_\infty^{(x_2,y_2)} \omega$ , then the following modified formulae will hold, by simply replacing $y_2$ with $-y_2$:

$$P_{11}(\beta) = \frac{F(x_1,x_2) + 2y_1y_2}{(x_1 - x_2)^2}, \quad P_{12}(\beta) = -x_1x_2, \quad P_{22}(\beta) = x_1 + x_2$$

$$P_{111}(\beta) = -2\frac{y_2\psi(x_1,\ x_2)+ y_1\psi(x_2,\ x_1)}{(x_1 - x_2)^3},\quad P_{112}(\beta) = 2\frac{y_1 x_2^2 + y_2 x_1^2}{x_1 - x_2},$$

$$P_{122}(\beta) = -2\frac{y_2 x_1 + y_1 x_2}{x_1 - x_2},\quad P_{222}(\beta) = 2\frac{y_1 + y_2}{x_1 - x_2}$$

The validity of these modified formulae will be demonstrated below, with the formulae being denoted as Eqs. (4.2.46) – (4.2.48) and (4.2.50) – (4.2.53). They will play an essential role in many practical calculations involving spin structure-independent parts.

**Proof**

First, we review how to obtain the standard solution of the Jacobi inversion formula, eq. (4.2.8), (4.2.9) ,(4.2.10). Once the logic of the proof has been understood, it is not difficult to prove Proposition 2 for the $\beta$ type cases.

We start from the "Corollary of the main formula" , eq.(2.19) :

$$\frac{F(x_r,x)+2y_r y}{(x_r-x)^2} = \sum_{i=1}^{g}\sum_{j=1}^{g} P_{ij}\left(\int_{\infty}^{P}\omega - u\right)x_r^{i-1}x^{j-1} \qquad (4.2.24)$$

where $u$ is the sum of g number of variables $u = \sum_{k=1}^{g}\int_{\infty}^{(x_k,y_k)}\omega$ and

$P = (x, y)$ , $P_k = (x_k, y_k)$

$x_r$ in (4.2.24) is any one of the coordinates $x_k$ $(k = 1,2,..g)$ .

Here we set $g = 2$ , and we first assume $r = 1$ . In eq.(4.2.24), take the limit $P_2 \to \infty$ and re-define $\bar{P} = (x,\ -y)$ as $P_2$ . Then we have

$$\frac{F(x_1,x_2)-2y_1y_2}{(x_1-x_2)^2} = \sum_{i=1}^{2}\sum_{j=1}^{2} P_{ij}\ (\alpha)x_1^{i-1}x_2^{j-1} \quad . \qquad (4.2.25)$$

where $\alpha = \int_{\infty}^{(x_1,y_1)}\omega + \int_{\infty}^{(x_2,y_2)}\omega.$

On the left-hand side of (4.2.25), the signature in front of $2y_1y_2$ is different compared with (4.2.24). That is, in the proof of standard Jacobi inversion theorem, the "involution re-interpretation" is used. As we will see below, in the modified version of the inversion theorem does not actually use involution.

The eq.(4.2.25) can be written as

$$\frac{F(x_1,x_2)-2y_1y_2}{(x_1-x_2)^2} = P_{11}\ (\alpha) + P_{12}\ (\alpha)x_2 + P_{21}\ (\alpha)x_1 + P_{22}\ (\alpha)x_1x_2 \qquad (4.2.26)$$

Next, backing to the eq.(4.2.24), divide the both sides by $x$ and take the limit $P \to \infty$ . The highest degree term of $x$ in $F(x_r,\ x)$ is $x_r^2x^3$ , and since $\frac{y}{x^3} \to 0$ ,

$$x_r^2 - \sum_{i=1}^{2} P_{i2}\ (\alpha)x_r^{i-1} = 0 \qquad (4.2.27)$$

for both of the case of $r = 1$ and $r = 2$ .

The equation (4.2.27) can be written as

$$x_1^2 - P_{22}(\alpha)x_1 - P_{12}(\alpha) = 0 \text{ and } x_2^2 - P_{22}(\alpha)x_2 - P_{12}(\alpha) = 0 \qquad (4.2.28)$$

The eqs.(4.2.26) and (4.2.28) give 3 equations to determine $P_{11}, P_{12}, P_{22}$ . We have

$$P_{11}(\alpha) = \frac{F(x_1,x_2)-2y_1y_2}{(x_1-x_2)^2} \qquad (4.2.29)$$

$$P_{12}(\alpha) = -x_1x_2 \qquad (4.2.30)$$

$$P_{22}(\alpha) = x_1 + x_2 \qquad (4.2.31)$$

for $\alpha = \int_\infty^{(x_1,y_1)} \omega + \int_\infty^{(x_2,y_2)} \omega$ .

Comparing (4.2.26) and (4.2.29), we see that the terms on the right hand side of (4.2.26), the sum of the last three terms gives zero, which is consistent with the results (4.2.30) and (4.2.31).

Later we will also use the following formulas

$$P_{111}(\alpha) = 2\frac{y_2\psi(x_1, x_2) - y_1\psi(x_2, x_1)}{(x_1 - x_2)^3} \qquad (4.2.32)$$

where

$$\psi(x, z) = (3x+z)x^3z + 4\mu_1 x^3z + \mu_2 (x+3z)x^2 + 2\mu_3 x(x+z) + \mu_4 (3x+z) + 4\mu_5 \qquad (4.2.33)$$

$$P_{112}(\alpha) = 2\frac{y_1x_2^2 - y_2x_1^2}{x_1 - x_2} \qquad (4.2.34)$$

$$P_{122}(\alpha) = 2\frac{y_2x_1 - y_1x_2}{x_1 - x_2} \qquad (4.2.35)$$

$$P_{222}(\alpha) = 2\frac{y_1 - y_2}{x_1 - x_2} \qquad (4.2.36)$$

Also, we note

$$y_k = P_{222}(\alpha)x_k + P_{122}(\alpha) \quad \text{for } k = 1,2 \qquad (4.2.37)$$

All these of (4.2.32) – (4.2.36) can be calculated from the equations (4.2.29) – (4.2.32) by using

$$\begin{bmatrix} \frac{\partial x_1}{\partial \alpha_1} & \frac{\partial x_1}{\partial \alpha_2} \\ \frac{\partial x_2}{\partial \alpha_1} & \frac{\partial x_2}{\partial \alpha_2} \end{bmatrix} = \begin{bmatrix} \frac{\partial \alpha_1}{\partial x_1} & \frac{\partial \alpha_1}{\partial x_2} \\ \frac{\partial \alpha_2}{\partial x_1} & \frac{\partial \alpha_2}{\partial x_2} \end{bmatrix}^{-1} = \begin{bmatrix} \frac{1}{2y_1} & \frac{1}{2y_2} \\ \frac{x_1}{2y_1} & \frac{x_2}{2y_2} \end{bmatrix}^{-1} = \frac{1}{x_2 - x_1} \begin{bmatrix} 2x_2y_1 & -2y_1 \\ -2x_1y_2 & 2y_2 \end{bmatrix} \qquad (4.2.38)$$

For later convenience, we differentiate $P_{222}(\alpha)$ one more time with respect to the variable $\alpha_2$ and see what $P_{2222}(\alpha)$ becomes. By (4.2.38),

$$P_{2222}(\alpha) = 2\frac{\partial}{\partial \alpha_2}\frac{y_1 - y_2}{x_1 - x_2} = 6(x_1^2 + x_2^2) + 8x_1x_2 + 4\mu_1(x_1 + x_2) + 2\mu_2 \ . \qquad (4.2.39)$$

In $P_{12}(\alpha)$, $P_{22}(\alpha)$, $P_{2222}(\alpha)$, *or in any Pe function*, $\alpha = 0$ corresponds to $x_1 = x_2 = \infty$ .
By (4.2.30) and (4.2.31),

$$P_{2222} = 6P_{22}^2 + 4P_{21} + 4\mu_1 P_{22} + 2\mu_2 \qquad (4.2.40)$$

which is the same form as that of the differential equation of $P_{2222}$ in eq.(3.19).

In the case that the variable is of $\beta$ type, we first assume $r=1$ in (4.2.24).
In eq.(4.2.24), take the limit $P_2 \to \infty$ and re-define $P=(x,\ y)$ as $P_2$ . Then we have

$$\frac{F(x_1,x_2)+2y_1y_2}{(x_1-x_2)^2} = \sum_{i=1}^{2}\sum_{j=1}^{2} P_{ij}\ (\beta)x_1^{i-1}x_2^{j-1} \tag{4.2.41}$$

where $\beta = \int_{\infty}^{(x_2,y_2)} \omega - \int_{\infty}^{(x_1,y_1)} \omega.$
The eq.(4.2.41) can be written as

$$\frac{F(x_1,x_2)+2y_1y_2}{(x_1-x_2)^2} = P_{11}\ (\beta) + P_{12}\ (\beta)x_2 + P_{21}\ (\beta)x_1 + P_{22}\ (\beta)x_1x_2 \tag{4.2.42}$$

Next, divide the both sides of eq.(4.2.24) by $x_r$ and take the limit $P_r \to \infty$ .

The highest degree term of $x_r$ in $F(x_r,\ x)$ is $x_r^3x^2$ , and since $\frac{y_r}{x_r^3} \to 0$ ,

$$x_s^2 - \sum_{j=1}^{2} P_{j2}\left(\int_{\infty}^{P} \omega - u_s\right)x_s^{j-1} = 0 \tag{4.2.43}$$

for $s \neq r$. If we redefine $P=(x,\ y)$ as $P_2$ and $P_s$ as $P_1$ ,
The eq. (4.2.43) can be written as

$$x_1^2 - P_{22}(\beta)x_1 - P_{12}(\beta) = 0 \quad . \tag{4.2.44}$$

We can also re-define $P=(x,\ y)$ as $P_1$ and $P_s$ as $P_2$ , and since Pe functions are even,

$$x_2^2 - P_{22}(\beta)x_2 - P_{12}(\beta) = 0 \quad . \tag{4.2.45}$$

Then, in the case that the variables are of $\beta$ type, we have

$$P_{11}(\beta) = \frac{F(x_1,x_2)+2y_1y_2}{(x_1-x_2)^2} \tag{4.2.46}$$

$$P_{12}(\beta) = -x_1x_2 \tag{4.2.47}$$

$$P_{22}(\beta) = x_1 + x_2 \tag{4.2.48}$$

In the $\beta$ type , eq.(4.2.38) is changed to be

$$\begin{bmatrix} \frac{\partial x_1}{\partial \beta_1} & \frac{\partial x_1}{\partial \beta_2} \\ \frac{\partial x_2}{\partial \beta_1} & \frac{\partial x_2}{\partial \beta_2} \end{bmatrix} = \begin{bmatrix} \frac{\partial \beta_1}{\partial x_1} & \frac{\partial \beta_1}{\partial x_2} \\ \frac{\partial \beta_2}{\partial x_1} & \frac{\partial \beta_2}{\partial x_2} \end{bmatrix}^{-1} = \begin{bmatrix} \frac{1}{2y_1} & \frac{-1}{2y_2} \\ \frac{x_1}{2y_1} & \frac{-x_2}{2y_2} \end{bmatrix}^{-1} = \frac{1}{x_2 - x_1}\begin{bmatrix} 2x_2y_1 & -2y_1 \\ 2x_1y_2 & -2y_2 \end{bmatrix} \tag{4.2.49}$$

By this, we can calculate the corresponding formulas of (4.2.32),(4.2.34),(4.2.35), (4.2.36) for the $\beta$ type as , replacing $y_2$ with $-y_2$,

$$P_{111}(\beta) = -2\frac{y_2\psi(x_1,\ x_2) + y_1\psi(x_2,\ x_1)}{(x_1-x_2)^3} \tag{4.2.50}$$

$$P_{112}(\beta) = \ 2\frac{y_1x_2^2 + y_2x_1^2}{x_1 - x_2} \quad (4.2.51)$$

$$P_{122}(\beta) = \ -2\frac{y_2x_1 + y_1x_2}{x_1 - x_2} \quad (4.2.52)$$

$$P_{222}(\beta) = \ 2\frac{y_1 + y_2}{x_1 - x_2} \quad (4.2.53)$$

**(spin structure dependent parts written in branch points)**

In the $\alpha$-type case, in the final stage of the calculations we will set $\alpha$ equal to

$\int_{\infty}^{(e_i,0)} \omega + \int_{\infty}^{(e_j,0)} \omega$ for any combinations of $e_i$, $e_j$ which we will denote $\Omega_\delta$.

In this case, by equations (4.2.29) - (4.2.31) ,

$$P_{11}(\Omega_\delta) \ = \frac{F(e_i,e_j)}{(e_i - e_j)^2} \quad (4.2.54)$$

$$P_{12}(\Omega_\delta) \ = P_{21}(\Omega_\delta) = \ -e_ie_j \quad (4.2.55)$$

$$P_{22}(\Omega_\delta) \ = \ e_i + e_j \quad (4.2.56)$$

The right-hand side of (4.2.54) leads to, after some calculations, by using the fact that coefficients $\mu_i$ are expressed by the elementary symmetric polynomials of five branch points,

$P_{11}(\Omega_\delta) = \ (e_p + e_q + e_r)e_ie_j + e_pe_qe_r$ where $p,q,r$ are all different from $i,j$. (4.2.57)

By the known results (4.2.55), (4.2.56), (4.2.57), and by the fact that the partition functions are also written by branch points, spin structure sum can be done by algebra of branch points, same as in the genus 1 case. These (4.2.55)-(4.2.57) can be regarded as a natural generalizations of the genus one fact that $P$ $(\Omega_\delta)$ can be expressed by branch point $e_\delta$ and the spin sum can be performed by the simple algebra of branch points.

**(spin structure independent parts written in x,y coordinates of the vertex inserting points)**

On the other hand, the $\beta$ part corresponds to the spin structure-independent parts, which use the same explicit form of Kleinian polar $F$ as in eq.(2.20). The spin structure-independent term of $N = 2$ is expressed as $-P_{11}(\beta)$, whose expression can be found in eq. (4.2.46). $P_{12}(\beta)$ and $P_{22}(\beta)$ are expressed using fundamental symmetric functions of the two x-coordinates of the vertex insertion points, $x_1$ and $x_2$. $P_{11}(\beta)$, $P_{12}(\beta)$ and $P_{22}(\beta)$ are purely derivatives of theta functions. However, using the modified solutions of the Jacobi's inversion theorem, these can be expressed in terms of the $x$ and $y$ coordinates of the points on a Riemann surface.

The total of the $N=2$ result (4.2.22) can be rewritten as

$$-S_\delta(z_1,z_2)^2 = \left[\ \frac{F(e_i,e_j)}{(e_i-e_j)^2} - \frac{F(x_1,x_2)+2y_1y_2}{(x_1-x_2)^2} - (x_1+x_2)e_ie_j + x_1x_2(e_i+e_j)\ \right]\omega_1(z_1)\omega_1(z_2) \tag{4.2.58}$$

where any of the choice of the two branch points $e_i, e_j$ out of 5 points is denoted $\delta$ as an even spin structure, and the coordinates of the vertex insertion points are $z_1(x_1,\ y_1),\ z_2(x_2,\ y_2)$ .

**(correspondence with the hyper-elliptic formulation )**

For future reference, we compiled a list of the formal correspondences with the results in [15], which completely decomposed the products of fermion correlation functions in terms of x and y co-ordinates. From eq.(4.2.16),

$$S_\delta(z_1,z_2)^2 = -\frac{\sigma(\Omega_\delta+\beta)\,\sigma(\Omega_\delta-\beta)}{\sigma(\Omega_\delta)^2\,\sigma(\beta)^2}\omega_1(z_1)\omega_1(z_2) = -\frac{\sigma(\Omega_\delta+\beta)\,\sigma(\Omega_\delta-\beta)}{\sigma(\Omega_\delta)^2\,\sigma(\beta)^2}\frac{dx_1}{2y_1}\frac{dx_2}{2y_2}$$

$$= -\{P_{11}(\Omega_\delta) - P_{11}(\beta) + P_{12}(\Omega_\delta)P_{22}(\beta) - P_{12}(\beta)P_{22}(\Omega_\delta)\}\omega_1(z_1)\omega_1(z_2)$$

$$= \{-P_{11}(\Omega_\delta) + P_{11}(\beta) - (x_1+x_2)P_{12}(\Omega_\delta) - x_1x_2P_{22}(\Omega_\delta)\}\frac{dx_1}{2y_1}\frac{dx_2}{2y_2} \tag{4.2.59}$$

This corresponds to eq.(3.27) of [15] (page21) :

$$C_\delta(1,2) = N_\delta(1,2)\frac{dx_1}{2x_{12}s_1}\frac{dx_2}{2x_{21}s_2} = L_\delta(1,2) + \frac{Z(1,2)+s_1s_2}{2x_{12}x_{21}}\frac{dx_1}{s_1}\frac{dx_2}{s_2}$$

$$= \left[\ \mathbb{L}_\delta(1,2) + \frac{Z(1,2)+s_1s_2}{2x_{12}x_{21}}\right]\frac{dx_1}{s_1}\frac{dx_2}{s_2}$$

$$= \left[\ \ell_\delta^{11} - (x_1+x_2)\,\ell_\delta^{12} + x_1x_2\ell_\delta^{22} + \frac{Z(1,2)+s_1s_2}{2x_{12}x_{21}}\right]\frac{dx_1}{s_1}\frac{dx_2}{s_2}$$

$$= \left[\ 4\ell_\delta^{11} - 4(x_1+x_2)\,\ell_\delta^{12} + 4x_1x_2\ell_\delta^{22} + \frac{2Z(1,2)+2s_1s_2}{x_{12}x_{21}}\right]\frac{dx_1}{2s_1}\frac{dx_2}{2s_2} \tag{4.2.60}$$

Noting that $C_\delta(1,2) = S_\delta(z_1,z_2)S_\delta(z_2,z_1) = -S_\delta(z_1,z_2)^2$ ,

$$P_{11}(\Omega_\delta) \leftrightarrow 4\ell_\delta^{11} \tag{4.2.61}$$

$$P_{12}(\Omega_\delta) \leftrightarrow -4\ell_\delta^{12} \tag{4.2.62}$$

$$P_{22}(\Omega_\delta) \leftrightarrow 4\ell_\delta^{22} \tag{4.2.63}$$

$$-P_{11}(\beta) \leftrightarrow 4\,\frac{Z(1,2)+s_1s_2}{2x_{12}x_{21}} = \frac{2Z(1,2)+2s_1s_2}{x_{12}x_{21}} = -\frac{2Z(1,2)+2s_1s_2}{(x_1-x_2)^2} \tag{4.2.64}$$

$$F(x_1,\ x_2) \leftrightarrow 2Z(1,\ 2) \tag{4.2.65}$$

$$C_\delta(1,\dots,n) = N_\delta(1,\dots,n)\prod_{i=1}^{n}\frac{dx_i}{2x_{i,i+1}s_i} = \frac{N_\delta(1,\dots,n)}{x_{1,2}x_{2,3}\cdots x_{n,1}}\prod_{i=1}^{n}\frac{dx_i}{2s_i}$$

$$\leftrightarrow \prod_{i=1}^{N} S_\delta(z_i,z_{i+1}) = \prod_{i=1}^{N}\frac{\sigma(\beta_i+\Omega_\delta)}{\sigma(\beta_i)\sigma(\Omega_\delta)}\cdot\prod_{i=1}^{N}\omega_1(z_i) = \prod_{i=1}^{N}\frac{\sigma(A(z_i-z_{i+1})+\Omega_\delta)}{\sigma(A(z_i-z_{i+1}))\sigma(\Omega_\delta)}\cdot\prod_{i=1}^{N}\frac{dx_i}{2y_i}$$

(4.2.66)

**(spin structure dependence in terms of theta constants)**

The following is what we already described in [11], as a conjecture base. Now that we have proved (4.2.22), we can state that it is a fact.

The eq.(4.2.22), $-S_\delta(z_1,z_2)^2 = \sum_{I,J}[\,P_{IJ}(\Omega_\delta) - P_{IJ}(\beta)\,]\,\omega_I(z_1)\omega_J(z_2)$, can be written as

$$-S_\delta(z_1,z_2)^2 = \sum_{I,J}\left[\partial_I\partial_J \ln\theta_R\left((2M)^{-1}A(z_1-z_2)\right) + [2\eta(2M)^{-1}]_{IJ}\,\right]\omega_I(z_1)\omega_J(z_2) + $$
$$+\sum_{I,J} P_{IJ}(\Omega_\delta)\omega_I(z_1)\omega_J(z_2) \qquad (4.2.67)$$

On the other hand, it holds that, as the generalisation of Weierstrass formula of genus one, in the arbitrary genus of the hyper-elliptic case,

$$[2\eta(2M)^{-1}]_{IJ} = -P_{IJ}(\Omega_\delta) - [\,((2M)^{-1})^T H(2M)^{-1}]_{IJ} \qquad (4.2.68)$$

where a matrix $H$ is defined as

$$H_{KL} = \frac{\partial_K\partial_L\,\theta[\delta](0)}{\theta[\delta](0)}$$

for $I,J = 1,2,\dots g$ . (See eq. (2.100) in [21], page 47 )

The $\delta$ represents any non-singular even spin structure. Eq.(4.2.68) states that, since the spin structure dependence of the simple product of fermion correlation functions under the cyclic condition is only through $P_{IJ}(\Omega_\delta)$, we can say that in terms of theta constants the spin structure dependence is only through one kind of constants $\frac{\partial_K\partial_L\,\theta[\delta](0)}{\theta[\delta](0)}$.

By this formula, we have from (4.2.67),

$$S_\delta(z_1,z_2)^2 = -\sum_{I,J}\partial_I\partial_J \ln\theta_R\left((2M)^{-1}A(z_1-z_2)\right)\omega_I(z_1)\omega_J(z_2) + $$
$$+\sum_{I,J}\omega_I(z_1)\omega_J(z_2)\,[\,((2M)^{-1})^T H(2M)^{-1}]_{IJ} \qquad (4.2.69)$$

In a sense this means that if we get rid of the bad behavior constant $[2\eta(2M)^{-1}]_{IJ}$ from $\partial_I\partial_J \ln\theta_R$ to make $P_{IJ}$ , and add it to the theta constant $\frac{\partial^I\partial^J\,\theta[\delta](0)}{\theta[\delta](0)}$ ,then the sum of two bad behavior constants becomes $P_{IJ}(\Omega_\delta)$ which can be expressed in branch points and those will have good behavior combined with partition functions. Then all of the spin sums can come down to the branch points algebra.

This is a natural generalization of the genus 1 case which is described around eq.(1.46) in [11].

**(an example of the spin sum $(N=2)\times(N=2)$)**

We provide a brief explanation of a method of spin sum in our framework here to ensure the description is self-contained.

The spin sum method of genus 2 superstring amplitudes was developed and investigated thoroughly in [23] and its related works. For example, in the case of N=4 point amplitudes with external massless bosons, the $(N=2)\times(N=2)$ product of the fermion correlation functions $S_\delta(z_1,z_2)^2 S_\delta(z_3,z_4)^2$ is an example of Wick contraction which gives a non-zero result. The contribution of this to the integrand of amplitude is

$$Z_\delta * S_\delta(z_1,z_2)^2 S_\delta(z_3,z_4)^2 * Boson\, field\, contraction$$

where $Z_\delta$ is the partition function of the external massless boson amplitudes for the Type I and Type II superstring models which has the form[23]

$$Z_\delta = \frac{\Xi_6[\delta]\theta[\delta](0)^4}{\Psi_{10}}$$

We need to perform spin sum in which only the factor $Z_\delta\, S_\delta(z_1,z_2)^2 S_\delta(z_3,z_4)^2$ is relevant.

The function $S_\delta(z_1,z_2)^2$ is expressed as in eq.(4.2.22), then $S_\delta(z_1,z_2)^2 S_\delta(z_3,z_4)^2$ is equal to

$$\left[\sum_{I,J=1}^{2}\left[P_{IJ}(\Omega_\delta)-P_{IJ}(\beta_1)\right]\omega_I(z_1)\omega_J(z_2)\right]\left[\sum_{I,J=1}^{2}\left[P_{IJ}(\Omega_\delta)-P_{IJ}(\beta_3)\right]\omega_I(z_3)\omega_J(z_4)\right] \quad (4.2.70)$$

where $\beta_1$, $\beta_3$ are defined as in eq.(2.30).

In our framework, we need to modify the method given in [23] since there are 5 branch points instead of 6. Expanding (4.2.70) as a polynomial of $P_{AB}(\Omega_\delta)$, we obtained the following calculation results using the forms of $P_{22}(\Omega_\delta)$, $P_{12}(\Omega_\delta)$, $P_{11}(\Omega_\delta)$ in terms of branch points [8].

If the Wick contraction results contain only $P_{KL}(\Omega_\delta)$ $(K,L=1\ or\ 2)$ independent terms or linear monomials of $P_{KL}(\Omega_\delta)$ $(K,L=1\ or\ 2)$, the spin sum results are zero :

$$\sum_\delta Z_\delta = 0,\quad \sum_\delta Z_\delta\, P_{KL}(\Omega_\delta) = 0 \ \ (for\ any\ combinations\ of\ K,L=1,2) \quad (4.2.71)$$

If the Wick contraction gives quadratic form of $P_{KL}(\Omega_\delta)$, such as $\sum_\delta Z_\delta\, P_{IJ}(\Omega_\delta)\, P_{KL}(\Omega_\delta)$ $(I,J,K,L=1,2)$, only the following combinations $I,J,K,L$ give non-zero results :

$$\sum_\delta Z_\delta\, P_{11}(\Omega_\delta)\, P_{22}(\Omega_\delta) = -8\pi^4 \quad (4.2.72)$$

$$\sum_\delta Z_\delta\, P_{12}(\Omega_\delta)\, P_{12}(\Omega_\delta) = 4\pi^4 \quad (4.2.73)$$

$$\sum_\delta Z_\delta\, P_{IJ}(\Omega_\delta)\, P_{KL}(\Omega_\delta) = 0 \quad for\ all\ of\ other\ combinations \quad (4.2.74)$$

Then, in the integrand of the amplitudes, we only need to replace $P_{11}(\Omega_\delta)P_{22}(\Omega_\delta)$ by $-8\pi^4$, $P_{12}(\Omega_\delta)P_{12}(\Omega_\delta)$ by $4\pi^4$, and all other degree zero, one, two polynomial terms of $P_{AB}(\Omega_\delta)$ by zero. This replacement is equivalent to performing the entire process of multiplying the partition functions and summing over spin structures, as demonstrated in [9] for massless boson external states. This procedure yields the sum of spin structures $\sum_\delta Z_\delta\, S_\delta(z_1,z_2)^2 S_\delta(z_3,z_4)^2$ .

Inserting (4.2.72), (4.2.73), (4.2.74) into the expansion of the product (4.2.70), the spin sum reproduces the well-known results of [23] :

$$-4\pi^4\, [\, \Delta(z_1,z_3)\Delta(z_2,z_4) + \Delta(z_1,z_4)\Delta(z_2,z_3)\, ]$$

where

$$\Delta(x,y) = \omega_1(x)\omega_2(y) - \omega_2(x)\omega_1(y)$$

For any Wick contraction results in N point calculations, owing to the discovery of trilinear relations found in [15], the degree of polynomials of $P_{AB}(\Omega_\delta)$ can be reduced to 2 in any amplitude calculation. In our framework, it can be achieved using the relations eq.(3.20) – eq.(3.29). There is no need to renew the branch point algebra; the same replacement (4.2.71)-(4.2.74) can be applied to the result. That is, after reducing the degree of the polynomials to 2, we can perform the spin sum by repeating the same process of replacing $P_{12}(\Omega_\delta)P_{12}(\Omega_\delta)$ $P_{11}(\Omega_\delta)P_{22}(\Omega_\delta)$ with non-zero c-number.

## 4-3 N = 3

Define the coefficients of $H^3_0$, $H^3_{11}$, $H^3_{12}$, $H^3_{22}$ ,$H^3_{111}$, $H^3_{112}$ ,$H^3_{122}$, $H^3_{222}$, $H^3_P$ by the following equation:

$$\prod_{i=1}^{3} \frac{\sigma(\alpha+\beta_i)}{\sigma(\beta_i)\sigma(\alpha)} = H^3_0 + H^3_{11}P_{11}(\alpha) + H^3_{12}P_{12}(\alpha) + H^3_{22}P_{22}(\alpha)$$

$$+H^3_{111}P_{111}(\alpha) + H^3_{112}P_{112}(\alpha) + H^3_{122}P_{122}(\alpha) + H^3_{222}P_{222}(\alpha) + H^3_P P(\alpha)\,, \qquad (4.3.1)$$

We performed similar calculations to those in the case of $N = 2$ and found that

$$\sigma(\alpha)^3 = +\alpha_1^{\,3} + O(4)$$

$$\sigma(\alpha)^3 P_{11}(\alpha) = +\alpha_1 - \frac{1}{3}\alpha_2^{\,3} + \frac{1}{6}\mu_3\alpha_1^{\,3} + O(4)$$

$$\sigma(\alpha)^3 P_{12}(\alpha) = -\alpha_1\alpha_2^2 + O(4)$$

$$\sigma(\alpha)^3 P_{22}(\alpha) = +2\alpha_1^{\,2}\alpha_2 + O(4)$$

$$\sigma(\alpha)^3 P_{111}(\alpha) = -2 + \frac{1}{2}\mu_3\alpha_1^2 + O(4)$$

$$\sigma(\alpha)^3 P_{112}(\alpha) = +2\alpha_2^{\,2} + O(4)$$

$$\sigma(\alpha)^3 P_{122}(\alpha) = -2\alpha_1\alpha_2 + O(4)$$

$$\sigma(\alpha)^3 P_{222}(\alpha) = +2\alpha_1^{\,2} + O(4)$$

$$\sigma(\alpha)^3 P(\alpha) = +2\alpha_2 - 2A_{3,2}\alpha_1^3 - 6A_{2,3}\alpha_1^{\,2}\alpha_2 - 12A_{1,4}\alpha_1\alpha_2^2 - 20A_{0,5}\alpha_2^{\,3} + O(4)$$

$$= +2\alpha_2 + \frac{\mu_4}{3}\alpha_1^{\,3} + \mu_3\alpha_1^{\,2}\alpha_2 + \mu_2\alpha_1\alpha_2^{\,2} + O(4) \qquad (4.3.2)$$

Here, $O(4)$ are the polynomial terms of $\alpha_1, \alpha_2$ in which the sum of the degrees of $\alpha_1$and $\alpha_2$ are 4 or greater.

Multiplying $\sigma(\alpha)^3$ on both sides of the eq. (4.3.1) , we obtain the following results by performing $\partial$ calculations :

( noting again the signature of the definitions $\zeta_I = \partial_I ln\sigma,\ P_{IJ} = -\partial_I \zeta_J$ )

Leave everything as it is and simply set $\alpha_1 = \alpha_2 = 0$ : $H^3_{111} = \frac{1}{2}$ (4.3.3)

$$\partial_1\, e_0^{L_3} = +H^3_{11} \quad (4.3.4)$$

$$\partial_2\, e_0^{L_3} = +2H^3_P \quad (4.3.5)$$

$$\partial_1\partial_1\, e_0^{L_3} = +4H^3_{222} + \mu_3 H^3_{111} \quad (4.3.6)$$

$$\partial_1\partial_2\, e_0^{L_3} = -2H^3_{122} \quad (4.3.7)$$

$$\partial_2\partial_2\, e_0^{L_3} = +4H^3_{112} \quad (4.3.8)$$

$$\partial_1\partial_1\partial_1\, e_0^{L_3} = +6H^3_0 + \mu_3 H^3_{11} + 2\mu_4 H^3_P \quad (4.3.9)$$

$$\partial_1\partial_1\partial_2\, e_0^{L_3} = +4H^3_{22} + 2\mu_3 H^3_P \quad (4.3.10)$$

$$\partial_1\partial_2\partial_2\, e_0^{L_3} = -2H^3_{12} + 2\mu_2 H^3_P \quad (4.3.11)$$

$$\partial_2\partial_2\partial_2\, e_0^{L_3} = -2H^3_{11} \quad (4.3.12)$$

There are $N^2 = 9$ basis functions, and $\partial$ calculations give us $\frac{(N+1)(N+2)}{2} = 10$

equations.

To obtain the product $S_\delta(z_1,z_2)\, S_\delta(z_2,z_3)\, S_\delta(z_3,z_1)$ , we will later set $\alpha$ equal to $\Omega_\delta$ in the eq.(4.3.1). Since $P_{ABC}(\Omega_\delta)$=0 for any values of $ABC$, we only need to know the formulae of $H^3_0,\ H^3_{11},\ H^3_{12},\ H^3_{22}\ , H^3_P$ from eqs. (4.3.4), (4.3.5), (4.3.9), (4.3.10), (4.3.11), (4.3.12) above. We can see that the type $H_{ABC}$ and the type $H_{DE}$ coefficients are not mixed in the same equation in the above list.

Also, when looking at the above equations, we can see that for the even basis functions, such as $P_{11}(\alpha),\ P_{12}(\alpha), P_{22}(\alpha), P(\alpha)$ , the degrees of the right-hand sides as polynomial functions of $\alpha_1, \alpha_2$ are odd. For the odd basis functions, such as $P_{111}(\alpha), P_{112}(\alpha)$, $P_{122}(\alpha), P_{222}(\alpha)$ , the degree of the right-hand sides as polynomial functions of $\alpha_1, \alpha_2$ are even. Therefore, after calculating the coefficients, the results from the even and odd functions do not mix.

**( consistency identities and the zeta part of the amplitudes)**

In (4.3.3) – (4.3.12), we see that $H^3_{11}$ has two equations that need to be satisfied : (4.3.4) and (4.3.12).

$H^3_{11}$ can be solved from (4.3.4)

$$H^3_{11} = +\partial_1\, e_0^{L_3} = \sum_{i=1}^3 \zeta_1\big(A(z_i - z_{i+1})\big) = \sum_{i=1}^3 \zeta_1(\beta_i)\ . \quad (4.3.13)$$

Substituting this into (4.3.12) where we denote 3 variables inside the functions as $u, v, w$ which satisfy $u + v + w = 0$ ,we obtain the following identity:

$$2[\,\zeta_1(u) + \zeta_1(v) + \zeta_1(w)\,] - [\,P_{222}(u) + P_{222}(v) + P_{222}(w)]$$
$$-\,3[P_{22}(u) + P_{22}(v) + P_{22}(w)][\,\zeta_2(u) + \zeta_2(v) + \zeta_2(w)\,] + [\,\zeta_2(u) + \zeta_2(v) + \zeta_2(w)]^3 = 0$$

(4.3.14)

This equation has exactly the same form as that described in [21] page 85, eq. (5.12). It is a theta identity that holds for 3 variables $u, v, w$ with the constraint $u + v + w = 0$ in genus 2.

In genus one, it is known that, under the condition $\beta_1 + \beta_2 + \beta_3 = 0$ ( here $\beta_i \in C^1$ ) on the 3 variables, the following identity holds :

$$[\sum_{i=1}^{3} \partial_{x_i}^{1} \ln \theta_1(\beta_i)\,]^2 = \sum_{i=1}^{3} P\,(\beta_i) \qquad (4.3.15)$$

or equivalently,

$$[\sum_{i=1}^{3} \zeta(\beta_i)\,]^2 = \sum_{i=1}^{3} P\,(\beta_i)$$

Equation (4.3.14) is one of the genus 2 generalized formulae of (4.3.15).

The sigma function relations of our results (4.3.3) – (4.3.12) are identities that do not hold in the general case. They are only valid when the cyclic condition is satisfied. Calculations based on this assumption result in formulas that express consistency conditions, such as eq.(4.3.14).

Here, eq.(4.3.5) gives the explicit form of $H_P^3$ :

$$H_P^3 = +\frac{1}{2}\,\partial_2\, e_0^{L_3} = \frac{1}{2}\sum_{i=1}^{3} \zeta_2(\beta_i) \qquad (4.3.16)$$

On the other hand, the following relationship is described in [21] (page 85, above eq.(5.12) ) under the constraint $u + v + w = 0$ :

$$\zeta_2(u) + \zeta_2(v) + \zeta_2(w) = -h_1 \qquad (4.3.17)$$

where

$$h_1 = -2\frac{\begin{vmatrix} P_{21}(v)-P_{21}(u) & P_{22}(v)P_{21}(u)-P_{22}(u)P_{21}(v) \\ P_{22}(v)-P_{22}(u) & P_{21}(v)-P_{21}(u) \end{vmatrix}}{\begin{vmatrix} P_{21}(v)-P_{21}(u) & P_{221}(v)-P_{221}(u) \\ P_{22}(v)-P_{22}(u) & P_{222}(v)-P_{222}(u) \end{vmatrix}} \qquad (4.3.18)$$

[ **Note**: In paper [21], on page 85, the (1,2) component of the determinant in the numerator of the right-hand side of eq. (4.3.18) is described as $P_{22}(u)P_{21}(v) - P_{22}(v)P_{21}(u)$, but I think this is a typo and it should correctly be $P_{22}(v)P_{21}(u) - P_{22}(u)P_{21}(v)$ . The considerations below ( please see "General consideration of the zeta part" a few pages later) suggest that this is correct. I assume that the description of eq.(4.3.18) here is correct. ]

If we naively understand that the "cyclic condition" to mean that the sum of the variables is zero, $u + v + w = 0$ , then eq.(4.3.18) says that $\zeta_2(u) + \zeta_2(v) + \zeta_2(w)$ can be non-zero in general. Then, by (4.3.16), $H_P^3$ is non-zero.

This is not a good result because the factors $P_{11}P_{22}$, ${P_{12}}^2$ in the combination $P = P_{11}P_{22} - {P_{12}}^2$ are those that give non-zero contributions in the spin sum process for external massless boson amplitudes in genus two, and they do not vanish in the combination $P_{11}P_{22} - {P_{12}}^2$ . If $H_P^3$ is not zero, this implies that the spin sum result of $N = 3$ decomposition formula becomes non-zero, even in the case of three external massless bosons in Type I and II superstring amplitudes. This contradicts the well-known results.

Suppose we add the further condition that the variables $u, v, w$ should be of the $\beta$ type Abel map, such that $u = A(z_2 - z_1)$, $v = A(z_3 - z_2), w = A(z_1 - z_3)$, and assume that the x and y coordinates of $u, v, w$ are $(x_1, y_1)$, $(x_2, y_2)$, $(x_3, y_3)$ respectively. This is actually the case for string amplitudes calculations. Then, by the modified inversion formula,

$$P_{21}(u) = -x_1 x_2, \quad P_{22}(u) = x_1 + x_2, \quad P_{21}(v) = -x_2 x_3, \quad P_{22}(v) = x_2 + x_3\,. \tag{4.3.19}$$

In this case, the determinant of the numerator of $h_1$ of (4.3.18) becomes zero by the direct calculation. In other words, if the variables $u, v, w$ are represented by $\beta$ type Abel map in genus two, then

$$\zeta_2(u) + \zeta_2(v) + \zeta_2(w) = 0. \tag{4.3.20}$$

and, from eq.(4.3.14),

$$\zeta_1(u) + \zeta_1(v) + \zeta_1(w) = \frac{1}{2}[\, P_{222}(u) + P_{222}(v) + P_{222}(w)] \tag{4.3.21}$$

By these, we conclude that from (4.3.16) that

$$H_P^3 = 0\ , \tag{4.3.22}$$

as it should be. Later we will generalize (4.3.20) and (4.3.21) for arbitrary $N$ in the case of the cyclic condition $z_{N+1} = z_1$ in a different way.

This observation means that, to obtain consistent theta function expressions for the fermion correlation functions, we need to properly include the fact that the variables in the spin structure independent factors are expressed by $\beta$-type Abel map, adding to the naïve condition that stated the sum of all of the variables is zero.

**(N=3 result expressed in terms of theta functions)**

By multiplying $\prod_{i=1}^{3} \omega_1(z_i)$ and setting $\alpha = \Omega_\delta$ in (4.3.1), we obtain the following result for $N = 3$:

$$S_\delta(z_1, z_2) S_\delta(z_2, z_3) S_\delta(z_3, z_1) = (H_{11}^3 P_{11}(\Omega_\delta) + H_{12}^3 P_{12}(\Omega_\delta) + H_{22}^3 P_{22}(\Omega_\delta) + H_0^3\,) \prod_{i=1}^{3} \omega_1(z_i) \tag{4.3.23}$$

By using the variables $\beta_i = A(z_i - z_{i+1}) = \int_\infty^{(x_i, y_i)} \omega - \int_\infty^{(x_{i+1}, y_{i+1})} \omega$, $z_4 \equiv z_1$, it shows

$$\begin{aligned}
H_{11}^3 &= +\frac{1}{2}[\, \textstyle\sum_{i=1}^{3} P_{222}(\beta_i)\,] \\
H_{12}^3 &= +\frac{1}{2}[\textstyle\sum_{i=1}^{3} P_{122}(\beta_i)\,] + \frac{1}{4}[\textstyle\sum_{i=1}^{3} P_{22}(\beta_i)][\,\textstyle\sum_{i=1}^{3} P_{222}(\beta_i)] \\
H_{22}^3 &= -\frac{1}{4}[\textstyle\sum_{i=1}^{3} P_{112}(\beta_i)] - \frac{1}{4}[\textstyle\sum_{i=1}^{3} P_{12}(\beta_i)][\textstyle\sum_{i=1}^{3} P_{222}(\beta_i)] \\
H_0^3 &= -\frac{1}{6}[\, \textstyle\sum_{i=1}^{3} P_{111}(\beta_i)] - \frac{1}{4}[\textstyle\sum_{i=1}^{3} P_{11}(\beta_i)][\,\textstyle\sum_{i=1}^{3} P_{222}(\beta_i)] \\
&\quad + \frac{1}{48}[\textstyle\sum_{i=1}^{3} P_{222}(\beta_i)]^3 - \frac{\mu_3}{12}\,[\textstyle\sum_{i=1}^{3} P_{222}(\beta_i)]
\end{aligned} \tag{4.3.24}$$

Eq.(4.3.23) expresses the $N = 3$ result in terms of theta functions, as defined in (2.1) with Riemann constant characteristics, through the definition of the sigma function, as given in (2.16) :

$$\sigma(u) = \mathrm{c}\cdot\exp\left(\frac{1}{2}\ {}^t u\eta M^{-1}u\right)\frac{\theta_R((2M\ \ )^{-1}u,\tau)}{\partial_{u_1}\theta_R(0,\tau)}\quad,\quad \tau = M^{-1}\,\widehat{M}$$

and the definitions of Pe-functions described in (2.17),(2.18).

All Pe-functions $P_{111}(\beta), P_{112}(\beta), P_{122}(\beta), P_{222}(\beta)$ also have the $x, y$ coordinate expressions. These situations are similar in the case of genus one.

**( correspondence to the hyper-elliptic formulation )**

By noting the modified Jacobi inversion theorem, the relationship between the results (4.3.23), (4.3.24) and those obtained in the hyper-elliptic formulation in [15] becomes clear.

Below, we define $x_{ij} = x_i - x_j$ .

In [15], $N_\delta(1,2,3)$ is defined as [ page 38, eq.(5.3)]

$$N_\delta(1,2,3) = -4s_1x_{23}^2\,\mathbb{L}_\delta(2,3) - 4s_2x_{31}{}^2\mathbb{L}_\delta(3,1) - 4s_3x_{12}^2\,\mathbb{L}_\delta(1,2)$$
$$+2s_1Z(2,3) + 2s_2Z(3,1) + 2s_3Z(1,2) + 2s_1s_2s_3$$

It relates to the product of the fermion correlation functions, as given in eq.(3.1) in [15] :

$$C_\delta(1,2,3) = N_\delta(1,2,3)\ \frac{dx_1}{2x_{12}s_1}\frac{dx_2}{2x_{23}s_2}\frac{dx_3}{2x_{31}s_3}$$

We write this $C_\delta(1,2,3)$ as

$$C_\delta(1,2,3) = \ \left[A\left(4\ell_\delta^{11}\right) + B\left(-4\ell_\delta^{12}\right) + C\left(4\ell_\delta^{22}\right)\right]\frac{dx_1}{2s_1}\frac{dx_2}{2s_2}\frac{dx_3}{2s_3} + other\ terms \qquad (4.3.25)$$

Then, by their results $A,\ B,\ C$ are , after rewriting $s_i$ as $y_i$ ,

$$A = -[\frac{x_{23}}{x_{12}x_{31}}\,y_1 + \frac{x_{31}}{x_{12}x_{23}}\,y_2 + \frac{x_{12}}{x_{23}x_{31}}\,y_3\,] \qquad (4.3.26)$$

$$B\ = -[\,\frac{x_{23}}{x_{12}x_{31}}\,y_1(x_2+x_3) + \frac{x_{31}}{x_{12}x_{23}}\,y_2(x_3+x_1) + \frac{x_{12}}{x_{23}x_{31}}\,y_3(x_1+x_2)] \qquad (4.3.27)$$

$$C\ = -[\,\frac{x_{23}}{x_{12}x_{31}}\,y_1(x_2x_3) + \frac{x_{31}}{x_{12}x_{23}}\,y_2(x_3x_1) + \frac{x_{12}}{x_{23}x_{31}}\,y_3(x_1x_2)\,] \qquad (4.3.28)$$

On the other hand, from the formulas in (4.2.50) – (4.2.53) ,

$$[\textstyle\sum P_{222}] = 2\frac{y_1+y_2}{x_1-x_2} + 2\frac{y_2+y_3}{x_2-x_3} + 2\frac{y_3+y_1}{x_3-x_1} = -2(\frac{x_{23}}{x_{12}x_{31}}\,y_1 + \frac{x_{31}}{x_{12}x_{23}}\,y_2 + \frac{x_{12}}{x_{23}x_{31}}\,y_3)$$

(4.3.29)

$$[\textstyle\sum P_{122}] = -2\frac{x_2y_1+x_1\,y_2}{x_1-x_2} - 2\frac{x_3y_2+x_2\,y_3}{x_2-x_3} - 2\frac{x_1y_3+x_3\,y_1}{x_3-x_1} \qquad (4.3.30)$$

$$= \ 2\left(\frac{x_{23}x_1}{x_{12}x_{31}}\,y_1 + \frac{x_{31}x_2}{x_{12}x_{23}}\,y_2 + \frac{x_{12}x_3}{x_{23}x_{31}}\,y_3\right) \tag{4.3.31}$$

$$[\,\textstyle\sum P_{112}] = 2\,\frac{x_2^2\,y_1 + x_1^2\,y_2}{x_2 - x_1} \; + \; 2\,\frac{x_3^2\,y_2 + x_2^2\,y_3}{x_3 - x_2} \; + \; 2\,\frac{x_1^2\,y_3 + x_3^2\,y_1}{x_1 - x_3}$$

$$= \ 2\left(\frac{x_{23}(x_2x_3 - x_1x_2 - x_1x_3)}{x_{12}x_{31}}\,y_1 + \frac{x_{31}(x_1x_3 - x_1x_2 - x_2x_3)}{x_{12}x_{23}}\,y_2 + \frac{x_{12}(x_1x_2 - x_3x_2 - x_3x_1)}{x_{23}x_{31}}\,y_3\right) \tag{4.3.32}$$

By comparing (4.3.26), (4.3.29)

$$A \ = \ +\tfrac{1}{2}[\,\textstyle\sum P_{222}] = \ H_{11}^3 \tag{4.3.33}$$

Also, noting $\sum P_{22} = 2(x_1 + x_2 + x_3\,)$ and $\sum P_{12} \ = \ -(x_1x_2 + x_2x_3 + x_3x_1\,)$, we can see that

$$B \ = \ +\tfrac{1}{2}[\,\textstyle\sum P_{122}] + \tfrac{1}{4}[\,\sum P_{22}][\,\sum P_{222}] = \ H_{12}^3 \tag{4.3.34}$$

$$C \ = \ -\tfrac{1}{4}[\,\textstyle\sum P_{112}] - \tfrac{1}{4}[\,\sum P_{12}][\,\sum P_{222}] = \ H_{22}^3 \tag{4.3.35}$$

Therefore, each of $H_{11}^3, H_{12}^3\,, H_{12}^3$ expressed in the modified Jacobi inversion formula is certainly the $A, B, C$ which is corresponding coefficient of the spin structure dependent factors $4\ell_\delta^{11}, -4\ell_\delta^{12}, 4\ell_\delta^{22}$ in [15], by using the results (4.2.46) – (4.2.53).

As for the spin-structure independent terms $H_0^3$ , it can be obtained by $\partial_1\partial_1\partial_1\, e_0^{L_3}$ after obtaining other coefficients $H_{11}^3,\ H_{12}^3\,, H_{22}^3$ , as a theta function expression.
It is also directly confirmed that the theta function expression of $H_0^3$ in (4.3.24)：

$$-\tfrac{1}{6}[\,\textstyle\sum P_{111}] - \tfrac{1}{4}[\,\sum P_{11}][\,\sum P_{222}] + \tfrac{1}{48}(\,\sum P_{222})^3 - \tfrac{\mu_3}{12}\,[\sum P_{222}]$$

can also be written as $D$ defined by

$$D \equiv \frac{y_1 F(x_2, x_3) + y_2 F(x_3, x_1) + y_3 F(x_1, x_2) + 2y_1y_2y_3}{x_{12}x_{23}x_{31}} \tag{4.3.36}$$

where $F\left(x_i, x_j\right)$ are Kleinian 2-polars defined in (2.20) .
Formula (4.3.36) corresponds to the result of $N = 3$ spin structure-independent term in the last term of eq.(5.4) with (5.5), on page 38 of [15].

The total $N = 3$ result (4.3.23) in the theta function expression can be written in terms of $x$ and $y$ coordinates as follows:

$$S_\delta(z_1, z_2)S_\delta(z_2, z_3)S_\delta(z_3, z_1) = (A \cdot P_{11}(\Omega_\delta) + B \cdot P_{12}(\Omega_\delta) + C \cdot P_{22}(\Omega_\delta) + D\,)\textstyle\prod_{i=1}^3 \omega_1(z_i) \tag{4.3.37}$$

where $A, B, C, D$ are expressed in (4.3.26), (4.3.27),(4.3.28), (4.3.36) respectively.
This is consistent with the result obtained first in [15].

In genus one, the decomposition of the case $N = 3$ has the form

$$S_\delta(\beta_1)S_\delta(\beta_2)S_\delta(\beta_3)$$

$$= \tfrac{1}{6}\left(\textstyle\sum_{i=1}^3 \partial_{x_i}^3 \ln\theta_1\,(\beta_i)\right) - \tfrac{1}{2}\left[\,\sum_{i=1}^3 \partial_{x_i}^1 \ln\theta_1\,(\beta_i)\ \right][\,\sum_{i=1}^3 P\,(\beta_i)]$$

$+\frac{1}{6}(\sum_{i=1}^{3}\partial_{x_i}^{1}\ln\theta_1(\beta_i))^3+\left[\sum_{i=1}^{3}\partial_{x_i}^{1}\ln\theta_1(\beta_i)\right]e_\delta$ (4.3.38)

The spin structure-independent terms ( $e_\delta$ independent terms) correspond to $H_0^3$ . Naively, in genus 2, the first 3 terms in eq.(4.3.24), $-\frac{1}{6}[\sum P_{111}]-\frac{1}{4}[\sum P_{11}][\sum P_{222}]+\frac{1}{48}(\sum P_{222})^3$ have a similar form to the genus one spin structure results, considering the signature of the Pe function and the fact that $\sum\frac{1}{2}P_{222}$ correspond to $\sum_{i=1}^{3}\partial_{x_i}^{1}\ln\theta_1(\beta_i)$ . However, an additional term $-\frac{\mu_3}{12}[\sum P_{222}]$ is necessary in genus 2 to reproduce the result (4.3.36).

The coefficient of $e_\delta$ in (1.50) corresponds to $H_{11}^3$ in genus 2 in (4.3.23).

**(general consideration on the zeta part)**

In previous results on superstring amplitudes in one-loop and two loop five-point calculations, we observed that the one-time derivatives of the logarithm of the odd theta functions remained non-zero after the spin sum. These can be expressed as $\partial_1\ln\sigma(\beta_i)$, $\partial_2\ln\sigma(\beta_i)$, and these are the $\zeta$ functions defined in eq. (2.17) in our framework. These functions often appear in the final form of the amplitudes after the spin sum. They are generally non-Abelian.

In the $N=3$ case, we saw that $\zeta_2(u)+\zeta_2(v)+\zeta_2(w)=0$ and $\zeta_1(u)+\zeta_1(v)+\zeta_1(w)=\frac{1}{2}[P_{222}(u)+P_{222}(v)+P_{222}(w)]$ in eqs.(4.3.20), (4.3.21). We can therefore use the function $P_{222}$ in the corresponding part instead of non-abelian one-time derivatives $\zeta_1$, $\zeta_2$.

Investigating this kind of consistency condition for the zeta function part step by step for every value of $N$ would be difficult. Instead, we may be able to derive a general result for this part as follows.

It is known that the following relations hold for the meromorphic differentials of the second kind, defined in (2.7) , if the variables are of the $\alpha$-type. ( See page.38, eq.(2.57) of [21] , [12])

Suppose that the $\alpha$ -type variable of the functions $\alpha=(\alpha_1,\alpha_2)$ has the form $\alpha_1=(x_1,y_1)$ , $\alpha_2=(x_2,y_2)$. Then

$$\int_\infty^{(x_1,y_1)}r_1+\int_\infty^{(x_2,y_2)}r_1=\zeta_1(\alpha)-\frac{1}{2}P_{222}(\alpha) \quad (4.3.39)$$

$$\int_\infty^{(x_1,y_1)}r_2+\int_\infty^{(x_2,y_2)}r_2=\zeta_2(\alpha) \quad (4.3.40)$$

where $P_{222}(\alpha)=2\frac{y_2-y_1}{x_2-x_1}$.

Here, $r_1$ , $r_2$ are the second kind differential defined in (2.7) :

$$r_1=-(\mu_2x+2\mu_1x^2+3x^3)\frac{dx}{2y} \qquad r_2=-x^2\frac{dx}{2y}$$

which are used in the definitions of $\eta_{IJ}$ , $\hat{\eta}_{IJ}$ in eq. (2.5 ).

Unlike in the case of genus one, for genus 2 and higher genera where functions have multiple variables, a non-zero function such as $-\frac{1}{2}\, P_{222}(\alpha)$ must be added to the right-hand side of (4.3.39) to satisfy the consistency condition

$\partial_2\zeta_1(\alpha) = \partial_1\zeta_2(\alpha)$ $\;(= -P_{12}(\alpha) = +x_1x_2)$ ( note again the signature ) (4.3.41)

Differentiating the left-hand side of (4.3.39) with respect to $\alpha_2$ for example, we have

$$-\frac{3x_1^3+2\mu_1x_1^2+\mu_2x_1}{2y_1}\,\frac{\partial x_1}{\partial\alpha_2} - \frac{3x_2{}^3+2\mu_1\,x_2^2+\mu_2x_2}{2y_2}\,\frac{\partial x_2}{\partial\alpha_2} \quad . \tag{4.3.42}$$

By the same differentiation, the right-hand side of (4.3.39) becomes

$$-P_{12}(\alpha) - \frac{1}{2}\, P_{2222}(\alpha) \tag{4.3.43}$$

Eq.(4.3.42) is equal to (4.3.43) by the formulas (4.2.29) – (4.2.36) and (4.2.39).

In case that the variable is of the $\beta$ type, we expect that eq.(4.3.39) will be modified as

$$\int_\infty^{(x_1,y_1)} r_1 \;-\; \int_\infty^{(x_2,y_2)} r_1 \;=\; \zeta_1(\beta) - \frac{1}{2}\, P_{222}(\beta) \tag{4.3.44}$$

$$\int_\infty^{(x_1,y_1)} r_2 \;-\; \int_\infty^{(x_2,y_2)} r_2 \;=\; \zeta_2(\beta) \tag{4.3.45}$$

with the replacement $y_2 \to -y_2$ where, in this case, $P_{222}(\beta) = 2\,\frac{y_1+y_2}{x_2-x_1}$

The consistent condition, which is obtained by differentiating the both sides of (4.3.44) and (4.3.45), can be satisfied in the same way as in the $\alpha$ -type case.

In case that the variables are of $\beta$ type, eq.(4.3.42) will be changed into

$$-\frac{3x_1^3+2\mu_1x_1^2+\mu_2x_1}{2y_1}\,\frac{\partial x_1}{\partial\beta_2} + \frac{3x_2{}^3+2\mu_1\,x_2^2+\mu_2x_2}{2y_2}\,\frac{\partial x_2}{\partial\beta_2} \quad . \tag{4.3.46}$$

And eq.(4.3.43) will be changed into $-P_{12}(\beta) - \frac{1}{2}\, P_{2222}(\beta)$ (4.3.47)

For eq.(4.3.46) and eq.(4.3.47) to be equal, eq.(4.3.39) and eq.(4.3.40) would need to be replaced by (4.3.44) and (4.3.45).

Therefore, if we also require that the cyclic condition $z_{N+1} = z_1$ to be satisfied,

$$0 = \sum_{i=1}^N\Big[\int_\infty^{(x_i,y_i)} r_1 - \int_\infty^{(x_{i+1},y_{i+1})} r_1\Big] = \sum_{i=1}^N \zeta_1(\beta_i) - \frac{1}{2}\sum_{i=1}^N P_{222}(\beta_i) \tag{4.3.48}$$

$$0 = \sum_{i=1}^N\Big[\int_\infty^{(x_i,y_i)} r_2 - \int_\infty^{(x_{i+1},y_{i+1})} r_2\Big] = \sum_{i=1}^N \zeta_2(\beta_i) \tag{4.3.49}$$

That is, in genus 2, if the cyclic condition $z_{N+1} = z_1$ is satisfied AND the variables corresponding to the vertex insertion points on the Riemann surface are of $\beta$ type, for any fixed value of $N$ , we have

$$\sum_{i=1}^N \zeta_1(\beta_i) = \frac{1}{2}\sum_{i=1}^N P_{222}(\beta_i) \tag{4.3.50}$$

$$\sum_{i=1}^N \zeta_2(\beta_i) = 0. \tag{4.3.51}$$

This means that the $\zeta$ part of the decomposed formulae ( i.e. one-time differentiation part of the ln of the sigma function) written using non-Abelian functions can be replaced by the Abelian function $P_{222}(\beta)$ under these conditions.

**( The case of arbitrary odd value of N for the zeta part)**

An example of this can be seen in the $N$=5 case. As in [15], eq. (5.12) on page 41,

$$C_\delta(1,2,3,4,5) = \left\{ \frac{x_{41}\, dx_5}{4x_{45}x_{51}} [L_\delta(1,2)L_\delta(3,4) + L_\delta(1,4)L_\delta(2,3)] + cycl(1,2,3,4,5) \right\} + otherterms$$

The coefficient of the $\ell_\delta^{11}\ell_\delta^{11}$ corresponds to the coefficient of $P_{11}(\Omega_\delta)P_{11}(\Omega_\delta)$ and is proportional to

$$\{\frac{s_5\, x_{41}}{x_{45}x_{51}} + cyclic(1,2,3,4,5)\}\ \frac{dx_1\, dx_2\, dx_3\, dx_4\, dx_5}{s_1 s_2 s_3 s_4 s_5}$$

Since $s_i$represents $y_i$ within the bracket in the above expression, it is equal to

$$\frac{y_5\, x_{41}}{x_{45}x_{51}} + cyclic(1,2,3,4,5) = \frac{y_1+y_2}{x_1-x_2} + \frac{y_2+y_3}{x_2-x_3} + \frac{y_3+y_4}{x_3-x_4} + \frac{y_4+y_5}{x_4-x_5} + \frac{y_5+y_1}{x_5-x_1} \quad . \qquad (4.3.52)$$

This is the same as our result $\partial_1 e_0^{L_5} = \sum_{i=1}^{5} \zeta_1(\beta_i) = \frac{1}{2} \sum_{i=1}^{5} P_{222}(\beta_i)$ , by using the modified inversion formula eq.(4.2.53), $P_{222}(\beta) = 2\frac{y_1+y_2}{x_1-x_2}$

The same equation applies to any odd $N$ case as shown in eq.(4.3.52):

$$\frac{y_N\, x_{N-1,1}}{x_{N-1,N}x_{N,1}} + cyclic(1,2,\dots N) = \frac{y_1+y_2}{x_1-x_2} + \frac{y_2+y_3}{x_2-x_3} \dots + \frac{y_N+y_1}{x_N-x_1} = \frac{1}{2}\sum_{i=1}^{N} P_{222}(\beta_i) \quad .$$

(4.3.53)

This shows one example of the validity of the equation (4.3.50), (4.3.51). For the terms from small number of differentiations, it is possible to derive such a general result by the simple $\partial$- calculations.

**(second example of spin sum $(\boldsymbol{N}=\mathbf{3}) \times (\boldsymbol{N}=\mathbf{2})$ )**

In the context of $N = 5$ point amplitudes with external massless bosons, the $(N = 3) \times (N = 2)$ product of the fermion correlation functions $S_\delta(z_1, z_2)S_\delta(z_2, z_3)S_\delta(z_3, z_1)S_\delta(z_4, z_5)^2$ serves as an illustration of a Wick contraction, yielding a non-zero result. As demonstrated in [24], this product is investigated in $J_1(z_1, z_2, z_3; z_4, z_5\,)$ in eq.(3.6) in [24].

In this product, the following results from (4.3.23), (4.3.37) will be utilized

$$\begin{aligned} S_\delta(z_1, z_2)S_\delta(z_2, z_3)S_\delta(z_3, z_1) &= (H_{11}^3 P_{11}(\Omega_\delta) + H_{12}^3 P_{12}(\Omega_\delta) + H_{22}^3 P_{22}(\Omega_\delta) + H_0^3\,) \textstyle\prod_{i=1}^{3} \omega_1(z_i) \\ &= (A \cdot P_{11}(\Omega_\delta) + B \cdot P_{12}(\Omega_\delta) + C \cdot P_{22}(\Omega_\delta) + D) \textstyle\prod_{i=1}^{3} \omega_1(z_i) \end{aligned}$$

(4.3.54)

where $A, B, C, D$ are expressed in (4.3.26) – (4.3.28), (4.3.33) – (4.3.35) as

$$A = -[\frac{x_{23}}{x_{12}x_{31}} y_1 + \frac{x_{31}}{x_{12}x_{23}} y_2 + \frac{x_{12}}{x_{23}x_{31}} y_3 ] = +\frac{1}{2}[\sum P_{222}] = H^3_{11}$$

$$B = -[\frac{x_{23}}{x_{12}x_{31}} y_1(x_2+x_3) + \frac{x_{31}}{x_{12}x_{23}} y_2(x_3+x_1) + \frac{x_{12}}{x_{23}x_{31}} y_3(x_1+x_2)]$$

$$= +\frac{1}{2}[\sum P_{122}] + \frac{1}{4}[\sum P_{22}][\sum P_{222}] = H^3_{12}$$

$$C = -[\frac{x_{23}}{x_{12}x_{31}} y_1(x_2x_3) + \frac{x_{31}}{x_{12}x_{23}} y_2(x_3x_1) + \frac{x_{12}}{x_{23}x_{31}} y_3(x_1x_2)]$$

$$= -\frac{1}{4}[\sum P_{112}] - \frac{1}{4}[\sum P_{12}][\sum P_{222}] = H^3_{22}$$

. By eq.(4.2.1) or (4.2.16), with noting that $S_\delta(z_4,z_5)S_\delta(z_5,z_4) = -S_\delta(z_4,z_5)^2$,

$$S_\delta(z_4,z_5)^2 = \{-P_{11}(\Omega_\delta) + P_{11}(\beta_4) - P_{12}(\Omega_\delta)P_{22}(\beta) + P_{12}(\beta_4)P_{22}(\Omega_\delta)\}\omega_1(z_4)\omega_1(z_5)$$
$$= \{-P_{11}(\Omega_\delta) + P_{11}(\beta_4) - (x_4+x_5)P_{12}(\Omega_\delta) - x_4x_5P_{22}(\Omega_\delta)\}\omega_1(z_4)\omega_1(z_5) \quad (4.3.55)$$

where $\beta_4$ , that depends on $z_4$ and $z_5$ , is an abbreviated expression of the variable

$\beta_i = (\beta_{i,1}, \beta_{i,2})$ with $\beta_{i,1} = \int_{(x_{i+1},y_{i+1})}^{(x_i,y_i)} \omega_1$ , , $\beta_{i,2} = \int_{(x_{i+1},y_{i+1})}^{(x_i,y_i)} \omega_2$ , as described in (2.28), (2.29), (2.30).

Following the multiplication of (4.3.54) and (4.3.55), it is possible to extract from the product solely those terms which are dependent on $P_{11}(\Omega_\delta)\, P_{22}(\Omega_\delta)$ and $P_{12}(\Omega_\delta)\, P_{12}(\Omega_\delta)$, where the spin sum result is nonzero.

With regard to x, y co-ordinate expressions, the following can be deduced from the modified Jacobi inversion formulae (that is, using $A, B, C$ and $x_4, x_5$ ) :

The sum of non-zero contribution terms

$$= -[(A\cdot x_4x_5 + C)\, P_{11}(\Omega_\delta)P_{22}(\Omega_\delta) + B\cdot(x_4+x_5)P_{12}(\Omega_\delta)P_{12}(\Omega_\delta)] \prod_{i=1}^{5} \omega_1(z_i)$$

$$= [\frac{x_{12}}{x_{23}x_{31}} y_3\omega_1(z_1)\omega_1(z_2)\omega_1(z_3)\{(x_4x_5 + x_1x_2)\, P_{11}(\Omega_\delta)P_{22}(\Omega_\delta)$$

$$+(x_1+x_2)(x_4+x_5)P_{12}(\Omega_\delta)P_{12}(\Omega_\delta)\} + (cyclic\ 1,2,3)]\, \omega_1(z_4)\omega_1(z_5) \quad (4.3.56)$$

[ The application of the cyclic sum is restricted exclusively to the indices $i$ in $x_i, y_i, \omega_1(z_i)$. $P_{11}(\Omega_\delta)P_{22}(\Omega_\delta)$, $P_{12}(\Omega_\delta)P_{12}(\Omega_\delta)$ do not receive the cyclic replacements 1,2,3 .]

As outlined in Chapter 4-1, in the integrand of the amplitudes, we need to replace $P_{11}(\Omega_\delta)P_{22}(\Omega_\delta)$ by $-8\pi^4$ , $P_{12}(\Omega_\delta)P_{12}(\Omega_\delta)$ by $4\pi^4$ , and all others such as degree zero, one, two polynomial terms of $P_{AB}(\Omega_\delta)$ by zero. This procedure automatically gives the result of the sum of spin structures. That is, this replacing is equivalent to performing the whole process of multiplying the partition functions and summing over spin structures in the case of massless boson external states. By the replacements and by using $\omega_1(z_3) = \frac{dx_3}{2y_3}$ , the spin sum result is derived, denoted J :

$$J = (4\pi^4)[\frac{x_{12}}{x_{23}x_{31}}\frac{dx_3}{2}\,\omega_1(z_1)\omega_1(z_2)\{-2x_4x_5 - 2x_1x_2 + (x_1+x_2)(x_4+x_5)\}$$

$$+(cyclic\ 1,2,3)\,]\,\omega_1(z_4)\omega_1(z_5)$$

$$= (-2\pi^4)[\frac{x_{12}dx_3}{x_{23}x_{31}}\{(x_4-x_1)(x_5-x_2)+(x_5-x_1)(x_4-x_2)\}\,\omega_1(z_1)\omega_1(z_2)$$

$$+(cyclic\ 1,2,3)\,]\,\omega_1(z_4)\omega_1(z_5) \qquad (4.3.57)$$

By $x_i\omega_1(z_i) = \omega_2(z_i)$, we see that

$$[(x_4-x_1)(x_5-x_2)+(x_5-x_1)(x_4-x_2)]\,\omega_1(z_1)\omega_1(z_2)\omega_1(z_4)\omega_1(z_5)$$

$$= \Delta(z_1,z_4)\Delta(z_2,z_5) + \Delta(z_1,z_5)\Delta(z_2,z_4) \qquad (4.3.58)$$

where $\Delta(x,y) = \omega_1(x)\omega_2(y) - \omega_2(x)\omega_1(y)$ .

Then,

$$J = -2\pi^4\frac{x_{12}dx_3}{x_{23}x_{31}}\{\,\Delta(z_1,z_4)\Delta(z_2,z_5) + \Delta(z_1,z_5)\Delta(z_2,z_4)\,\} + (cycl\ 1,2,3) \qquad (4.3.59)$$

This result is consistent with that of $J_1(z_1,z_2,z_3;z_4,z_5)$, as shown in eq.(3.7) of [24]. It can thus be concluded that, by means of the pole structure argument presented in Appendix E of [24], the right-hand of (4.3.59) can be rewritten in a form in which the arbitrary odd theta function used in the prime form and the holomorphic Abelian differentials, as presented in eq.(3.14), [24], can be used.

In terms of theta functions with Rieman constants characteristic, the sum of non-zero contributions (4.3.56) can also be expressed as follows :

$$-[\,(-H^3_{11}\cdot P_{12}(\beta_4) + H^3_{22})\,P_{11}(\Omega_\delta)P_{22}(\Omega_\delta) + H^3_{12}\cdot P_{22}(\beta_4)P_{12}(\Omega_\delta)P_{12}(\Omega_\delta)]\prod_{i=1}^{5}\omega_1(z_i) \qquad (4.3.60)$$

Replacing $P_{11}(\Omega_\delta)P_{22}(\Omega_\delta)$ by $-8\pi^4$ , $P_{12}(\Omega_\delta)P_{12}(\Omega_\delta)$ by $4\pi^4$, we obtain the spin sum result in another form :

$$J = (-4\pi^4)[\,\left(\textstyle\sum_{i=1}^{3}P_{222}(\beta_i)\right)\cdot P_{12}(\beta_4)$$
$$+\{\tfrac{1}{2}\left(\textstyle\sum_{i=1}^{3}P_{122}(\beta_i)\right) + \tfrac{1}{4}\left(\textstyle\sum_{i=1}^{3}P_{22}(\beta_i)\right)\left(\textstyle\sum_{i=1}^{3}P_{222}(\beta_i)\right)\}\,P_{22}(\beta_4)$$
$$+\tfrac{1}{2}\left(\textstyle\sum_{i=1}^{3}P_{112}(\beta_i)\right) + \tfrac{1}{2}\left(\textstyle\sum_{i=1}^{3}P_{12}(\beta_i)\right)\left(\textstyle\sum_{i=1}^{3}P_{222}(\beta_i)\right)\,]\prod_{i=1}^{5}\omega_1(z_i) \qquad (4.3.61)$$

## 4-4 N = 4

For the $N = 4$ case, we write the expansion as

$$\prod_{i=1}^{4}\frac{\sigma(\alpha+u_i)}{\sigma(u_i)\sigma(\alpha)} = H^4_0 + H^4_{11}P_{11}(\alpha) + H^4_{12}P_{12}(\alpha) + H^4_{22}P_{22}(\alpha)$$

$$+H^4_{111}P_{111}(\alpha) + H^4_{112}P_{112}(\alpha) + H^4_{122}P_{122}(\alpha) + H^4_{222}P_{222}(\alpha) + H^4_{P}P(\alpha)$$
$$+H^4_{1111}P_{11}(\alpha)P_{11}(\alpha) + H^4_{1112}P_{11}(\alpha)P_{12}(\alpha) + H^4_{11,22}P_{11}(\alpha)P_{22}(\alpha)$$

$$+H^4_{1222}P_{12}(\alpha)P_{22}(\alpha)+H^4_{2222}P_{22}(\alpha)P_{22}(\alpha)\ +H^4_{P\alpha 1}\frac{\partial P}{\partial\alpha_1}+H^4_{P\alpha 2}\frac{\partial P}{\partial\alpha_2}$$

(4.4.1)

where $P=P_{11}P_{22}-P_{12}{}^2$, and in the last two terms of (4.4.1), $\frac{\partial P}{\partial\alpha_1}$ and $\frac{\partial P}{\partial\alpha_2}$ are defined as derivatives of $P$ as given in (2.44).

After some calculations, we have the following results in the calculations up to $O(4)$ level for $N=4$ :

$\sigma(\alpha)^4=\ \alpha_1^4+O(5)$ (4.4.2)

$\sigma(\alpha)^4P_{11}(\alpha)=\ \alpha_1^2-\frac{2}{3}\alpha_2{}^3\,\alpha_1+\frac{1}{3}\mu_3\alpha_1^4+O(5)$ (4.4.3)

$\sigma(\alpha)^4P_{12}(\alpha)=\ -\alpha_1^2\alpha_2^2\ +O(5)$ (4.4.4)

$\sigma(\alpha)^4P_{22}(\alpha)=+2\alpha_1^3\alpha_2\ +O(5)$ (4.4.5)

$\sigma(\alpha)^4P_{111}(\alpha)=\ -2\sigma(\alpha)-\frac{1}{2}\mu_3\alpha_1^3=\ +2\alpha_1-\frac{2}{3}\alpha_2{}^3-\frac{1}{6}\mu_3\alpha_1^3\ \ +O(5)$ (4.4.6)

$\sigma(\alpha)^4P_{112}(\alpha)=\ -2\alpha_1\alpha_2{}^2\ +O(5)$ (4.4.7)

$\sigma(\alpha)^4P_{122}(\alpha)=\ +2\alpha_1{}^2\alpha_2\ +O(5)$ (4.4.8)

$\sigma(\alpha)^4P_{222}(\alpha)=\ -2\alpha_1{}^3+O(5)$ (4.4.9)

$$\begin{aligned}\sigma(\alpha)^4P(\alpha)\ &=2\alpha_1\alpha_2-\frac{2}{3}\alpha_2{}^4+\frac{1}{3}\mu_3\alpha_1^3\alpha_2-\textstyle\sum_{k+l=5}l(l-1)A_{kl}\alpha_1^{k+1}\alpha_2^{l-2}+O(5)\\ &=2\alpha_1\alpha_2-\frac{2}{3}\alpha_2{}^4+\frac{1}{3}\mu_3\alpha_1^3\alpha_2-12A_{14}\alpha_1^2\alpha_2{}^2-6A_{23}\alpha_1^3\alpha_2-2A_{32}\alpha_1^4+O(5)\\ &=2\alpha_1\alpha_2-\frac{2}{3}\alpha_2{}^4+\frac{4}{3}\mu_3\alpha_1^3\alpha_2+\mu_2\alpha_1^2\alpha_2{}^2+\frac{1}{3}\mu_4\alpha_1^4+O(5)\end{aligned}\qquad(4.4.10)$$

where

$A_{5,0}=+\frac{(\mu_3^2+2\mu_2\mu_4)}{5!}$, $A_{4,1}=-\frac{8\mu_5}{4!}$, $A_{3,2}=-\frac{\mu_4}{6}$, $A_{2,3}=-\frac{\mu_3}{6}$, $A_{1,4}=-\frac{\mu_2}{12}$, $A_{0,5}=0$,

as in (2.22). Also,

$$\begin{aligned}\sigma(\alpha)^4P_{11}(\alpha)P_{11}(\alpha)&=\ 1+\frac{1}{6}\mu_3^2\alpha_1{}^4+\frac{2}{3}\mu_3\alpha_1\alpha_2^3+\textstyle\sum_{k+l=5}2k(3-k)A_{kl}\alpha_1^{k-1}\alpha_2^l+O(5)\\ &=1-\frac{1}{3}\mu_2\mu_4\alpha_1^4-\frac{1}{3}\mu_2\alpha_2^4+\frac{8}{3}\mu_5\alpha_1^3\alpha_2\ \ +O(5)\end{aligned}\qquad(4.4.11)$$

$$\begin{aligned}\sigma(\alpha)^4P_{11}(\alpha)P_{12}(\alpha)&=-\alpha_2{}^2-\frac{1}{2}\mu_3\alpha_1{}^2\alpha_2{}^2-3A_{23}\alpha_1^2\alpha_2{}^2+4A_{32}\alpha_1^3\alpha_2-3A_{41}\alpha_1^4\ +O(5)\\ &=-\alpha_2{}^2+\frac{2}{3}\mu_4\alpha_1^3\alpha_2+\mu_5\alpha_1^4+O(5)\end{aligned}\qquad(4.4.12)$$

$$\begin{aligned}\sigma(\alpha)^4P_{11}(\alpha)P_{22}(\alpha)&=2\alpha_1\alpha_2+\frac{1}{3}\alpha_2{}^4+\frac{1}{3}\mu_3\alpha_1^3\alpha_2-2A_{32}\alpha_1^4-6A_{23}\alpha_1^3\alpha_2-12A_{14}\alpha_1^2\alpha_2^2+\\ &\quad+O(5)\\ &=2\alpha_1\alpha_2+\frac{1}{3}\alpha_2{}^4+\frac{4}{3}\mu_3\alpha_1^3\alpha_2+\frac{1}{3}\mu_4\alpha_1^4+\mu_2\alpha_1^2\alpha_2^2\ +O(5)\end{aligned}\qquad(4.4.13)$$

$\sigma(\alpha)^4P_{12}(\alpha)P_{22}(\alpha)=-2\alpha_1\alpha_2^3\ +O(5)$ (4.4.14)

$\sigma(\alpha)^4P_{22}(\alpha)P_{22}(\alpha)=+4\alpha_1^2\alpha_2^2+O(5)$ (4.4.15)

$\sigma(\alpha)^4P_{\alpha 1}(\alpha)=\ +6\alpha_2+O(3)$ (4.4.16)

$\sigma(\alpha)^4P_{\alpha 2}(\alpha)=\ -6\alpha_2{}^3+O(5)$ (4.4.17)

As was evidenced in the $N = 3$ case, all of the right-hand sides of the even basis functions are of even degree as polynomials of $\alpha_1, \alpha_2$, and all of the right-hand sides of the odd basis functions are of odd degree. It can be thus concluded that when we do $\partial$-calculations and all variables are to zero, the results of the odd basis functions are not influenced by those of the even basis functions. Moreover, when the condition $\alpha = \Omega_\delta$ is imposed, it can be deduced that the odd basis function contributed coefficients are rendered null.

The right-hand side of (4.4.16), where the polynomials are of odd degree, includes complicated $O(3)$ contributions. We have abbreviated the expression because these do not affect our results for the aforementioned reason, $P_{\alpha 1}(\Omega_\delta) = 0$.

Also, calculating $\sigma(\alpha)^4 P_{12}(\alpha) P_{12}(\alpha)$, we have

$$\sigma(\alpha)^4 P_{12}(\alpha) P_{12}(\alpha) = +\alpha_2^4 + O(5) \tag{4.4.18}$$

Therefore, as illustrated by the results above, it can be concluded that the set of basis functions $P_{11}(\alpha)P_{22}(\alpha)$ and $P(\alpha)$ can be replaced by $P_{11}(\alpha)P_{22}(\alpha)$ and $P_{12}(\alpha)P_{12}(\alpha)$. The coefficient of $P_{12}(\alpha)P_{12}(\alpha)$ is hereby defined as $H^4_{12,12}$. In this study, the coefficients $H^4_{11,22}$ and $H^4_{12,12}$ will be utilized in lieu of the coefficients $H^4_{11,22}$ and $H^4_P$, as previously defined in eq.(4.4.1). In the following, the term $P(\alpha)$ is not utilized

**(results)**

The following section will exclusively address the even function basis function part. The result of the $\partial$ calculations is as follows:

Leave everything as it is and simply set set $\alpha_1 = \alpha_2 = 0$ : $H^4_{1111} = 1$ (4.4.19)

$$\partial_1\partial_1 e_0^{L_4} = 2H^4_{11} \tag{4.4.20}$$

$$\partial_1\partial_2 e_0^{L_4} = +2H^4_{1122} \tag{4.4.21}$$

$$\partial_2\partial_2 e_0^{L_4} = -2H^4_{1112} \tag{4.4.22}$$

$$\partial_1\partial_1\partial_1\partial_1 e_0^{L_4} = 24H^4_0 + 8\mu_3 H^4_{11} + +8\mu_4 H^4_{1122} + 24\mu_5 H^4_{1112} - 8\mu_2\mu_4 H^4_{1111} \tag{4.4.23}$$

$$\partial_1\partial_1\partial_1\partial_2 e_0^{L_4} = 12H^4_{22} + 8\mu_3 H^4_{1122} + 4\mu_4 H^4_{1112} + 16\mu_5 H^4_{1111} \tag{4.4.24}$$

$$\partial_1\partial_1\partial_2\partial_2 e_0^{L_4} = -4H^4_{12} + 16H^4_{2222} + 4\mu_2 H^4_{1122} \tag{4.4.25}$$

$$\partial_1\partial_2\partial_2\partial_2 e_0^{L_4} = -12H^4_{1222} - 4H^4_{11} \tag{4.4.26}$$

$$\partial_2\partial_2\partial_2\partial_2 e_0^{L_4} = 24H^4_{12,12} + 8H^4_{1122} - 8\mu_2 H^4_{1111} \tag{4.4.27}$$

Originally there are $N^2 = 16$ basis functions, and $\partial$ calculations give us $\frac{(N+1)(N+2)}{2} =$ 15 equations. For the 10 coefficients of even-basis functions, there are 9 equations obtained by the $\partial$ calculations.

Using the equations

$\sum_{i=1}^{N}\zeta_1(\beta_i) = \frac{1}{2}\sum_{i=1}^{N}P_{222}(\beta_i)$ , $\sum_{i=1}^{N}\zeta_2(\beta_i) = 0$ for $N = 4$ in this subsection under $\sum_{i=1}^{N}\beta_i = 0$ which are obtained in Chapter 4-2, we have the following results for the coefficients defined in (4.4.0), from the (4.4.19) – (4.4.27) , except $H_{12}^4$ and $H_{2222}^4$ in eq.(4.4.25) :

$$H_{1111}^4 = 1 \qquad (4.4.19)$$

$$H_{11}^4 = \frac{1}{2}\partial_1\partial_1\, e_0^{L_4} = +\frac{1}{8}[\sum_{i=1}^4 P_{222}(\beta_i)]^2 - \frac{1}{2}[\sum_{i=1}^4 P_{11}(\beta_i)] \qquad (4.4.28)$$

$$H_{1122}^4 = \frac{1}{2}\,\partial_1\partial_2\, e_0^{L_4} = -\frac{1}{2}[\sum_{i=1}^4 P_{12}(\beta_i)] = \frac{1}{2}(x_1x_2 + x_2x_3 + x_3x_4 + x_4x_1) \qquad (4.4.29)$$

$$H_{1112}^4 = -\frac{1}{2}\,\partial_2\partial_2\, e_0^{L_4} = +\frac{1}{2}[\sum_{i=1}^4 P_{22}(\beta_i)] = x_1 + x_2 + x_3 + x_4 \qquad (4.4.30)$$

$$H_{12,12}^4 = \frac{1}{24}\partial_2\partial_2\partial_2\partial_2\, e_0^{L_4} - \frac{1}{3}H_{1122}^4 + \frac{1}{3}\mu_2\, H_{1111}^4$$

$$= -\frac{1}{24}[\sum_{i=1}^4 P_{2222}(\beta_i)] + \frac{1}{8}\,[\sum_{i=1}^4 P_{22}(\beta_i)\,]^2 + \frac{1}{6}[\sum_{i=1}^4 P_{12}(\beta_i)] + \frac{1}{3}\mu_2 \qquad (4.4.31)$$

On the last line of (4.4.31), we use the differential equation (3.19):

$P_{2222} = 6\,P_{22}^2 + 2\mu_2 + 4\,\mu_1 P_{22} + 4P_{21}$ ($\mu_1 = 0$ *here*), then some terms cancel each other out and we are left with

$$H_{12,12}^4 = -\frac{1}{4}\,[\sum_{i=1}^4 (P_{22}(\beta_i)\,)^2] + \frac{1}{8}\,[\sum_{i=1}^4 P_{22}(\beta_i)\,]^2$$

$$= -\frac{1}{4}[(x_1+x_2)^2 + (x_2+x_3)^2 + (x_3+x_4)^2 + (x_4+x_1)^2\,] + \frac{1}{8}[2(x_1+x_2+x_3+x_4)]^2$$

$$= \frac{1}{2}(x_1+x_2)(x_3+x_4) + \frac{1}{2}(x_1+x_4)(x_2+x_3) \qquad (4.4.32)$$

Also, from (4.4.26),

$$H_{1222}^4 = -\frac{1}{12}\partial_1\partial_2\partial_2\partial_2\, e_0^{L_4} - \frac{1}{3}H_{11}^4$$

$$= \frac{1}{12}[\sum_{i=1}^4 P_{1222}(\beta_i)] - \frac{1}{4}[\sum_{i=1}^4 P_{22}(\beta_i)]\,[\sum_{i=1}^4 P_{12}(\beta_i)\,] + \frac{1}{6}[\sum_{i=1}^4 P_{11}(\beta_i)] \qquad (4.4.33)$$

On the last line of (4.4.33), we use the differential equation (3.18):

$P_{1222} = 6\,P_{22}P_{12} + 4\,\mu_1 P_{12} - 2P_{11}$ ($\mu_1 = 0$ *here*) , so we are left with

$$H_{1222}^4 = +\frac{1}{2}\,[\sum_{i=1}^4 P_{22}(\beta_i)P_{12}(\beta_i)] - \frac{1}{4}[\sum_{i=1}^4 P_{22}(\beta_i)]\,[\sum_{i=1}^4 P_{12}(\beta_i)\,]$$

$$= \frac{1}{2}(x_1+x_2)(x_3x_4) + \frac{1}{2}(x_3+x_4)(x_1x_2) + \frac{1}{2}(x_1+x_4)(x_2x_3) + \frac{1}{2}(x_2+x_3)(x_1x_4) \qquad (4.4.34)$$

From (4.4.24)

$$H_{22}^4 = \frac{1}{12}\,\partial_1\partial_1\partial_1\partial_2\, e_0^{L_4} - \frac{2}{3}\mu_3 H_{1122}^4 - \frac{1}{3}\mu_4 H_{1112}^4 - \frac{4}{3}\mu_5 H_{1111}^4$$

$$= -\frac{1}{12}[\sum_{i=1}^4 P_{1112}(\beta_i)] - \frac{1}{8}[\sum_{i=1}^4 P_{112}(\beta_i)]\,[\sum_{i=1}^4 P_{222}(\beta_i)\,]$$

$$+\frac{1}{4}[\sum_{i=1}^4 P_{11}(\beta_i)]\,[\sum_{i=1}^4 P_{12}(\beta_i)\,] - \frac{1}{16}[\sum_{i=1}^4 P_{12}(\beta_i)]\,[\sum_{i=1}^4 P_{222}(\beta_i)]^2$$

$$+\frac{1}{3}\mu_3[\sum_{i=1}^4 P_{12}(\beta_i)] - \frac{1}{6}\mu_4\,[\sum_{i=1}^4 P_{22}(\beta_i)] - \frac{4}{3}\mu_5$$

On the last line above, we use the differential equation (3.16) :

$P_{1112} = 6P_{11}P_{12} - 4\,\mu_5 - 2\,\mu_4 P_{22} + 4\mu_3 P_{21}$, so some terms cancel each other out and we are left with

$$H_{22}^4 = -\frac{1}{2}[\sum_{i=1}^4 P_{11}(\beta_i)\cdot P_{12}(\beta_i)] - \frac{1}{8}[\sum_{i=1}^4 P_{112}(\beta_i)]\,[\sum_{i=1}^4 P_{222}(\beta_i)\,]$$

$$+\frac{1}{4}[\sum_{i=1}^4 P_{11}(\beta_i)]\,[\sum_{i=1}^4 P_{12}(\beta_i)\,] - \frac{1}{16}[\sum_{i=1}^4 P_{12}(\beta_i)]\,[\sum_{i=1}^4 P_{222}(\beta_i)]^2 \qquad (4.4.35)$$

From (4.4.23),

$$H_0^4 = \frac{1}{24}\partial_1\partial_1\partial_1\partial_1\, e_0^{L_4} - \frac{1}{24}\mu_3[\sum_{i=1}^4 P_{222}(\beta_i)]^2 + \frac{1}{6}\mu_3[\sum_{i=1}^4 P_{11}(\beta_i)])$$

$$+\frac{1}{6}\mu_4[\,\sum_{i=1}^4 P_{12}(\beta_i)]\ -\frac{1}{2}\mu_5[\sum_{i=1}^4 P_{22}(\beta_i)\,]+\frac{1}{3}\mu_2\mu_4 \qquad (4.4.36)$$

where $\frac{1}{24}\partial_1\partial_1\partial_1\partial_1\, e_0^{L_4}$ equals to

$$\frac{1}{24}\partial_1\partial_1\partial_1\partial_1\, e_0^{L_4} = -\frac{1}{24}[\,\sum_{i=1}^4 P_{1111}(\beta_i)] - \frac{1}{12}[\,\sum_{i=1}^4 P_{111}(\beta_i)]\,[\sum_{i=1}^4 P_{222}(\beta_i)\,]$$
$$+\frac{1}{8}[\,\sum_{i=1}^4 P_{11}(\beta_i)\,]^2 - \frac{1}{16}[\,\sum_{i=1}^4 P_{11}(\beta_i)]\,[\,\sum_{i=1}^4 P_{222}(\beta_i)]^2 + \frac{1}{24\times 16}[\,\sum_{i=1}^4 P_{222}(\beta_i)\,]^4$$

By using the differential equation (3.15) ($\mu_1 = 0$ *here*)

$$P_{1111} = 6P_{11}^2 - 8\,\mu_5\mu_1 + 2\,\mu_4\mu_2 - 12\mu_5 P_{22} + 4\mu_4 P_{21} + 4\,\mu_3 P_{11}\ ,$$

$H_0^4$ can be written as

$$H_0^4 = -\frac{1}{4}[\,\sum_{i=1}^4 (P_{11}(\beta_i))^2] - \frac{1}{12}[\,\sum_{i=1}^4 P_{111}(\beta_i)]\,[\sum_{i=1}^4 P_{222}(\beta_i)\,]$$
$$+\frac{1}{8}[\,\sum_{i=1}^4 P_{11}(\beta_i)\,]^2 - \frac{1}{16}[\,\sum_{i=1}^4 P_{11}(\beta_i)]\,[\,\sum_{i=1}^4 P_{222}(\beta_i)]^2$$
$$+\ \frac{1}{24\times 16}[\,\sum_{i=1}^4 P_{222}(\beta_i)\,]^4 - \ \frac{\mu_3}{24}[\,\sum_{i=1}^4 P_{222}(\beta_i)\,]^2 \qquad (4.4.37)$$

**(correspondence with the results in hyper-elliptic formulation)**

We consider the corresponding results in [15] here.

$H_{11}^4$ can be written as

$$H_{11}^4 = \ +\frac{1}{8}[\,\sum_{i=1}^4 (\,P_{222}(\beta_i))^2] + \frac{1}{4}\sum_{i\neq j} P_{222}(\beta_i)\ P_{222}(\beta_j)\ -\frac{1}{2}[\,\sum_{i=1}^4 P_{11}(\beta_i)] \quad (4.4.38)$$

Using the differential equation of $(P_{222}(\beta_i))^2$ and $P_{222}(\beta_i) = \ 2\frac{y_i + y_{i+1}}{x_i - x_{i+1}}$,

eq.(4.4.38) can be modified as

$$H_{11}^4 = \frac{1}{2}[\,\mu_3 - x_1x_2(x_1+x_2) + \mu_2\,(x_1+x_2) + \,(x_1+x_2)^3 + \mu_1\,(x_1+x_2)^2 + cyclic\,]$$
$$+\left[\frac{y_{12}y_{23}}{x_{12}x_{23}} + cyclic\right] + \left[\frac{y_{12}y_{34}}{x_{12}x_{34}} + \frac{y_{23}y_{43}}{x_{23}x_{41}}\right] \qquad (4.4.39)$$

( In our notation, $\mu_1 = 0$, and $y_{ij}$ is defined as $y_i - y_j$ . )

The coefficient of the $4\ell_\delta^{11}$ term in eq. (5.9) of [15] is

$$\frac{1}{8x_{12}x_{34}}\left(W_2^+(2,3) + W_2^+(4,1) + W_2^-(1,3) + W_2^-(2,4)\right) +$$

$$\frac{1}{4x_{12}x_{23}}\,(W_2^+(1,2) + W_2^+(2,3) + W_2^-(1,3) + cycl(1,2,3,4)\,) \qquad (4.4.40)$$

After some calculations, it can be shown that this is the same as the $H_{11}^4$ expression of eq.(4.4.39) , when $W_2^+$, $W_2^-$ are rewritten in our notation.

We have not yet carried out similar considerations for $H_{22}^4$, $H_0^4$ .

**( $H_{2222}^4$ and $H_{12}^4$ )**

With regard to $H_{2222}^4 (= x_1x_2x_3x_4)$ in eq.(4.4.25), it appears to be a challenging task to obtain the theta expression using the method employed.

The following factors are likely to be contributing this situation. The $\partial$-calculation can thus be regarded as a method that has been modified for application to genus 2 , by which the method originally employed for genus 1 is employed, with using only differentiations with respect to the variables. This method can be applied to the periodic functions on the Jacobian variety, without the assumption that the variables are restricted to those moving only on the genus 2 Riemann surface. The method gives excessively symmetrical results with regard to the differential indices. A similar situation arises in the basis functions construction, where the derivatives of the Pe function alone are insufficient to span the entire function space. To a certain extent, the $\partial$-calculation is unable to account for the correlated effects of non-neighbour insertion points. The expansion coefficients $H$ are thought to be expressed as polynomials of seven beta-type Pe functions $P_{11}(\beta)$, $P_{12}(\beta)$, $P_{22}(\beta)$, $P_{111}(\beta), P_{112}(\beta)$, $P_{122}(\beta), P_{222}(\beta)$ . It is noted that there are numerous types of polynomials that cannot be obtained by $\partial$-calculations in general. In the context of $\partial$-calculations, it can be demonstrated that each factor in the coefficient results is of the form $\sum_{i=1}^{N} P_{AB}(\beta_i)$ or $\sum_{i=1}^{N} P_{ABC}(\beta_i)$. It is important to note that the sum is always from 1 to $N$, and not from 1 to some number less than $N$. This has not been problematic in genus one, but the situation is different in genus two.

The following methodology is to be applied in order to obtain the concrete form in theta functions for the two coefficients $H_{2222}^4$ and $H_{12}^4$.

In [15] page 40, it is described that

$$C_\delta(1,2,3,4) = \frac{1}{2}L_\delta(1,2)L_\delta(3,4) + \frac{1}{2}L_\delta(1,4)L_\delta(2,3) + \textit{other terms} \quad \text{[p.40 eq.(5.9)]} \quad (4.4.41)$$

$$C_\delta(1,2,3,4,5,6) = \frac{1}{2}L_\delta(1,2)L_\delta(3,4)L_\delta(5,6) + \frac{1}{2}L_\delta(2,3)L_\delta(4,5)L_\delta(6,1) + \textit{other terms} \quad (4.4.42)$$

$$C_\delta(1,2,\dots,k) = \frac{1}{2}L_\delta(1,2)L_\delta(3,4)..L_\delta(k-1,k) + \frac{1}{2}L_\delta(2,3)L_\delta(4,5)..L_\delta(k,1) + \textit{other terms}$$

for even k (4.4.43)

By picking up the coefficients of $4\ell_\delta^{11}, 4\ell_\delta^{12}, 4\ell_\delta^{22}$ , which correspond to $P_{11}(\alpha), P_{12}(\alpha), P_{22}(\alpha)$ , the following $H_{ABCD}^{4,CORES}$ will be the accurate solutions for the expansion coefficients in the x,y-coordinate expression:

$$H_{1111}^{4,CORRES} = 1 \quad (4.4.44)$$

$$H_{1122}^{4,CORRES} = +\frac{1}{2}(x_1x_2 + x_2x_3 + x_3x_4 + x_4x_1) \quad (4.4.45)$$

$$H_{1112}^{4,CORRES} = x_1 + x_2 + x_3 + x_4 \quad (4.4.46)$$

$$H_{12,12}^{4,CORRES} = \frac{1}{2}(x_1 + x_2)(x_3 + x_4) + \frac{1}{2}(x_1 + x_4)(x_2 + x_3) \quad (4.4.47)$$

$$H_{1222}^{4,CORRES} = \frac{1}{2}(x_1 + x_2)(x_3x_4) + \frac{1}{2}(x_3 + x_4)(x_1x_2) + \frac{1}{2}(x_1 + x_4)(x_2x_3) + \frac{1}{2}(x_2 + x_3)(x_1x_4) \quad (4.4.48)$$

$$H_{2222}^{4,CORRES} = x_1x_2x_3x_4 \quad (4.4.49)$$

To express $H_{2222}^4$ in terms of the theta function, we adopt the following logic in addition to the $\partial$-calculation.

We note that eq.(4.4.41) can be expressed in terms of $S_\delta(z_i, z_j)$ as follows:

$$S_\delta(z_1,z_2)S_\delta(z_2,z_3)S_\delta(z_3,z_4)S_\delta(z_4,z_1) = \tfrac{1}{2}S_\delta^2(z_1,z_2)\,S_\delta^2(z_3,z_4) + \tfrac{1}{2}S_\delta^2(z_1,z_4)\,S_\delta^2(z_2,z_3) + other\ terms \qquad (4.4.50)$$

This can be regarded as one of the additional restrictions obtained by reflecting the fact that the variables in the functions can only vary over the Riemann surface, not over the whole of the Jacobian variety, as explained in the last part of Section 4-1.

In $\frac{1}{2}L_\delta(1,2)L_\delta(3,4) + \frac{1}{2}L_\delta(1,4)L_\delta(2,3)$ of the right hand side of eq.(4.4.41), the spin structure independent terms, denoted $P_{11}(\beta_i)$ in this paper, are not included. However, this is inconsequential in the present context. It can be hypothesized that the terms containing $P_{11}(\beta_i)$ are packaged and included under the category of "other terms" with an opposite sign.

The argument is that the degree 2 basis functions (highest degree in N=4) as monomials of $P_{IJ}(\alpha)$:

$P_{11}(\alpha)P_{11}(\alpha)$, $P_{11}(\alpha)P_{22}(\alpha)$, $P_{11}(\alpha)P_{12}(\alpha)$, $P_{12}(\alpha)P_{12}(\alpha)$ $P_{12}(\alpha)P_{22}(\alpha)$ $P_{22}(\alpha)P_{22}(\alpha)$

are exclusively encompassed within the initial two terms of the right hand side of (4.4.50). It is noteworthy that "other terms" do not incorporate these highest degree basis functions. Therefore, it is sufficient to consider the only first two terms on the right-hand side of (4.4.50) to obtain the coefficients of the highest degree basis functions. In higher even values of $N$, the same observation can be applied, as seen from (4.4.42), (4.4.43).

The $S_\delta^2(z_i,z_j)$ is to be replaced by its theta function formula, as in (4.2.59) :

$$S_\delta(z_i,z_j)^2 = -\{P_{11}(\Omega_\delta) - P_{11}(\beta_i) + P_{12}(\Omega_\delta)P_{22}(\beta_i) - P_{12}(\beta_i)P_{22}(\Omega_\delta)\}\omega_1(z_i)\omega_1(z_j)$$

and the first two terms on the right hand side of (4.4.50) are to be expanded. It is evident that the coefficients of the 6 highest-degree basis functions can be expressed in the explicit theta function form. The coefficients are denoted by $H_{ABCD}^{4,HIGHEST}$ , and we obtain

$$H_{1111}^{4,HIGHEST} = 1$$
$$H_{1122}^{4,HIGHEST} = -\tfrac{1}{2}[\textstyle\sum_{i=1}^4 P_{12}(\beta_i)]$$
$$H_{1112}^{4,HIGHEST} = +\tfrac{1}{2}[\textstyle\sum_{i=1}^4 P_{22}(\beta_i)]$$
$$H_{12,12}^{4,HIGHEST} = \tfrac{1}{2}P_{22}(\beta_1)P_{22}(\beta_3) + \tfrac{1}{2}P_{22}(\beta_4)P_{22}(\beta_2)$$
$$H_{1222}^{4,HIGHEST} = -\tfrac{1}{2}P_{22}(\beta_1)P_{12}(\beta_3) - \tfrac{1}{2}P_{22}(\beta_3)P_{12}(\beta_1) - \tfrac{1}{2}P_{22}(\beta_4)P_{12}(\beta_2) + \tfrac{1}{2}P_{22}(\beta_2)P_{12}(\beta_4)$$
$$H_{2222}^{4,HIGHEST} = \tfrac{1}{2}P_{12}(\beta_1)P_{12}(\beta_3) + \tfrac{1}{2}P_{12}(\beta_4)P_{12}(\beta_2)$$

It can be posited that the $H_{ABCD}^{4,HIGHEST}$ coefficients are all correct, and it can be demonstrated that they all give the same results as $H_{ABCD}^{4,CORRES}$ in their coordinate expressions.

Despite the apparent disparity in the theta function expressions $H^4_{12,12}$ and $H^4_{1222}$ in comparison to those of $H^{4,HIGHEST}_{12,12}$ , $H^{4,HIGHEST}_{1222}$ , it can be observed that each of these expressions are equivalent. This can be concluded by (4.4.32) and (4.4.34), due to modified Jacobi inversion formula in proposition 2. It can be concluded that the equations $H^4_{12,12} = H^{4,HIGHEST}_{12,12}$ and $H^4_{1222} = H^{4,HIGHEST}_{1222}$ are theta identities under the cyclic condition, in addition to the fact that the variables are of $\beta$ type.

It is found that only $H^4_{2222}$ ( and hence $H^4_{12}$ too, from(4.4.25) ) do not have theta function expression of the form of combinations of $\sum_{i=1}^{N} P_{AB}(\beta_i)$ and $\sum_{i=1}^{N} P_{ABC}(\beta_i)$ where the sum is from 1 to $N(=4)$. We conclude that $H^4_{2222}$ is obtained by the theta functions as

$$H^4_{2222} = \frac{1}{2} P_{12}(\beta_1) P_{12}(\beta_3) + \frac{1}{2} P_{12}(\beta_4) P_{12}(\beta_2) \tag{4.4.51}$$

From eq.(4.4.25), we also have the solution of $H^4_{12}$ in the form of theta functions from. Using the differential equation (3.17): $P_{1122} = 4P_{21}^2 + 2\, P_{22}P_{11} + 2\mu_2 P_{21}$ , we have

$$\begin{aligned} H^4_{12} = & \sum_{i=1}^{4} (P_{12}(\beta_i))^2 - \frac{1}{2} \left[ \sum_{i=1}^{4} P_{12}(\beta_i) \right]^2 + \frac{1}{4} \left[ \sum_{i=1}^{4} P_{222}(\beta_i) \right] \left[ \sum_{i=1}^{4} P_{122}(\beta_i) \right] \\ & + \frac{1}{16} \left[ \sum_{i=1}^{4} P_{222}(\beta_i) \right]^2 \left[ \sum_{i=1}^{4} P_{22}(\beta_i) \right] + \frac{1}{2} \left[ \sum_{i=1}^{4} P_{22}(\beta_i) P_{11}(\beta_i) \right] \\ & - \frac{1}{4} \left[ \sum_{i=1}^{4} P_{22}(\beta_i) \right] \left[ \sum_{i=1}^{4} P_{11}(\beta_i) \right] + 2P_{12}(\beta_1) P_{12}(\beta_3) + 2P_{12}(\beta_4) P_{12}(\beta_2) \end{aligned} \tag{4.4.52}$$

As a summary of this subsection, the 4-point fermion correlation function with cyclic condition $S_\delta(z_1, z_2) S_\delta(z_2, z_3) S_\delta(z_3, z_4) S_\delta(z_4, z_1)$ is written as

$$\begin{aligned} & [\, H^4_0 + H^4_{11} P_{11}(\Omega_\delta) + H^4_{12} P_{12}(\Omega_\delta) + H^4_{22} P_{22}(\Omega_\delta) + H^4_{1111} P_{11}(\Omega_\delta) P_{11}(\Omega_\delta) \\ & + H^4_{1112} P_{11}(\Omega_\delta) P_{12}(\Omega_\delta) + H^4_{11,22} P_{11}(\Omega_\delta) P_{22}(\Omega_\delta) + H^4_{12,12} P_{12}(\Omega_\delta) P_{12}(\Omega_\delta) \\ & + H^4_{1222} P_{12}(\Omega_\delta) P_{22}(\Omega_\delta) + H^4_{2222} P_{22}(\Omega_\delta) P_{22}(\Omega_\delta) \,] \prod_{i=1}^{4} \omega_1(z_i) \end{aligned} \tag{4.4.53}$$

where the 10 coefficients $H^4_0$, $H^4_{AB}, H^4_{ABCD}$ are given in (4.4.19), (4.4.28), (4.4.29), (4.4.30), (4.4.32), (4.4.34), (4.4.35), (4.4.37), (4.4.51), (4.4.52).

**( one more example of spin sum )**

N=4 point amplitudes with external massless boson case also include the Wick contraction results $S_\delta(z_1, z_2) S_\delta(z_2, z_3) S_\delta(z_3, z_4) S_\delta(z_4, z_1)$.

Since only $P_{11}(\Omega_\delta) P_{22}(\Omega_\delta)$ and $P_{12}(\Omega_\delta) P_{12}(\Omega_\delta)$ remain non-zero by summing over spin structures, the (4.4.53) is effectively equal to

$$\cong \left[\, H^4_{1122} P_{11}(\Omega_\delta) P_{22}(\Omega_\delta) + H^4_{12,12} P_{12}(\Omega_\delta) P_{12}(\Omega_\delta) \right] \prod_{i=1}^{4} \omega_1(z_i)$$

When $H^4_{11,22}$ and $H^4_{12,22}$ are expressed in the x-coordinates, this equals to

$$\begin{aligned} = & \left[ \frac{1}{2} (x_1 x_2 + x_3 x_4 + x_1 x_4 + x_2 x_3) P_{11}(\Omega_\delta) P_{22}(\Omega_\delta) + \right. \\ & \left. + \frac{1}{2} \{ (x_1 + x_2)(x_3 + x_4) + (x_1 + x_4)(x_2 + x_3) \} P_{12}(\Omega_\delta) P_{12}(\Omega_\delta) \right] \prod_{i=1}^{4} \omega_1(z_i) \end{aligned} \tag{4.4.54}$$

As explained in the last subsection of Chapter 4-2, replacing $P_{11}(\Omega_\delta)\,P_{22}(\Omega_\delta) \to -8\pi^4$, $P_{12}(\Omega_\delta)\,P_{12}(\Omega_\delta) \to 4\pi^4$ is equivalent to the whole process of multiplying partition functions and summing over spin structures. By this replacement, (4.4.54) gives

$$2\pi^4[\,(x_1-x_2)(x_3-x_4)-(x_1-x_4)(x_2-x_3)\,]\prod_{i=1}^{4}\omega_1(z_i) \tag{4.4.55}$$

Using the definition of holomorphic 1-forms $x_i\omega_1(z_i)=\omega_2(z_i)$, we see that the whole of (4.4.54) is equal to

$$2\pi^4[\,\Delta(z_1,z_2)\Delta(z_3,z_4)-\Delta(z_1,z_4)\Delta(z_2,z_3)\,] \tag{4.4.56}$$

where $\Delta(x,y)=\omega_1(x)\omega_2(y)-\omega_2(x)\omega_1(y)$.

Eq.(4.4.56) is the same as that given in the original result in [23].

It is possible to write in terms of theta function and write the result as

$$2\pi^4\left[2[\,\textstyle\sum_{i=1}^{4}P_{12}(\beta_i)]-\frac{1}{2}\,[\sum_{i=1}^{4}(P_{22}(\beta_i)\,)^2]\;+\frac{1}{4}\,[\sum_{i=1}^{4}P_{22}(\beta_i)\,]^2\right]\prod_{i=1}^{4}\omega_1(z_i) \tag{4.4.57}$$

## 4-5 N = 6 and higher N

In the $N=6$ case, the total number of basis functions is $N^2=36$. However, as evidenced by the lower $N$ cases, there is a strong probability that incorporating odd basis functions is unnecessary. It is evident that a total of the following 20 basis functions will be required in order to expand the product of the corresponding sigma-function：

10 basis functions we already saw in N=4： 1, $P_{11}(\alpha), P_{12}(\alpha)$, $P_{22}(\alpha)$, $P_{11}(\alpha)P_{11}(\alpha)$,
$P_{11}(\alpha)P_{12}(\alpha)$, $P_{11}(\alpha)P_{22}(\alpha)$, $P_{12}(\alpha)P_{12}(\alpha)$, $P_{12}(\alpha)P_{22}(\alpha)$, $P_{22}(\alpha)P_{22}(\alpha)$ ,

10 degree-3 ( highest degree ) new basis functions which appear for the first time in $N=6$ of the form $P_{AB}P_{CD}P_{EF},\ each\ of\ A,B,C,D,E,F\ \ takes\ values\ 1\ or\ 2$ ：

$P_{11}(\alpha)P_{11}(\alpha)P_{11}(\alpha)$, $P_{11}(\alpha)P_{11}(\alpha)P_{12}(\alpha)$, $P_{11}(\alpha)P_{11}(\alpha)P_{22}(\alpha)$, $P_{11}(\alpha)P_{12}(\alpha)P_{12}(\alpha)$,
$P_{11}(\alpha)P_{12}(\alpha)P_{22}(\alpha)$, $P_{11}(\alpha)P_{22}(\alpha)P_{22}(\alpha)$ , $P_{12}(\alpha)P_{12}(\alpha)P_{12}(\alpha)$, $P_{12}(\alpha)P_{12}(\alpha)P_{22}(\alpha)$,
$P_{12}(\alpha)P_{22}(\alpha)P_{22}(\alpha)$, $P_{22}(\alpha)P_{22}(\alpha)P_{22}(\alpha)$.

The products $P_{ABC}(\alpha)P_{DEF}(\alpha)$ are also even functions. Initially, it appears that these should be incorporated into the basis functions at this order. Nevertheless, by the differential equations of Pe functions (3.5) – (3.14), these can be expressed by the basis functions that were previously encountered in $N=4$, and thus do not constitute novel basis functions.

It is also possible to make use of the fermion correlation function decomposition result in [15] for $N=6$ , the equivalent formula of which is quoted as (4.4.42) from [15]：

$$\begin{aligned}&S_\delta(z_1,z_2)S_\delta(z_2,z_3)S_\delta(z_3,z_4)S_\delta(z_4,z_5)S_\delta(z_5,z_6)S_\delta(z_6,z_1)\\&=\frac{1}{2}S_\delta^2(z_1,z_2)\,S_\delta^2(z_3,z_4)S_\delta^2(z_5,z_6)+\frac{1}{2}S_\delta^2(z_2,z_3)\,S_\delta^2(z_4,z_5)S_\delta^2(z_6,z_1)\\&+other\ terms\end{aligned} \tag{4.5.1}$$

As shown in section 5.5 in [15], in the expression of $C_\delta(1,2,3,4,5,6)$ in eq.(4.4.42), any monomial formed by selecting three from $\ell_\delta^{11}, \ell_\delta^{12}, \ell_\delta^{22}$ (allowing duplicates) and multiplying them appears only in the first two terms $\frac{1}{2}L_\delta(1,2)L_\delta(3,4)L_\delta(5,6) + \frac{1}{2}L_\delta(2,3)L_\delta(4,5)L_\delta(6,1)$ on the right hand side of (4.4.42), and not in the "other terms". If we reinterpret (4.4.42) as (4.5.1), the highest-degree basis functions $P_{AB}(\alpha)P_{CD}(\alpha)P_{EF}(\alpha)$ are included only in the first two terms on the right-hand side of (4.5.1) without any change from (4.4.42), although the contents of the "other terms" in (4.5.1) differ from those in (4.4.42) with regards to the non-highest degree basis functions terms. The situation is the same as that described in the N=4 case below (4.4.50).

Substituting the N=2 result:

$$S_\delta(z_1,z_2)^2 = -\{P_{11}(\Omega_\delta) - P_{11}(\beta_1) + P_{12}(\Omega_\delta)P_{22}(\beta_1) - P_{12}(\beta_1)P_{22}(\Omega_\delta)\}\omega_1(z_1)\omega_1(z_2)$$

into (4.5.1) and expanding term by term immediately reveals the explicit forms of the coefficients of the 10 highest degree basis functions ( $P_{AB}P_{CD}P_{EF}$ type basis functions ) in theta function expressions.

In the N=6 case, there are effectively 10 other unknown coefficients.

Conversely, the $\partial$-calculations employed in this chapter will yield the following number of equations :

Doing nothing and setting $\alpha = 0$ : 1 equation, which gives the coefficient of $P_{11}(\alpha)P_{11}(\alpha)P_{11}(\alpha)$ as 1.

$\partial_1\partial_1$, $\partial_1\partial_2$, $\partial_2\partial_2$ : 3 equations

$\partial_A\partial_B\partial_C\partial_D$ , *each of* $A,B,C,D$ *takes values* 1 *or* 2 : 5 equations

$\partial_A\partial_B\partial_C\partial_D\partial_E\partial_F$ , *each of* $A,B,C,D,E,F$ *takes values* 1 *or* 2 : 7 equations

Subsequently, we have in total $1 + 3 + 5 + 7 = 16$ equations.

Consequently, there is a possibility of obtaining all of the coefficients in terms of theta functions, as evidenced in the case $N = 4$, upon repeating the same consideration at the higher expansion level.

It is evident that, for higher $N$, an elementary consideration leads to the following conclusion.

For the arbitrary even $N$ , the total number of the even-basis functions is $\frac{1}{48}(N^3 + 12N^2 + 44N + 48)$ , and the number of the highest-degree basis functions is $\frac{1}{48}(6N^2 + 36N + 48)$. The difference of these two, $\frac{1}{48}(N^3 + 6N^2 + 8N)$ , which is also equal to the total number of the even basis functions for $N - 2$, corresponds to the number of unknown coefficients of the basis functions at this level.

In even $N$, the $\partial$-calculations give $\frac{1}{4}(N^2 + 4N + 4)$ equations, with the restriction that the basis functions are limited to those that are even.

The number of unknown coefficients increases at an order $N^3$,while the number of equations increases at an order $N^2$. We see that, for even $N$,

$$\frac{1}{4}(N^2+4N+4)-\frac{1}{48}(N^3+6N^2+8N)<0 \quad if \;\; N>\;10$$

Consequently, at least for $N$ greater than or equal to 12, the strategy does not provide a sufficient number of equations.

In any case, it is interesting that there is a high likelihood of being able to compute a 6-point function in a concrete theta function form by the method we described.

## 5. Summary

Let $z_i(x_i,y_i)$ $(i=1,2,\ldots,N+1)$ be the $N+1$ number of points on the genus 2 Riemann surface representing the vertex insertion points. The Szegö kernel ( fermion correlation function ) is given by

$$S_\delta(z_i,\; z_{i+1}\,)=\frac{\theta[\delta](\beta_i)}{\theta[\delta](0)E(z_i,z_{i+1})}$$

where $\beta_i=(\beta_{i,1}\,,\beta_{i,2}),\ \beta_{i,a}=\int_\infty^{(x_i,y_i)}\omega_a\ -\int_\infty^{(x_{i+1},y_{i+1})}\omega_a$ , $a=1,2$ , and $E(z_i,z_{i+1})$ are the prime forms, $\theta[\delta]$ is the theta function with a non-singular and even characteristics $\delta$ . When we calculate string amplitudes, we are sometimes interested in the simple product of the fermion correlation function, $\Pi_{i=1}^{N}\frac{\theta[\delta](\beta_i)}{\theta[\delta](0)E(z_i,z_{i+1})}$ , subject to the cyclic condition $z_{N+1}=z_1$

In this paper, we attempted to decompose this product in a manner that extends equation (1.4) to the genus 2 case. We set $e_6=\infty$ , $\mu_1=e_1+e_2+e_3+e_4+e_5=0$ , and use the genus 2 theta function $\theta_R(u)$ whose characteristics are the Riemann constants as the unique function representing fermion field contributions to string amplitudes after the spin sum. This is a direct generalization of the case genus 1.

In genus 2, a natural generalisation of the half periods $\Omega_\delta\ \in C^2$ corresponding to non-singular and even spin structures can be given as $\Omega_\delta=\int_\infty^{(e_i,0)}\omega+\int_\infty^{(e_j,0)}\omega\ \in C^2$ as in eq.(2.26) (2.27), for any pair of branch points $e_i$ and $e_j$ out of 5 branch points, with a total of 10 half periods. $\delta$ denotes one choice of the pair $i,j$ .

In genus 1, selecting one branch point $e_\delta$ out of 3 branch points $e_1,\ e_2\,,e_3$ and calculating $\Omega_\delta=\int_\infty^{(e_\delta,0)}\omega_1\ \in C^1$ gives a non-singular and even half period. As is well known, in the standard notation the integral gives $\Omega_1=\frac{1}{2},\ \ \Omega_2=-\frac{1+\tau}{2},\ \ \Omega_3=\frac{\tau}{2}$ .

We have the following formula, expressed in eq.(2.32) :

$$\frac{\theta[\delta](u)}{\theta[\delta](0)E(z,w)}=\ \exp\!\big(2\pi i(\textstyle\sum_{k=1}^{g}a_{i_k})\cdot u\big)\,\frac{\theta_R(u+\Omega_\delta)}{\theta_R(\Omega_\delta)E(z,w)}\qquad .$$

where $a_i$ are defined in Eq.(2.11) and $\theta_R$ is the theta function whose characteristics are the Riemann constants in genus 2.

We can therefore say that, since the product of the exponential factor will be 1 under the cyclic condition (see (2.33)),

$$\prod_{i=1}^{N} S_\delta(z_i,\ z_{i+1}) = \prod_{i=1}^{N} \frac{\theta[\delta](\beta_i)}{\theta[\delta](0)E(z_i,z_{i+1})} = \prod_{i=1}^{N} \frac{\theta_R(\beta_i+\Omega_\delta)}{\theta_R(\Omega_\delta)E(z_i,z_{i+1})} \ .$$

By setting the odd theta function in prime form to $\theta_R$, and using the normalization and Taylor expansion form of the genus 2 sigma function, as well as the cyclic condition, the above product can be written as, after setting $\Omega_\delta$ to a $C^2$ parameter $\alpha$ ,

$$\prod_{i=1}^{N} \frac{\theta_R(\beta_i+\alpha)}{\theta_R(\alpha)E(z_i,z_{i+1})} = \prod_{i=1}^{N} \frac{\sigma(\beta_i+\alpha)}{\sigma(\beta_i)\sigma(\alpha)} \cdot \prod_{i=1}^{N} \omega_1(z_i) \ . \qquad (5\text{-}1)$$

In this expression, the prime forms are factorised as $\prod_{i=1}^{N} \omega_1(z_i)$ where $\omega_2(z_i)$ are not included. This simplification of the product only applies to genus 2 , originating from the form of the Schur-Weierstrass polynomial in the expansion of the sigma function, which states that $\partial_2\theta_R(0) = 0$ . This has a formal correspondence to the hyperelliptic representation given in [15] :

$$C_\delta(1,\dots,n) = \frac{N_\delta(1,\dots,n)}{x_{1,2}x_{2,3}\cdots x_{n,1}} \prod_{i=1}^{n} \frac{dx_i}{2s_i} \leftrightarrow \prod_{i=1}^{N} S_\delta(z_i,\ z_{i+1}) = \prod_{i=1}^{N} \frac{\sigma(\beta_i+\Omega_\delta)}{\sigma(\beta_i)\sigma(\Omega_\delta)} \cdot \prod_{i=1}^{N} \frac{dx_i}{2y_i}$$

From what has been stated thus far, the similarity between the genus 1 and the genus 2 formulas should be evident.

For all the genus 2 formulas described above in this chapter, by replacing the variables $\alpha,\ \beta_i,\ \Omega_\delta\ \in C^2$ , $\theta_R$ with the corresponding $\alpha,\ \beta_i,\ \Omega_\delta\ \in C^1$ , $\theta_1$ and applying the cyclic condition $z_{N+1} = z_1$ , we can derive all the corresponding genus 1 formulas, which are reviewed in Chapter1. In genus 1, it is possible to formally use the prime form to write the genus 1 formula corresponding to (5.1) using the genus 1 sigma function under the cyclic condition as

$$\prod_{i=1}^{N} \frac{\theta_1(\beta_i+\alpha)}{\theta_1(\alpha)E(z_i,z_{i+1})} = \prod_{i=1}^{N} \frac{\sigma(\beta_i+\alpha)}{\sigma(\beta_i)\sigma(\alpha)} \cdot \prod_{i=1}^{N} \omega_1(z_i) \ ,$$

Therefore, the similarity has been perfectly manifest up to this step.

In genus 1, the product $\prod_{i=1}^{N} \frac{\sigma(\beta_i+\alpha)}{\sigma(\beta_i)\sigma(\alpha)}$ is doubly periodic with respect to $\alpha \in C^1$ under the cyclic condition. Therefore, it can be expanded using $N$ basis functions : $1, P(\alpha), P^{(1)}(\alpha), P^{(2)}(\alpha), \dots P^{(N-2)}(\alpha)$ . In genus 2, the product is periodic in the Jacobian variety. Since the product has theta divisors in the denominator and satisfies Abel's theorem as a function of $\alpha \in C^2$, it can be expanded using $N^2$ polynomials of the Pe function and their derivatives, as described in references [12] and [13]. After completing the expansion, setting $\alpha = \Omega_\delta$ yields the decomposition formula.

**( trilinear relations )**

The Pe functions in genus one and two satisfy the following two types of differential equations :

$$P_{ABC}P_{DEF} = P_{GH}P_{IJ}P_{KL} + lower\ degree\ terms\ of\ P \quad ,$$

and $\quad P_{ABCD} = P_{EF}\,P_{GH} + lower\ \ degree\ terms\ of\ P \quad .$

These equations are not independent of each other.

In genus 1, the indices $A, B, C, \ldots$ takes a value of 1 only. The former type of equation is

$$\{P_{111}\}^2 = 4P_{11}^3 - g_2P_{11} - g_3$$

and the latter type of equation, which can be derived from the former, is

$$P_{1111} = 6P_{11}^2 - g_2/2 \ .$$

Setting the variable equal to $\Omega_\delta \in C^1$ in the former type, the trilinear relation is obtained [18] as $4e^3 = g_2e + g_3$ .

In genus 2, the indices $A, B, C, \ldots$ take the values 1 or 2.

Considering all possible combinations of indices of $P_{ABC}P_{DEF}$ and $P_{ABCD}$ on the left-hand sides of the equations, there should be 10 equations of the former type, and 5 equations of the latter type.

Setting the variable equal to $\Omega_\delta = \int_\infty^{(e_i,0)} \omega + \int_\infty^{(e_j,0)} \omega \ \in C^2$ in the former 10 equations, the trilinear relations are obtained as listed up in Chapter 3 because the necessary and sufficient number of differential equations of Pe functions are known. These ensure that the degree 3 monomials $P_{GH}(\Omega_\delta)P_{IJ}(\Omega_\delta)P_{KL}(\Omega_\delta)$ can be written as degree two or lower polynomials of $P_{QR}(\Omega_\delta)$. There are 10 possible combinations of $P_{ABC}P_{DEF}$ , $P_{GH}P_{IJ}P_{KL}$ as well as non-singular even half periods $\Omega_\delta$. For the latter type, there are 5 possible combinations of $P_{ABCD}$ and 6 possible combinations of $P_{EF}\,P_{GH}$ in genus 2. Therefore, there are polynomials of Pe functions for which it is not possible to find a representation using only the Pe function itself and its derivatives. This relates to the fact that the function space dimension we are concerning with is $N^2$ in genus 2 and $N$ in genus 1.

**( standard and modified inversion theorems in genus 2 )**

In our framework, we use the genus 2 theta function whose characteristics are the Riemann constants as the odd theta function in the prime form. One advantage of this choice is that it enables us to apply straightforward logic of solving the Jacobi inversion theorem and its modifications. This logic is useful for expressing both spin structure-dependent and -independent parts in $x, y$ coordinates.

For genus $g$ hyperelliptic surfaces, consider the variables of the form

$$u_i = \sum_{k=1}^{g} \int_{\infty}^{(x_k, y_k)} \omega_i \quad \text{for } i = 1,2,..,g ,$$

where $\omega_i = \frac{x^{i-1}dx}{2y}$ and write $u = (u_1, u_2, \dots\ u_g)$ .

The standard inversion theorem says that $g$ number of $x$ coordinates $x_k$ are obtained as the solutions of the following degree $g$ equation of $x$ :

$$x^g - P_{gg}(u)x^{g-1} - P_{g,g-1}(u)x^{g-2} - \cdots - P_{g,1}(u) = 0 ,$$

and $y_k$ can be given as

$$y_k = P_{ggg}(u){x_k}^{g-1} + P_{gg,g-1}(u){x_k}^{g-2} + \cdots + P_{gg,1}(u) \quad \text{for } k = 1,2, \dots g .$$

In the case $g = 1$, the above equations are reduced to the formula:

$$x_1 = P_{11}(u_1) = P(u_1) \text{ and } \ y_1 = P_{111}(u_1) = P^{(1)}(u_1) \quad \text{for any } x_1 \in C^1, \ y_1 \in C^1 .$$

Setting $u_1$ to the half periods : $\Omega_\delta = \int_{\infty}^{(e_\delta, 0)} \frac{dx}{2y} \in C^1$ , we have, as a result of the inversion formula, $P(\Omega_\delta) = e_\delta$ and $P^{(1)}(\Omega_\delta) = 0$ , which are the basic facts to be used to represent spin structure dependent part of the amplitudes in terms of 3 branch points.

Similarly, for genus two, the $x_1, x_2$ are the solutions of the quadratic equation

$$x^2 - P_{22}(u)x - P_{21}(u) = 0, \tag{5.2}$$

and $\quad y_1 = P_{222}(u)x_1 + P_{221}(u), \quad y_2 = P_{222}(u)x_2 + P_{221}(u).$

Setting $u$ to the half periods , for any pair of $i, j$ denoted as $\delta$ ,

$$\Omega_\delta = \int_{\infty}^{(e_i, 0)} \omega + \int_{\infty}^{(e_j, 0)} \omega \quad \in C^2$$

The inversion formula gives, since (5.2) has two roots $e_i$ and $e_j$ ,

$$P_{12}(\Omega_\delta) = P_{21}(\Omega_\delta) = -e_i e_j , \qquad P_{22}(\Omega_\delta) = e_i + e_j$$

and also

$$P_{11}(\Omega_\delta) = \frac{F(e_i, e_j)}{(e_i - e_j)^2} = (e_p + e_q + e_r)e_i e_j + e_p e_q e_r \quad (p, q, r \text{ are all different from } i, j) ,$$

$$P_{222}(\Omega_\delta) = P_{221}(\Omega_\delta) = 0 .$$

Given these known results and by the fact that the partition functions are also expressed in terms of branch points, the sum over spin structures can be performed using the algebra of branch points in genus 2. Spin structure dependent parts can be treated in such a way in genus 1 and 2.

As for spin structure independent parts in genus 2, the corresponding theta functions will have a form such as $P_{ABCD\ldots}\left(\int_{\infty}^{(x_1,y_1)}\omega - \int_{\infty}^{(x_2,y_2)}\omega\right)$ where the variables inside the function is the difference of the two inserting points of the vertex inserting points on the Riemann surface. In genus 2 , the number of variables $u_i$, $u_i = \sum_{k=1}^{g}\int_{\infty}^{(x_k,y_k)}\omega_i$ is also two. Due to this special circumstance, the functions such as $P_{ABCD\ldots}\left(\int_{\infty}^{(x_1,y_1)}\omega - \int_{\infty}^{(x_2,y_2)}\omega\right)$ can be written by modified version of Jacobi inversion formula as shown in Chapter 4 of this paper. In the case $g > 2$ , by seeing the proof of the normal version of Jacobi inversion theorem, it would be difficult to apply the same logic to have the co-ordinate expressions of theta functions, even in the hyper-elliptic cases.

The spin structure independent parts of the decomposition formula will be given as polynomials of seven Pe functions only :

$$P_{11}(\beta),\ P_{12}(\beta),\ P_{22}(\beta),\ P_{111}(\beta),\ P_{112}(\beta),\ P_{122}(\beta),\ P_{222}(\beta)\ . \qquad (5.3)$$

Pe functions with higher derivatives will be re-written by the differential equations to the derivatives of Pe functions with at most 3 indices $P_{ABC}$ . Unlike the spin structure dependent part, 3 indices tensors $P_{ABC}$ are generally non-zero.
These are defined by theta functions, and they also have the following $x, y$ coordinates expressions as given in (4.2.46) – (4.2.53) :

$$P_{11}(\beta_i) = \frac{F(x_i,x_{i+1})+2y_iy_{i+1}}{(x_i-x_{i+1})^2},\ \ P_{12}(\beta_i) = -x_ix_{i+1}\ ,\ P_{22}(\beta_i) = x_i + x_{i+1}\ ,$$

$$P_{111}(\beta_i) = -2\frac{y_{i+1}\psi(x_i,\ x_{i+1})+ y_i\psi(x_{i+1},\ x_i)}{(x_i - x_{i+1})^3}\ ,\ P_{112}(\beta_i) = 2\frac{y_ix_{i+1}^2+ y_{i+1}x_i^2}{x_i - x_{i+1}}$$

$$P_{122}(\beta_i) = -2\frac{y_{i+1}x_i+ y_ix_{i+1}}{x_i - x_{i+1}},\quad P_{222}(\beta_i) = 2\frac{y_i+ y_{i+1}}{x_i - x_{i+1}}$$

For example, the decomposition formula of the case $N = 2$ can be written by using theta functions such as
( noting that $S_\delta(z_1,z_2)S_\delta(z_2,z_1) = -S_\delta(z_1,z_2)^2,\ \beta = \int_{\infty}^{(x_1,y_1)}\omega - \int_{\infty}^{(x_2,y_2)}\omega$ )

$$\begin{aligned} -S_\delta(z_1,z_2)^2 &= [P_{11}(\Omega_\delta) - P_{11}(\beta) + P_{12}(\Omega_\delta)P_{22}(\beta) - P_{12}(\beta)P_{22}(\Omega_\delta)\ ]\ \omega_1(z_1)\omega_1(z_2) \\ &= \sum_{I,J=1}^{2}[\ P_{IJ}(\Omega_\delta) - P_{IJ}(\beta)]\omega_I(z_1)\omega_J(z_2) \end{aligned}$$

which can be re-written in terms of coordinate variables only :

$$= [\frac{F(e_i,e_j)}{(e_i-e_j)^2} - \frac{F(x_1,x_2)+2y_1y_2}{(x_1-x_2)^2} - (x_1 + x_2)e_ie_j + x_1x_2(e_i + e_j)]\omega_1(z_1)\omega_1(z_2)$$

Similar expressions were explicitly derived for $N = 3$, and $N = 4$ in terms of theta functions. Modified Jacobi inversion formulae are therefore key to relating these two expressions within our framework and linking the theta function expressions of spin structure-independent parts in the decomposition formulas to the coordinate expressions for arbitrary $N$. On the other hand, the calculation method obtained as a direct generalization of that of genus one case (denoted $\partial$-calculation) does not provide enough number of equations to determine all the coefficients $H$ for $N > 3$.

As outlined in Section 4-1, the underlying causes of this phenomenon can be attributed to two primary factors. In Chapter2, the $N^2$ number of basis functions for genus 2 is discussed. It is important to acknowledge that these $N^2$ polynomials of the Pe functions and their derivatives constitute basis functions for expanding the periodic functions on the Jacobian variety of genus 2 which is the complex 2-dimensional torus, and are not limited for the functions on the Riemann surface embedded in the Jacobian variety.

Additionally, the direct generalization of the genus 1 method of calculating the coefficients of the expansion, termed "$\partial$- calculation" there, serves as a general tool to obtain the coefficients of the functions on the Jacobian variety. This is achieved by regarding the product $\prod_{i=1}^{N} \frac{\sigma(\alpha+\beta_i)}{\sigma(\beta_i)\sigma(\alpha)}$ as a periodic function of Jacobian variety whose variable being $\alpha$. In carrying out our calculations, the issue is that the Jacobian variety is not isomorphic to the Rieman surface except the case of genus 1.

In our case, the periodicity is ensured by the cyclic condition on the variables of the vertex-inserting points, which are only on the Riemann surface. It is evident that, in the context of our investigation, the product $\prod_{i=1}^{N} \frac{\sigma(\alpha+\beta_i)}{\sigma(\beta_i)\sigma(\alpha)}$ can be expanded by the basis functions described in Chapter 2. The calculations presented in Chapter 4 will yield accurate equations, as no additional or alternative basis functions are required for the expansion of this product as long as the function is periodic in the Jacobian variety in genus 2. Nonetheless, it is conceivable that, in general, it is not possible to ascertain all of the expansion coefficients by means of this $\partial$- calculation method alone. This is because we did not fully utilize the fact that vertex inserting points with the cyclic condition are only on the Rieman surface.

In order to ascertain all of the expansion coefficients $H_{IJ}$, $H_{IJK}$, .... on the right-hand side of equation (4.1.4):

$$\prod_{i=1}^{N} \frac{\sigma(\alpha+\beta_i)}{\sigma(\beta_i)\sigma(\alpha)} = H_0 + \sum_{I,J=1,2} H_{IJ} P_{IJ}(\alpha) + \sum_{I,J,K=1,2} H_{IJK} P_{IJK}(\alpha) + \ \dots\ , \quad (5.4)$$

it is imperative to additionally seek out restrictions as new equations, derived by reflecting the fact that the variables in the functions can only vary over the Riemann surface, not the entire Jacobian variety, for larger values of $N$.
In Chapter 4, it was demonstrated that only even functions of basis functions are necessary for the calculation of coefficients in this instance. However, this is insufficient and further measures are required in the form of new restrictions.

In Section 4-4, as one of such additional conditions, new equations were supplemented by drawing on some of the results of the decomposition formula in [15], and calculations were completed where $N = 4$. The novel condition for $N = 4$ comes from (4.4.50), given in [15] :

$$S_\delta(z_1,z_2)S_\delta(z_2,z_3)S_\delta(z_3,z_4)S_\delta(z_4,z_1) = \frac{1}{2}S_\delta^2(z_1,z_2)\,S_\delta^2(z_3,z_4) + \frac{1}{2}S_\delta^2(z_1,z_4)\,S_\delta^2(z_2,z_3) + other\ terms\ . \qquad (5.5)$$

As for the coefficients of highest degree (degree 2 for $N = 4$) monomials of genus 2 Pe functions of the form $P_{AB}(\alpha)P_{CD}(\alpha)$, since those are included only in the first two terms in the right-hand side of eq.(5.5) with $\alpha$ being set equal to $\Omega_\delta$, all theta function expressions can be calculated by inserting the concrete form of $S_\delta^2(z_i,z_j)$ in the right hand side of (5.5) and expand the terms. The similar calculations can be done for arbitrary even number of $N$ .

Although these equations may sometimes yield solutions that overlap with those obtained by the $\partial$-calculation to determine the coefficients of the decomposition formula, as seen for $H^4_{12,12}$ and $H^4_{1222}$ in (4.4.32) and (4.4.34), it has been verified for the case of $N = 4$ that the solutions from both methods coincide exactly, by means of the modified version of the Jacobi inversion theorem : $H^4_{12,12} = H^{4,HIGHEST}_{12,12}$ and $H^4_{1222} = H^{4,HIGHEST}_{1222}$ . For sufficiently large $N$, the number of equations required to determine all of the coefficients will not be enough again even after the addition of these supplementary conditions. Consequently, this approach cannot be applied to the case of an arbitrary $N$ and is, in a sense, a stopgap measure. In any case, it makes it highly likely that we will be able to perform the calculations for the case where $N = 6$.

If details of the forms of “other terms” in eq.(5.5) and in the corresponding terms in higher $N$ cases can be determined, the decomposition in terms of theta functions will be completed for arbitrary $N$ in genus 2. For example, as for $N = 4$, “other terms” may have the form

$$linear\ combinations\ of\ X_{i,j} * S_\delta^2(z_i,z_j)\ \ plus\ \ X_0 \qquad (5.6)$$

where $X_{i,j}$ and $X_0$ are written by the $x, y$- coordinates of the vertex insertion points, or ideally by the spin structure independent seven Pe functions in (5.3). Once we can perform this procedure for arbitrary even value of $N$, we will not need to use the general but insufficient method, $\partial$- calculation, from the beginning.

To determine "other terms" for arbitrary $N$ may be equivalent to rewrite the results of Chapter 5 of [15] only using $W_2^+(i,j)$ and without using $W_2^-(i,j)$, $Z(i,j)$, with coefficients similar to those of $X_{i,j}$ and $X_0$ in (5.6).

Alternatively, it would be also very interesting if it is possible to construct recursion relations given in [16] for the theta function expression in genus 2.

**($\zeta$ function part)**

Equations (4.3.39) and (4.3.40) are known as formulas that link the $\zeta$ functions to integrals of the second kind differentials. The integral of the second kind differential (4.3.39) must contain a term $-\frac{1}{2}\, P_{222}(\alpha)$ on the right-hand side, to satisfy $\partial_2\zeta_1(\alpha) = \partial_1\zeta_2(\alpha)$. When the variables are in the form of $\beta_i$ and the cyclic condition $z_{N+1} = z_1$ is satisfied, the following relationships are derived as in (4.3.50) and (4.3.51) :

$$\sum_{i=1}^{N}\zeta_1(\beta_i) = \frac{1}{2}\sum_{i=1}^{N}P_{222}(\beta_i)$$

$$\sum_{i=1}^{N}\zeta_2(\beta_i) = 0$$

In deriving these two, the modified Jacobi inversion formulae play an essential role. Based on these two conditions, when $N = 3$, the initial calculation results (4.3.3) – (4.3.12) , which initially appeared to be completely disjointed, become a compact expression of (4.3.23) in terms of theta functions.

These identities demonstrate that the one-time derivative of $\ln\sigma(\beta_i)$ can be expressed using the Abelian function $P_{222}(\beta_i)$ alone. Such identities will appear among multi-variable theta functions under the cyclic conditions in any framework. These conditions reproduce some of the results of the one-time derivatives of the decomposition formula for arbitrary odd values of $N$. For even values of $N$, these are also used to derive some accurate calculation results in Chapter 4 which always reproduce coordinate expressions obtained in [15] from the theta function expressions.

**(open issues )**

The most immediate task would be to perform the calculations for $N = 6$ using the method described in this paper. Since calculations would be extensive, they have not yet been completed. A computer calculation might be the most suitable approach.

The most important challenge is to generalize the results obtained here to an arbitrary value of $N$, although it is unclear whether this is mathematically possible or not. A promising approach to achieving this goal may be to establish a connection between the decomposition formula described in 6-1 of [15] and the theta function expression by means of rigorous logic, as delineated above. As pole structures are all determined as described in [15], it may be possible to derive theta function expressions for arbitrary $N$ at the genus 2 level in a natural and elegant way. At this juncture, the solution is not readily apparent to the author.

If such a solution is found, the results obtained using the framework described in this paper will extend the genus 1 decomposition formula of the product of fermion correlation functions with cyclic condition to the genus 2 case in a simple and intuitive manner.

**Appendix A**

**(Notations of genus one )**

Half periods $\Omega_\delta$ are related to moduli parameter $\tau$ as : $\Omega_1 = \frac{1}{2}, \quad \Omega_2 = -\frac{1+\tau}{2}, \quad \Omega_3 = \frac{\tau}{2}$ (A.1)

Periods: $A_{m,n} = 2m\Omega_1 + 2n\Omega_3$ (A.2)

where $m, n$ are integers.

Starting from the infinite product representation of sigma function at genus 1,

$$\sigma(\alpha) = \alpha \cdot \prod'_{m,n} \left(1 - \frac{\alpha}{A_{m,n}}\right) exp\left[\frac{\alpha}{A_{m,n}} + \frac{\alpha^2}{2{A_{m,n}}^2}\right] \tag{A.3}$$

taking ln of sigma function and expanding the terms which do not have singular part at $\alpha = 0$ , we have

$$\ln \sigma(\alpha) = \ln \alpha - \sum_{k=2}^{\infty} \frac{G_{2k}}{2k} \alpha^{2k} \tag{A.4}$$

where $G_{2k}$ are holomorphic Eisenstein series defined as

$$G_{2k}(\tau) \equiv \sum_{(m,n)\neq(0,0)} A_{m,n}^{-2k} \tag{A.5}$$

The unique odd theta function on the torus can be written as, adding $G_2$ ,

$$\ln \theta_1(\alpha) = \ln \sigma(\alpha) - \frac{G_2}{2}\alpha^2 + \ln\theta_1^{(1)}(0) = \ln \alpha - \sum_{k=1}^{\infty} \frac{G_{2k}}{2k} \alpha^{2k} + \ln\theta_1^{(1)}(0) \tag{A.6}$$

Also, we define

$$\varsigma(z) = \frac{d \ln \sigma(z)}{dz} \tag{A.7}$$

$$P(\alpha) \equiv -\frac{d^2 \ln \sigma}{d\alpha^2} = -2\eta_1 - \frac{\partial^2}{\partial \alpha^2} \ln \theta_1(\alpha, \tau) = \alpha^{-2} + \sum_{m=1}^{\infty}(2m+1)G_{2m+2}(\tau)\alpha^{2m} \tag{A.8}$$

$$P^{(2n)}(\alpha) = \frac{(2n+1)!}{\alpha^{2n+2}} + (2n+1)!\, G_{2n+2}(\tau) + O(\alpha)\ldots\ldots \tag{A.9}$$

The values of Pe function at half periods are the branch points of the curve:

$$e_\delta = P(\Omega_\delta) \qquad (\delta = 1,2,3) \tag{A.10}$$

The branch points $e_\delta$ are related to theta constants as

$$e_1 = \frac{\pi^2(\theta_4^4(0)+\theta_3^4(0))}{3} \; e_2 = \frac{\pi^2(\theta_2^4(0)-\theta_4^4(0))}{3} \; e_3 = \frac{-\pi^2(\theta_3^4(0)+\theta_2^4(0))}{3} \tag{A.11}$$

and

$$e_1 + e_2 + e_3 = 0 \quad e_1e_2 + e_2e_3 + e_3e_1 = -\frac{g_2}{4} \quad e_1e_2e_3 = \frac{g_3}{4} \tag{A.12}$$

Also, we use

$$\eta_\delta = \varsigma(\Omega_\delta) \quad (\delta = 1,2,3). \tag{A.13}$$

$\eta_1$ is related to appropriately regularized form of $G_2(\tau)$ as

$$G_2 = 2\eta_1 \tag{A.14}$$

The $g_2, g_3$ are classical notations of modular forms which are related to holomorphic Eisenstein series $G_{2k}(\tau)$ as

$$g_2 = 60 \sum_{m,n} A_{m,n}^{-4} \qquad g_3 = 140 \sum_{m,n} A_{m,n}^{-6} \tag{A.15}$$

$$P^{(2n)}(z) = \frac{d^{2n}P(z)}{dz^{2n}} = polynomial\ of\ P(z)\ of\ degree\ \ n+1 \tag{A.16}$$

$$P^{(2n+1)}(z) = P^{(1)}(z) * [polynomial\ of\ P(z)\ ofdegree\ n\ ] \tag{A.17}$$

$$P^{(1)}(\Omega_\delta) = 0 = P^{(ODD)}(\Omega_\delta) \tag{A.18}$$

A few examples of (A.16) are:

$P^{(2)} = 6P^2 - \frac{1}{2}g_2$

$P^{(4)} = 120P^3 - 18g_2P - 12g_3$

This series of polynomials is consecutively constructed by differentiating

$\{P^{(1)}(z)\}^2 = 4\{P(z)\}^3 - g_2P(z) - g_3$

The coefficient of the highest degree term in $P^{(2n-2)}$ is $(2n-1)!$ .

$$P^{(2n-2)} = (2n-1)!\,P^{(n)} + \cdots$$

**(some spin sum results in genus one)**

We denote the Pe function in genus 1 as $P$ ; in the below the variables in $P$ are arbitrary. We abbreviate writing the variables. We write the differential equation of $P$

$$\{P^{(1)}\}^2 = 4P^3 - g_2P - g_3 = 4(P^3 - aP - b) \tag{A.19}$$

where we defined $a$ and $b$ as

$$a = \frac{g_2}{4} = 15G_4\ , \quad b = \frac{g_3}{4} = 35G_6 \quad . \tag{A.20}$$

We first note the following three simple facts :

1) From the relation $e_\delta^3 - ae_\delta - b = 0$, we have $e_\delta^{K+1} - ae_\delta^{K-1} - be_\delta^{K-2} = 0$ , which can be regarded as the trilinear relation at genus 1. $f(K)$ is defined in (1.17) as

$$f(K) = \frac{(e_2-e_3)e_1^K + (e_3-e_1)e_2^K + (e_1-e_2)e_3^K}{(e_1-e_2)(e_2-e_3)(e_3-e_1)} \quad .$$

Simply using $e_\delta^{K+1} = a e_\delta^{K-1} + b e_\delta^{K-2}$ to the form of $f(K)$, we have (1.18),

$f(K+1) = af(K-1) + bf(K-2)$ .

Using the trilinear equation is equivalent to using this recurrence relation in genus one, in the case of external massless bosons amplitudes in Type I and II superstring theories.

2) To have a recurrence relation of $f(K)$ only, 1) is enough. The following description and the result of all of $A_K$, $B_K$, $C_K$ defined below may be useful in the case of calculating spin sums in the Heterotic string.

Divide the monomial $P^K$ by $4(P^3 - aP - b)$ , and write

$$P^K = 4(P^3 - aP - b)L(P) + A_K P^2 + B_K P + C_K \qquad \text{(A.21)}$$

where $L(P)$, $A_K$, $B_K$, $C_K$ are defined in this equation.

Multiply $P$ one more time, we have

$$P^{K+1} = 4(P^3 - aP - b)L(P)P + A_K(P^3 - aP - b) + B_K P^2 + (aA_K + C_K)\, P + bA_K \qquad \text{(A.22)}$$

Therefore, by seeing the last 3 terms in (A.22),

$$A_{K+1} = B_K, \quad B_{K+1} = aA_K + C_K, \quad C_{K+1} = bA_K \,. \qquad \text{(A.23)}$$

By these, we have

$$A_{K+1} = aA_{K-1} + bA_{K-2} \;. \qquad \text{(A.24)}$$

When $A_K$ are determined, $B_K$ and $C_K$ are all determined.

Also, each of $B_K$ and $C_K$ satisfy the same recurrence relations as that of $A_K$ , as is easily confirmed. The only difference between $A_K$, $B_K$, $C_K$ is in their initial term values.

We set $A_0 = B_0 = C_0 = 0$, and we see

$A_1 = 0,\; A_2 = 1, A_3 = 0,\; A_4 = a,$

$B_1 = 1,\; B_2 = 0, B_3 = a,\; B_4 = b,$

$C_1 = 0,\; C_2 = 0, C_3 = b,\; C_4 = 0.$

(A.24) is a recurrence relation among 4 terms ( the coefficient of $A_K$ is zero), and there is a standard way to have a general solution.

If the roots of the characteristic equation of (A.24):

$X^3 - aX - b = 0$ are denoted as $\alpha, \beta, \gamma$ , the solution is

$$A_K = \frac{\Phi(\alpha,\beta,\gamma)+\Phi(\beta,\gamma,\alpha)+\Phi(\gamma,\alpha,\beta)}{(\alpha-\beta)(\beta-\gamma)(\gamma-\alpha)} \qquad \text{(A.25)}$$

where

$$\Phi(x,y,z) = (x-y)z^{K-1}(A_3 - (x+y)A_2 + xyA_1) \,. \qquad \text{(A.26)}$$

The roots of the characteristic equation in our case are nothing but the branch points $e_1, e_2, e_3$ , then (A.25) becomes

$$A_K = \frac{(e_2-e_3)e_1^K+(e_3-e_1)e_2^K+(e_1-e_2)e_3^K}{(e_1-e_2)(e_2-e_3)(e_3-e_1)} = f(K) \qquad \text{(A.27)}$$

Then $f(K)$ satisfies the relation (A.24).

The argument here has nothing to do with string theory. The $A_K$ , defined in (A.21) as the coefficient of the quadratic term in the remainder when the monomial $P^K$ is divided by $P^3 - aP - b$ for an arbitrary variable, is equal to the result of multiplying the partition functions of superstring theories for the massless external bosons by $e_\delta^K$ $(= P^K(\Omega_\delta))$ and summing them.

3) More directly, multiply $\frac{(e_2-e_3)}{(e_1-e_2)(e_2-e_3)(e_3-e_1)}$, $\frac{(e_3-e_1)}{(e_1-e_2)(e_2-e_3)(e_3-e_1)}$, $\frac{(e_1-e_2)}{(e_1-e_2)(e_2-e_3)(e_3-e_1)}$

on the both side of (A.21) and sum over and set the variable of $P$ to the half period $\Omega_\delta$, we have

$$\frac{(e_2-e_3)e_1^K+(e_3-e_1)e_2^K+(e_1-e_2)e_3^K}{(e_1-e_2)(e_2-e_3)(e_3-e_1)} = A_K \qquad \text{(A.28)}$$

In actual calculations of the spin sum, as described in the Chapter1, we need to rewrite $P^{(2K-2)}(\Omega_\delta)$ into polynomials of $e_\delta$ , which requires cumbersome calculations for large values of $K$ . In genus 1, it is possible to adopt monomials of Pe functions $P$ , $P^2$, $P^3$, ... (and the terms multiplied by one-time derivative of Pe, $P^{(1)}$, for each of these monomials) as basis functions of the expansion, instead of the Pe function and its derivatives. In that case, the spin sum will take the simple form of $f(K)$ . However, as a trade-off fact, in this case the decomposition formula becomes very complicated and cannot be written in the simple form of eq.(1.12) nor eq.(1.14).

As is clear from the above argument, using trilinear relations is equivalent to use the recurrence relations (A.24) by adopting $P$, $P^{(1)}$, $P^2$, $P^{(1)}P^2$, $P^3$, ... as basis functions. In genus 2 we need to adopt this, and the result of the decomposition formula will not have a simple form.

Define the spin sum result of the external massless amplitudes in type I and II models as

$$s_{2n} = \frac{1}{(2n+1)!}\frac{(e_2-e_3)P^{(2n)}(\Omega_1)+(e_3-e_1)P^{(2n)}(\Omega_2)+(e_1-e_2)P^{(2n)}(\Omega_3)}{(e_1-e_2)(e_2-e_3)(e_3-e_1)} \qquad \text{(A.31)}$$

Since

$$P^{(2n)} = \frac{d^{2n}P(z)}{dz^{2n}} = polynomial\ of\ P\ of\ degree\ \ n+1 \ ,$$

in the calculation of $s_{2n}$ , the $f(K)$ appears up to $K = 0, 1, \dots n+1$ .

The author has tried to develop an efficient method of calculating $s_{2n}$ sometimes in the past, but have never succeeded. Here we have to adopt the following naïve approach.

Differentiating $2n-3$ times the both side of $P^{(3)} = 12P \cdot P^{(1)}$ and setting the variable equal to $\Omega_\delta$, we have

$$\frac{P^{(2n)}(\Omega_\delta)}{(2n+1)!} = \frac{12}{(2n+1)!} \sum_{k=0}^{n-2} \binom{2n-3}{2k} P^{(2k)}(\Omega_\delta)\, P^{(2n-2k-2)}(\Omega_\delta) \qquad \text{(A.32)}$$

We will use this relation iteratively to obtain $\frac{P^{(2n)}(\Omega_\delta)}{(2n+1)!}$ expressed as a polynomial of $P(\Omega_\delta)$ for higher values of $n$. After some calculations, we have

$s_0 = 0$

$s_2 = 1$ ( corresponds to $N \geq 4$ point amplitudes )

$s_4 = 0$ ( corresponds to $N \geq 6$ point amplitudes )

$s_6 = \frac{1}{5}a = 3G_4$ ( corresponds to $N \geq 8$ point amplitudes )

$s_8 = \frac{2}{7}b = 10G_6$

$s_{10} = \frac{2}{25}a^2 = 18{G_4}^2 = 42\, G_8$

$s_{12} = \frac{24}{11\times5\times3}ab = \frac{24\times35}{11}G_4G_6 = 168G_{10}$

$s_{14} = \frac{1}{5^3\times7^2\times3\times13}(127\cdot7^2\cdot a^3 + 5^3\cdot3^2\cdot19b^2) = \frac{127\times3^2}{13}{G_4}^3 + \frac{3\times5^2\times19}{13}{G_6}^2$

$s_{16} = \frac{23\times6}{5^2\times7\times11}a^2b = \frac{23\times6\times3\times15}{11}{G_4}^2G_6 = 23\times3^2\times13\, G_{14}$ ( corresponds to $N \geq 18$ point amplitudes )

where we used identities among holomorphic Eisenstein series

$G_8 = \frac{3}{7}{G_4}^2$ , $G_{10} = \frac{5}{11}G_4G_6$ , $G_{14} = \frac{30}{13\times11}{G_4}^2G_6$

By considering dimension of the functional space of holomorphic Eisenstein series, $s_6$, $s_8$, $s_{10}$, $s_{12}$, $s_{16}$ are proportional to $G_4$, $G_6$, $G_8$, $G_{10}$, $G_{14}$. However, $s_{14}$ and $s_{2n}$ for $2n > 16$, the results will be complicated polynomials of $G_4$ and $G_6$ ( or polynomials of other holomorphic Eisenstein series using the identities) in general.

**Acknowledgements**

The author is very grateful to the reviewer for his/her extremely helpful comments and for pointing out my misunderstandings and numerous errors. The author also thanks to Prof. A. Morozov for his kind help in the process of submitting [8] to arXiv. I am sincerely grateful to Professors Kozo Kuchitsu, C. Iso, N. Sakai, and T. Yoneya for their warm and kind guidance during my student days.

I am also grateful to Masumi & Masako & Yoshiaki T. and Shara-Ni for their help.

A.G.Tsuchiya email colasumi_0506@yahoo.co.jp